\documentclass[fleqn,usenatbib]{mnras}

\usepackage{newtxtext,newtxmath}

\usepackage[T1]{fontenc}

\DeclareRobustCommand{\VAN}[3]{#2}
\let\VANthebibliography\thebibliography
\def\thebibliography{\DeclareRobustCommand{\VAN}[3]{##3}\VANthebibliography}

\usepackage{graphicx}	
\usepackage{amsmath}	
\usepackage[flushleft]{threeparttable}
\usepackage{booktabs}
\usepackage{pdflscape}

\newcommand{\ms}{\,$\text{m}\,\text{s}^{-1}\,$}	
\newcommand{\kms}{\,$\text{km}\,\text{s}^{-1}$}	

\newcommand{\rhk}{\,$\log R'_{HK}$\,}

\title[The HD 85426 system]{HARPS-N, TESS, and CHEOPS\thanks{This article uses data from the CHEOPS Guaranteed Time Observation programme CH\_PR100024.} discover a transiting sub-Neptune and two outer companions around the bright solar analogue HD~85426}

\author[F. Lienhard et al.]{F. Lienhard,$^{1,2}$
A.~Mortier,$^{3}$
A.~Collier~Cameron,$^{4}$
M.~Cretignier,$^{5}$
L.~Borsato,$^{6}$
A.~Anna~John,$^{3,4}$
\newauthor J.~A.~Egger,$^{7}$
M.~Stalport,$^{8,9}$
T.~G.~Wilson,$^{10}$
A.~Deline,$^{11}$
A.~Fortier,$^{7,12}$
D.~W.~Latham,$^{13}$
L.~Malavolta,$^{6,14}$
\newauthor P.~F.~L.~Maxted,$^{15}$
S.~G.~Sousa,$^{16}$
S.~L.~Grimm,$^{17,18}$
L.~Buchhave,$^{19}$
Y.~Alibert,$^{12,7}$
B.~S.~Lakeland,$^{20,3}$
\newauthor X.~Dumusque,$^{36}$
J.~Cabrera,$^{21}$
L.~Naponiello,$^{22}$
A.~C.~M.~Correia,$^{23}$
F.~Rescigno,$^{20,3}$
L.~Fossati,$^{24}$
\newauthor A.~Sozzetti,$^{22}$
R.~Alonso,$^{25,26}$
T.~Bárczy,$^{27}$
D.~Barrado,$^{28}$
S.~C.~C.~Barros,$^{16,29}$
W.~Baumjohann,$^{24}$
W.~Benz,$^{7,12}$
\newauthor N.~Billot,$^{11}$
A.~Brandeker,$^{30}$
C.~Broeg,$^{7,12}$
K.~Collins,$^{13}$
Sz.~Csizmadia,$^{21}$
P.~E.~Cubillos,$^{24,31}$
M.~B.~Davies,$^{32}$
\newauthor M.~Deleuil,$^{33}$
O.~D.~S.~Demangeon,$^{16,29}$
B.-O.~Demory,$^{12,34,7}$
A.~Derekas,$^{35}$
B.~Edwards,$^{37}$
\newauthor D.~Ehrenreich,$^{11,38}$
A.~Erikson,$^{21}$
M.~Fridlund,$^{39,40}$
D.~Gandolfi,$^{41}$
K.~Gazeas,$^{42}$
M.~Gillon,$^{9}$
M.~Güdel,$^{43}$
\newauthor M.~N.~Günther,$^{44}$
R.~Haywood,$^{20}$
A.~Heitzmann,$^{11}$
Ch.~Helling,$^{24,45}$
K.~G.~Isaak,$^{44}$
J.~M.~Jenkins,$^{46}$
\newauthor L.~L.~Kiss,$^{47,48}$
J.~Korth,$^{11}$
K.~W.~F.~Lam,$^{21}$
J.~Laskar,$^{48}$
A.~Lecavelier~des~Etangs,$^{49}$
A.~Leleu,$^{11,7}$
\newauthor M.~Lendl,$^{11}$
D.~Magrin,$^{6}$
A.~F.~Mart\'inez~Fiorenzano,$^{51}$
B.~Merín,$^{52}$
C.~Mordasini,$^{7,12}$
V.~Nascimbeni,$^{6}$
\newauthor G.~Olofsson,$^{30}$
H.~P. Osborn,$^{12,1}$
R.~Ottensamer,$^{43}$
I.~Pagano,$^{53}$
L.~Palethorpe,$^{54,55}$
E.~Pallé,$^{25,26}$
G.~Peter,$^{21}$
\newauthor D.~Piazza,$^{56}$
G.~Piotto,$^{6,14}$
D.~Pollacco,$^{10}$
D.~Queloz,$^{1,2}$
R.~Ragazzoni,$^{6,14}$
N.~Rando,$^{44}$
H.~Rauer,$^{21,57}$
\newauthor I.~Ribas,$^{58,59}$
K.~Rice,$^{54,55}$
N.~C.~Santos,$^{16,29}$
G.~Scandariato,$^{53}$
D.~Ségransan,$^{11}$
A.~E.~Simon,$^{7,12}$
\newauthor A.~M.~S.~Smith,$^{21}$
S.~Sulis,$^{33}$
Gy.~M.~Szabó,$^{35,60}$
S.~Udry,$^{11}$
S.~Ulmer-Moll,$^{61,8}$
V.~Van~Grootel,$^{8}$
\newauthor J.~Venturini,$^{11}$
E.~Villaver,$^{25,26}$
N.~A.~Walton,$^{62}$
T.~Zingales,$^{14,6}$\\
Affiliations are listed at the end of the paper
}

\date{Accepted 2025 October 20. Received 2025 September 19; in original form 2025 June 13}

\pubyear{2024}

\begin{document}
\label{firstpage}
\pagerange{\pageref{firstpage}--\pageref{lastpage}}
\maketitle

\begin{abstract}
We provide a detailed characterisation of the planetary system orbiting HD~85426 (TOI-1774). This bright G-type star ($M_{\ast}$: 0.99~$\text{M}_{\odot}$; $R_{\ast}$: 1.13~$\text{R}_{\odot}$; age: 7.4~Gyr; V mag: 8.25) hosts a transiting sub-Neptune, HD~85426~b, with an orbital period of 16.71 days and a blackbody equilibrium temperature of $824^{+11}_{-11}$ K. By jointly analysing HARPS-N RVs, {\it TESS}, and {\it CHEOPS} photometric data and using two different stellar activity mitigation techniques, we constrain planet b's mass to $6.0^{+1.5}_{-1.6}$~$\text{M}_{\oplus}$ and $8.5^{+1.3}_{-1.4} $~$\text{M}_{\oplus}$, depending on the mitigation technique. We investigate the dependence of these results on the priors, data selection, and inclusion of other Keplerians in the modelling. Using this approach, we identify the presence of two non-transiting planetary companions with minimum masses near 10~$\text{M}_{\oplus}$ and orbital periods of 35.7 and 89 days. Additionally, we reject the initial hypothesis that the 35.7-day periodic signal was due to stellar activity.
We also determine HD~85426~b's radius to be $2.78^{+0.05}_{-0.04}$~$\text{R}_{\oplus}$ and compute a transmission spectroscopy metric in the range of 82 to 115, making this planet a highly valuable target for atmospheric characterisation.
\end{abstract}




\section{Introduction}
The synergy between radial velocity (RV) instruments on the ground and photometric satellites, such as the {\it Kepler} space telescope \citep{Borucki_2010} or the Transiting Exoplanet Survey Satellite \citep[{\it TESS};][]{Ricker_2015}, has enabled the precise characterisation of the mass, radius, and orbital parameters of numerous planets \citep[e.g.][]{Teske_2021,Chontos_2022,Bonomo_2025}. These analyses are essential for atmospheric characterisation, e.g., with the James Webb Space Telescope \citep[{\it JWST};][]{Gardner_2006} and provide target information for future missions. The combination of planetary mass and radius information, together with stellar host properties, allows us to draw conclusions about the interior composition of the planets \citep[e.g.][]{Zeng_2008} and to probe planetary formation and evolution mechanisms.

In this study, we analyse the planetary system orbiting the solar-type star HD~85426. This star hosts a transiting planet, HD~85426~b \citep{Giacalone_2021}, which belongs to the class of sub-Neptune planets. These planets are of particular interest because they represent a very common class of planets \citep[e.g.][]{Fulton_Petigura_2018}, yet their properties are still debated. With no equivalent in the Solar System, sub-Neptunes are typically defined and characterised by their distribution in the radius-period diagram. In this diagram, the sub-Neptunes sit just above the radius valley, which is located around 1.5--2 R$_{\oplus}$ \citep{Fulton_2017,VanEylen_2018} and separates sub-Neptunes from the smaller super-Earths \citep[e.g.][]{Bean_2021}. The dearth of planets in the radius valley occurs prominently for solar-type stars such as HD~85426 and is an active topic of research \citep[e.g.][]{Bean_2021,Parc_2024}. The mechanisms proposed to explain the origin of the radius valley are linked to the composition of super-Earths and sub-Neptunes. However, the latter occupy a degenerate space in the mass-radius diagram, meaning that different compositions can account for their bulk densities. 
According to one model, the bulk densities of the sub-Neptunes could be explained by a solid rock/iron core with a primordial H/He rich atmosphere \citep[e.g.][]{Lopez_Fortney_2014,Benneke_2019,Rogers_2023}. In this case, the observed radius gap between sub-Neptunes and super-Earths is mainly thought to be due to photoevaporation \citep[e.g.][]{Owen_2017,Jin_2018} and core-powered mass loss \citep{Ginzburg_2018,Gupta_2019}, stripping the atmospheres of lower mass planets, whereas cooler, more massive planets retain their primordial atmospheres. 
Alternatively, sub-Neptunes' bulk densities can be due to a water-rich composition with a steam atmosphere, in which case the super-Earths' smaller radii are thought to be due to a lower water content \citep[e.g.][]{Leger_2004,Mousis_2020,Aguichine_2021,Burn_2024}, with photoevaporation playing a critical role in shaping the radius valley \citep{Venturini_2020,Burn_2024}.

To advance the study of sub-Neptunes, we conducted a detailed analysis of the sub-Neptune HD~85426~b (also known as TOI-1774~b), along with its planetary system and host star. We gathered spectra of HD~85426 with HARPS-N \citep{Cosentino_2012} to analyse the RV and activity indicator time series and characterise the star. 

In addition to the available {\it TESS} data, we obtained observations with the CHaracterising ExOPlanet Satellite \citep[{\it CHEOPS};][]{Benz_2021,Fortier_2024} to refine the characterisation of the transiting planet and search for potential transit timing variations.
The spectral and photometric data were analysed jointly to estimate the mass and radius of HD~85426~b and search for planetary companions. We applied two independent stellar activity mitigation techniques to the RVs and tested the dependence of our inferred masses and orbital parameters on various methodological choices.
Our results highlight the need to investigate to what degree the identification and characterisation of planets is affected by the activity mitigation, the selection of data, or the priors.
By applying an ensemble of methods, we derive an accurate mass range for the transiting planet. This approach is in the spirit of the findings that stellar activity is very challenging to mitigate to date \citep[e.g.][]{Crass_2021,Zhao_2022} and there can be a significant dependence of the inferences on the chosen priors \citep[e.g.][]{Osborne_2025}.

This study is structured as follows. Section \ref{Data} presents the collected data and describes the processing methods. The properties of the host star are detailed in Section \ref{StellarCharacterisation}. Section \ref{RVModelling} outlines the modelling of the RV signatures of planet b and other signals using various techniques. We conclude about the existence of massive, long-period outer planets in Section \ref{lpps} and search for Transit Timing Variations in Section \ref{TTVs}. The results of the stellar characterisation and the joint RV and photometric modelling are used in Section \ref{InteriorComposition} to constrain the planetary properties. Finally, suitability for atmospheric follow-up observations is evaluated in Section \ref{sec:atmo_follow_up} and our results are summarised in Section \ref{Discussion}.

\section{Data} \label{Data}
The dataset analysed in this study includes space-based photometric observations described in Sections \ref{photo1} and \ref{photo2}, as well as ground-based spectroscopic measurements described in Section \ref{hndata}.

\subsection{\textit{TESS} photometry} \label{photo1}
In January 2020, {\it TESS} captured two transit-like events in the light curve of the bright G-star HD~85426 in sector 21. This target was subsequently upgraded from {\it TESS} Input Catalog object 4897275 (TIC 4897275) \citep{Stassun_2018} to {\it TESS} Object of Interest 1774 (TOI-1774).
The observations were processed by the Science Processing Operation Center (SPOC) pipeline \citep{Jenkins_2016} at NASA Ames Research Center, which detected the transits with a noise-compensating matched filter \citep{Jenkins_2002,Jenkins_2010,Jenkins_2020}, were fitted with an initial limb-darkened transit model \citep{Li_2019}, and passed the suite of diagnostic tests \citep{Twicken_2018}, including the difference image centroiding test, which located the host star to within 5.3$\pm$2.6 arcsec of the transit source. The data validation results were reviewed by the TESS Science Office at MIT and were alerted to the public on 12 March 2020 \citep{Guerrero_2021}.
The transiting planetary companion of HD~85426 was statistically validated in \citet{Giacalone_2021}, ruling out other transit-producing scenarios.  
The star was reobserved in {\it TESS}'s sector 48 in 2022, which remains the last observation by {\it TESS} until at least September 2026. Each sector is observed for two successive orbits of the spacecraft. In the middle of the sector's time series, at orbit perigee, the data are downlinked to Earth, producing a gap in the light curve \citep{Ricker_2015}. 
In this analysis, we used the 2-minute cadence Presearch Data Conditioning Simple Aperture Photometry flux (PDCSAP) light curves, which are corrected for instrumental systematics \citep{Smith_2012,Stumpe_2012,Stumpe_2014}. The data were retrieved from the MAST data archive\footnote{Mikulski Archive for Space Telescopes, \url{https://archive.stsci.edu/missions-and-data/tess}.} using the Python package \texttt{Lightkurve} \citep{lightkurve_2018}.

\begin{figure}
    \centering
    \includegraphics[width=0.99\columnwidth]{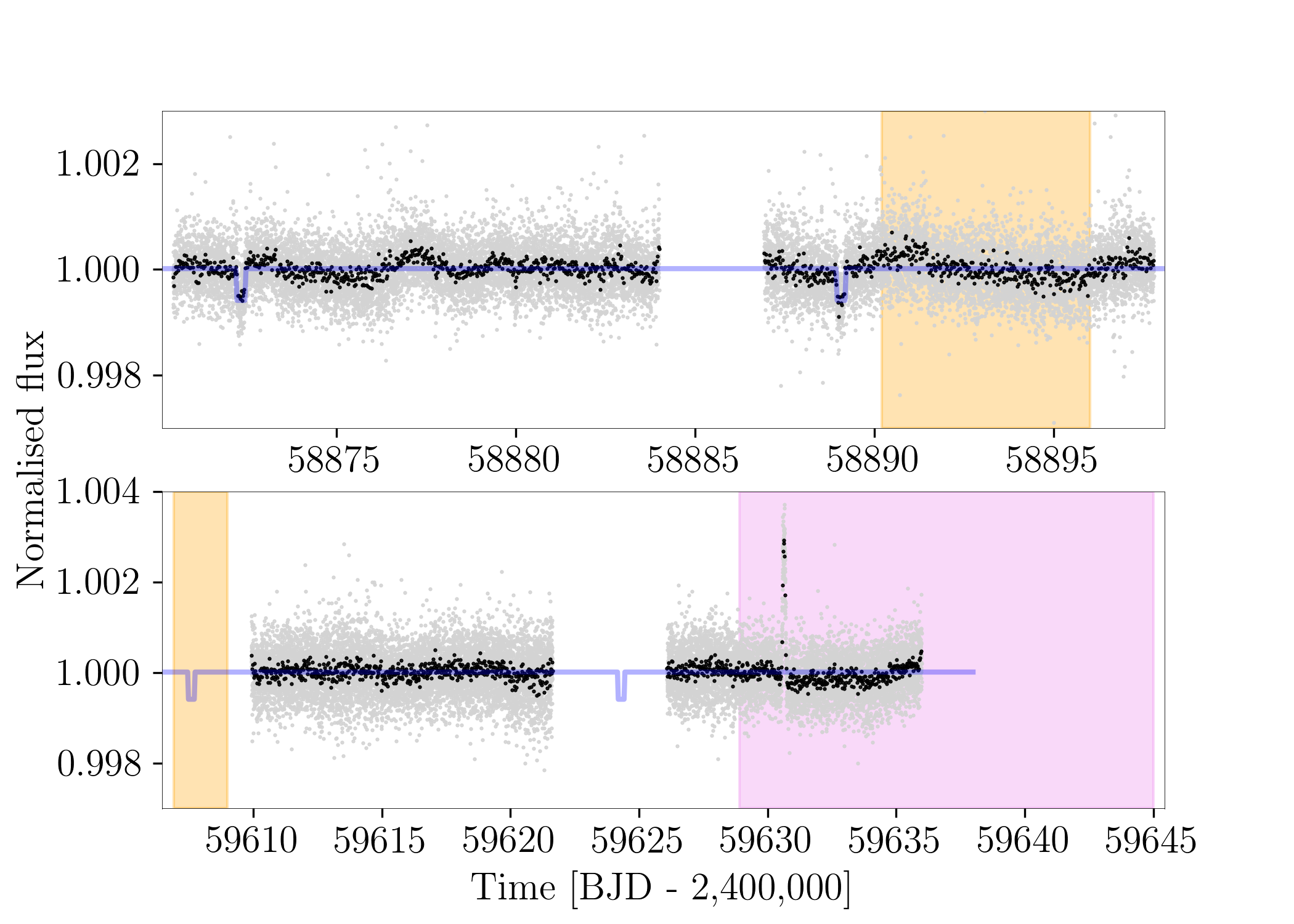}
    \caption{Photometric 2-minute cadence PDCSAP time series of TIC 4897275 (HD~85426) in {\it TESS} sectors 21 (top panel) and 48 (bottom panel) are shown in grey, with 30-minute binned data overlaid in black. The transits of HD~85426~b are indicated by the blue box transits. The rise in flux shortly after BJD 2,459,630 is caused by a passing asteroid. The predicted conjunction time windows for planet c and planet candidate d are shaded in orange and violet, respectively. The width of these conjunction time windows was set to twice the predicted uncertainty.}
    \label{fig:tesslc}
\end{figure}

Two transits of HD~85426~b were captured in sector 21, whereas the planetary transit occurred in the gap in the middle of the light curve of sector 48, as shown in Fig. \ref{fig:tesslc}. Therefore, there are no recorded transits in sector 48. The increased flux near BJD 2,459,630 is caused by a secondary object passing by the target star. Propagating the orbits of all sufficiently bright objects recorded in the Minor Planet Center\footnote{IAU Minor Planet Center, \url{https://www.minorplanetcenter.net}.} database to the time of the flux peak, we identify this object as the asteroid 581 Tauntonia, orbiting the Sun at about 3.2 AU in the outer region of the asteroid belt, as detailed in Appendix \ref{appendix:otherobj}.

\subsection{{\it CHEOPS} photometry} \label{photo2}
The first transit was successfully recovered by {\it CHEOPS} on 4 April 2022 at a cadence of 1 min. The star was reobserved on 13 and 30 January 2023, and 21 March 2023 at the same cadence. These observations were made under the {\it CHEOPS} Guaranteed Time Observation (GTO) programme CH\_PR100024 and are listed in Table \ref{tab:cheops_obs}. 
 
The {\it CHEOPS} data were reduced with Version 13.0 of the {\it CHEOPS} data reduction pipeline \citep{Hoyer2020A&A...635A..24H}, using the default aperture of 25~px, and detrended individually for each of the four visits with \texttt{pycheops} \citep{Maxted_2022}. Simultaneously with the transit fit, we detrended against first and second-order sinusoidal fits (i.e. $\sin\phi,\ \cos\phi,\ \sin(2\phi),\ \cos(2\phi)$) to the spacecraft roll angle $\phi$ and a linear trend in time. 
This correction is necessary because CHEOPS is in a sun-synchronous nadir-locked orbit, which means the field-of-view rotates once per 98 minutes, resulting in modulations in flux as a function of roll angle and other parameters.
Detrending against background, contamination by neighbouring stars, CCD smear due to nearby bright stars, and a thermal ramp were investigated and determined not to be necessary. An initial fit with the Levenberg-Marquardt algorithm was used as the starting point for computing the posterior probability distributions for all fitting parameters using the affine-invariant Markov-chain Monte Carlo sampler {\sc emcee} \citep{emcee}. The five detrending vectors were subsequently scaled by the mean values of their posterior distributions, and subtracted from the original flux light curves. These detrended light curves were used for further analysis. 

\begin{table*}
    \centering
    \begin{tabular}{c|c|c|c|c|c}
            ID  & Start date & Duration & File key  & Efficiency  & Planet  \\
                & [UTC]      & [h]      &           & [\%]            &         \\
                \hline \\
             1  & 2022-04-04T08:51:18           &  26.77        & CH\_PR100024\_TG015001\_V0200          & 60.7            & b       \\
             2  & 2023-01-13T11:22:37           &  24.75        & CH\_PR100024\_TG015002\_V0200          & 55.7            & b       \\
             3  & 2023-01-30T07:08:17           &  25.57        & CH\_PR100024\_TG015003\_V0200          & 59.4            & b       \\
             4  & 2023-03-21T09:03:18           &  25.25        & CH\_PR100024\_TG016001\_V0200          & 60.2            & b       \\
    \end{tabular}       
    \caption{Log of {\it CHEOPS} observations.}
    \label{tab:cheops_obs}
\end{table*}

\subsection{HARPS-N spectroscopy} \label{hndata}
HARPS-N is a high-precision, pressure- and temperature-stabilised, cross-dispersed echelle spectrograph installed at the Telescopio Nazionale Galileo in the Canary Islands. This spectrograph produces intensity spectra in the wavelength range of 383 to 690 nm, with a spectral resolution of R = 115,000.

The HARPS-N collaboration initiated a radial velocity follow-up campaign within the HARPS-N GTO programme to further characterise the transiting planet, measuring its mass and orbital parameters. HARPS-N observed HD~85426 in three observing seasons, with the first observation on 20 December 2020 and the last on 17 April 2023. In total, 151 HARPS-N spectra were taken over 141 nights, with a median exposure time of 15 min. The mean SNR in order 50 (wavelength range between about 5690 and 5740 \AA) is 133. 

There was an instrumental issue in May 2021 during the first season of observations. The impact of this issue is visible in the RV time series of other stars, such as the HARPS-N standard star HD 127334 or HD 152843 \citep{Nicholson_2023}. HD 127334 shows an anomalous RV increase between 8 and 11 May 2021, with no observations directly before or after these dates. This issue was attributed to a problem with the guiding system that tracks the star. We inspected the tracking images for the observations of HD~85426 around the relevant period, finding strong brightness asymmetries from 7 to 11 May, consistent with the diagnosed issue. All 8 observations taken during this period were removed from the data set, with 143 spectra taken over 137 nights remaining in the set.

\subsubsection{DRS CCF RVs} \label{1774ccfrvs}

Spectra, cross-correlation function (CCF) profiles, and CCF RVs were extracted with the HARPS-N Data Reduction System (DRS) version 3.0.1, which was adapted from the ESPRESSO pipeline \citep{Dumusque_2021} using the G2 mask.
The standard deviation of these RVs is 4.10 \ms, and the mean uncertainty is 0.84 \ms. The DRS pipeline also computes the standard activity indicators, i.e. the Full Width at Half Maximum (FWHM), contrast, and bisector inverse slope (BIS) of the CCF as well as the S-index. For the stellar activity indicators, there was a clear offset between the first and second observing seasons. We removed this offset by splitting the activity time series where the offset occurs and median-normalised both parts separately.

\subsubsection{\texttt{YARARA} RVs}

\texttt{YARARA} \citep{Cretignier_2021} is a post-processing pipeline for high-resolution spectra producing improved RV time series. One of its main objectives is to remove the impact of diverse contaminations, such as cosmic rays, telluric lines, stellar activity, and instrumental systematics (interference patterns, variations in the point spread function (PSF), contamination from fibre B, and ghosts). The code operates on one-dimensional order-merged spectra generated by the official DRS that are continuum normalised using the publicly available code \texttt{RASSINE} \citep{Cretignier_2020}.

A master spectrum is produced by aggregating individual spectra and serves to compute the residual spectra. Flux variations in this space are corrected through multilinear regressions in either the stellar or terrestrial rest frame.
Stellar activity is partially corrected by fitting a scaled version of the S-index to each wavelength column of the spectra time series matrix, as stellar lines exhibit first-order variations similar to the S-index \citep{Cretignier_2021}. Correction of the PSF follows the approach outlined in \citet{Stalport_2023}, where symmetric variations of the PSF of the CCFs are extracted, decorrelated from the S-index. Lastly, the RVs are extracted using the CCF technique with a tailored line selection based on the master spectrum.

Absorption of planetary signals in the cleaning process can be reduced by shifting the spectra according to a pre-fitted Keplerian solution. We pre-fitted planet b using the period and phase information from the photometry described in Section \ref{photometry_priors}.

\subsubsection{\texttt{TWEAKS}} \label{sss:tweaks_descr}

\texttt{TWEAKS} (Time and Wavelength-domain stEllar Activity mitigation using \texttt{kima} and \texttt{SCALPELS}) described in \citet{Cameron_2021,John_2022,John_2023} is a pipeline that aims to distill the planetary contribution out of a CCF.
More specifically, this pipeline makes use of the \texttt{SCALPELS} (Self-Correlation Analysis of Line Profiles for Extracting Low-amplitude Shifts) basis vectors computed from the CCF to distinguish between planetary shift-driven RVs and RV contributions produced by variations of the CCF shape induced by stellar variability. This separation is enabled by computing orthogonal modes of variation in the autocorrelation function (ACF) of the CCF. Since the ACF is independent of translational shifts, this step allows isolating shape variations.
However, because planetary RV contributions are not guaranteed to be perfectly orthogonal to the \texttt{SCALPELS} basis vectors within these limited and irregularly sampled data sets, some of the planetary RV contribution may be absorbed in the decorrelation process.
Therefore, the modelling of the Keplerian signals and the separation of the shift and shape-driven RV components is performed simultaneously by joining \texttt{SCALPELS} with the Keplerian solver \texttt{kima} \citep{Faria_2018}. This combination is called the \texttt{TWEAKS} method.

The current version of \texttt{SCALPELS} reorders the principal components into the sequence that gives the fastest decrease in the Bayesian Information Criterion (BIC) of the fit to the radial-velocity time series, as described by \cite{Cameron_2021} and \cite{OuldElhkim_2023}. 
For HD~85426, the four leading principal components after reordering were sufficient to achieve optimal detrending without overfitting noise. This corresponds to the solution that minimises the BIC. The fact that the BIC reaches its minimum for four principal components demonstrates, by construction of the BIC, that there are measurable RV contributions of non-planetary origin to the CCFs, which are removed by the \texttt{SCALPELS} algorithm.

\subsubsection{Data selections}

We ran \texttt{YARARA} and \texttt{TWEAKS} \citep{Cameron_2021,John_2022,John_2023} on the 143 spectra remaining in our set after the rejection of observations affected by the guiding issue. \texttt{YARARA}, and \texttt{TWEAKS} have different rejection criteria based on the RV, RV uncertainty, and CCF. First, we fed all 143 spectra (called set 0 hereafter) to both codes and analysed the output. Due to the different rejection criteria, \texttt{YARARA} included 134 nightly-binned RVs in the analysis, while \texttt{TWEAKS} made use of 128 nightly-binned observations. 

Since we noticed some differences in the output of the two codes, we created a new set (set 1) of data passing all rejection criteria and ran \texttt{YARARA} and \texttt{TWEAKS} on this set, thus including the same observations. This is done to ensure that any differences in the output are due to the codes themselves, not the different data selections.
The latest observation, which was taken 194 days after the penultimate one, was also removed in the last \texttt{TWEAKS} run. Therefore, we also removed it from the \texttt{YARARA} set for consistency. This approach is warranted because a single measurement taken about half a year after the other observations is not expected to aid the analysis, given the limited stability of RV instruments and the star's variability. Therefore, set 1 consists of 127 nightly binned observations. 

Most analyses in this study are based on set 1 because it is least likely to contain problematic data. For set 1, the standard deviation of the DRS RVs is 3.9 \ms and 3.2 \ms for the \texttt{YARARA} RVs, thus 18 per cent lower for the \texttt{YARARA} RVs. The mean uncertainty of the DRS RVs is 0.8 \ms, whereas the mean uncertainty of the \texttt{YARARA} RVs is 0.6 \ms. 38 observations were gathered in the first observing season (December 2020 to June 2021), 47 in the second season (December 2021 to June 2022), and 42 in the last observing season (October 2022 to April 2023).

\begin{figure}
    \centering
    \includegraphics[width=0.99\columnwidth]{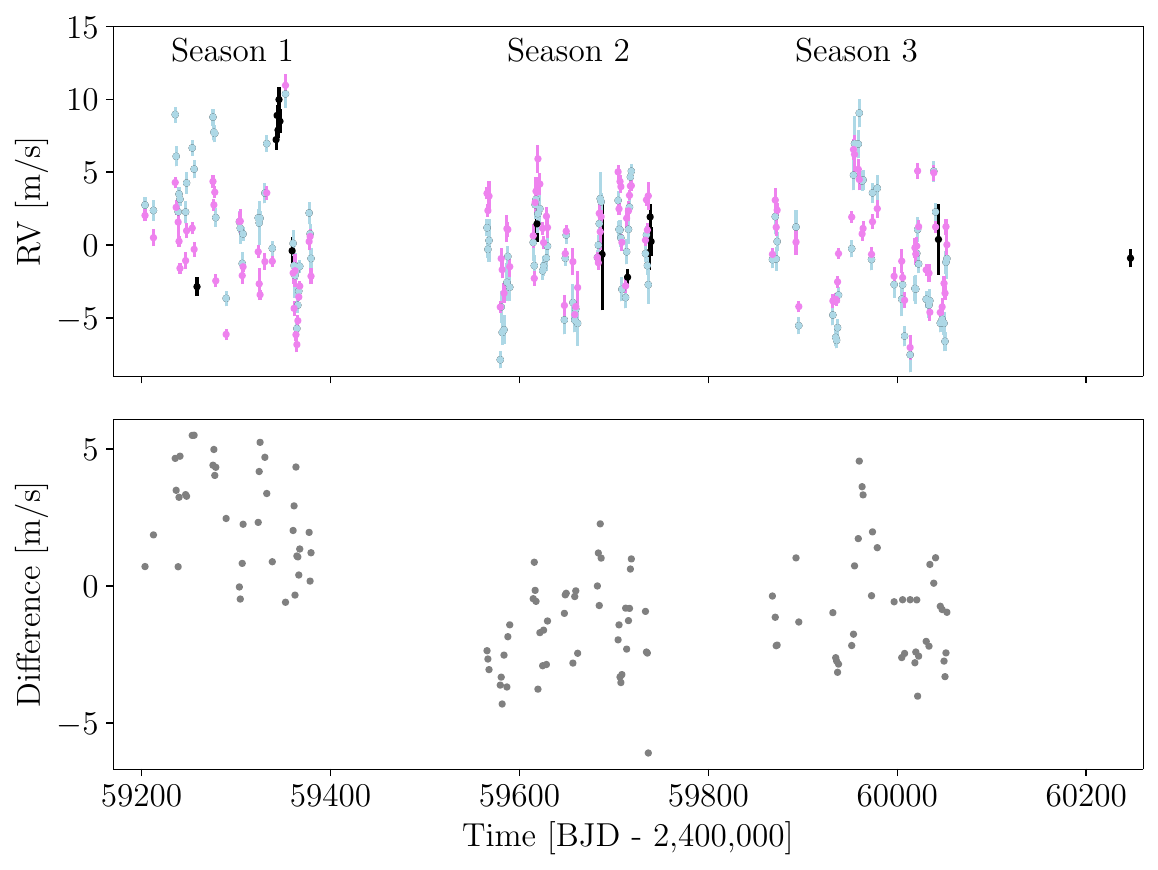}
    \caption{In the top panel, the HARPS-N DRS 3.0.1 RVs (light blue) and \texttt{YARARA} RVs (violet) are shown. The observations marked in black were rejected due to instrumental issues or rejection by either activity mitigation algorithm. The difference between the two RV sets are shown in the bottom panel.}
    \label{fig:allrvs}
\end{figure}

The RVs are shown in Fig. \ref{fig:allrvs}. Visually, there is no strong indication of an RV offset between the first and second observing seasons in the DRS or the \texttt{YARARA} RVs. However, \texttt{YARARA} removed the offset in some stellar activity indicators, such as the contrast and the FWHM.

\section{Stellar characterisation} \label{StellarCharacterisation}

It is indispensable to characterise the host star to derive planetary masses, radii, surface conditions, and internal structure. In this Section, we use information from various external sources, as specified in the text, and the HARPS-N DRS spectra to characterise HD~85426. We derived the stellar atmospheric parameters using the Stellar Parameter Classification (SPC) code \citep{Buchhave_2012}, CCFPams\footnote{\url{https://github.com/LucaMalavolta/CCFpams}} results, and ARES+MOOG \citep{Sousa_2014,Sousa_2015} using the HARPS-N DRS spectra. The derived parameters were then used separately as input to the \texttt{isochrone} code, together with the stellar parallax and broadband photometric magnitudes. Using both the Dartmouth \citep{Dotter_2008} and the MIST stellar evolution models \citep{Dotter_2016}, we then derived stellar masses, radii, and ages. These results were condensed into a final set of parameters following the methods detailed in \citet{Mortier_2020} and are shown in Table \ref{tab:stellarparameters_1774_other_sources}. 

HD~85426 is very similar to the Sun in effective temperature and metallicity. However, with an age of $7.4^{+0.9}_{-1.1}$~Gyr, it is significantly older and fits within the definition of a solar analogue \citep{deStrobel_1996,Soderblom_1998}.
The \textit{Gaia} Renormalized Unit Weight Error (RUWE) is equal to 0.96, suggesting that this is indeed a single star \citep{gaia1,gaia3}.

In addition, we derived the galactic velocities of HD~85426 using \textit{Gaia} DR3 \citep{gaia3} data. The three velocity components $U$, $V$, $W$, reported in Table \ref{tab:stellarparameters_1774_other_sources}, are calculated following \citet{Johnson_1987}. Note that these values are not in the Local Standard of Rest. The galactic velocities of a star can hint at membership to different galactic populations. Following \citet{Reddy_2006}, we deduced a probability of 57.99 $\pm$ 0.26 per cent that HD~85426 belongs to the thin disc, 41.69 $\pm$ 0.26 per cent probability of thick disc membership, and a probability of  0.32 $\pm$ 0.01 per cent that the star is a part of the galactic halo. Kinematically, the case is therefore not clear-cut. However, based on the star's solar metallicity, lack of alpha enhancement, and age, thin disc membership is more likely \citep{Gilmore_1995,Robin_2003,Duong_2018}.

Converting the S-index to \rhk following \citet{Noyes_1984}, we find a mean value of \rhk of -4.92. Using the relation between \rhk and the rotation periods, as a function of the convective turnover time computed via the colour index B-V, given in \citet{Noyes_1984}, we estimate a rotation period of about 25 days. We obtain the same result using the relation in \citet{Mamajek_2008}.
The rotation period estimates from \citet{Noyes_1984} and \citet{Mamajek_2008} are based on population fits and therefore provide a rough estimate of the rotation period but not an accurate value.

\begin{table}
\centering
    \caption{Stellar parameters of HD~85426 and method used for the derivation or the external source, such as Gaia DR3 \citep{gaia1,gaia3}, 2MASS \citep{Cutri_2003}, and AllWise \citep{Cutri_2014}.}
    \begin{threeparttable}
    \begin{tabular}{lrc}
    Parameter & Value & Source\\ [2 pt]
    \hline \\
    \multicolumn{2}{l|}{{\it Designations and coordinates}}\\[2 pt]
    TIC ID & 4897275\\[2 pt]
    TOI ID    & 1774\\[2 pt]
    2MASS ID  & J09523847+3506422\\[2 pt]
    {\it Gaia} DR3 ID & 796063843195758208\\[2 pt]
    RA (J2016) [h:m:s] & 09:52:39 & \textit{Gaia} DR3\\[2 pt]
    Dec (J2016) [d:m:s] & +35:06:40 & \textit{Gaia} DR3 \\[0.25cm]
    \multicolumn{2}{l|}{{\it Magnitudes and astrometric solution}}\\[2 pt]
        B & 8.913 $\pm$ 0.03 & (1)\\[2 pt]
        V & 8.25 $\pm$ 0.025 & (2)\\[2 pt]
        J & $7.055\pm0.024$ & 2MASS\\[2 pt]
        H & $6.728\pm0.015$ & 2MASS\\[2 pt]
        K & $6.684\pm0.021$ & 2MASS\\[2 pt]
        W1 & $6.653\pm0.081$ & AllWise\\[2 pt]
        W2 & $6.650\pm0.021$ & AllWise\\[2 pt]
        W3 & $6.682\pm0.018$ & AllWise\\[2 pt]
        Distance [pc] &              $53.76^{+0.08}_{-0.08}$   & (3)\\[2 pt]
        $\pi$ [mas] & $18.57 \pm 0.02$ & \textit{Gaia} DR3 \\[2 pt]
        U [\kms] &                  $-22.86^{+0.08}_{-0.08}$  & (4)\\[2 pt]
        V [\kms] &                  $-90.80^{+0.12}_{-0.12}$  & (4)\\[2 pt]
        W [\kms] &                  $-11.95^{+0.08}_{-0.08}$  & (4)\\[0.25cm]
    {\it Stellar parameters}\\[2 pt]
        $T_{\text{eff}}$ [K]& 5746 $\pm$ 59 & (5)\\[2 pt]
        [Fe/H] & -0.02 $\pm$ 0.05& (5)\\[2 pt]
        [Mg/H] & 0.03 $\pm$ 0.02& (6)\\[2 pt]
        [Si/H] & 0.00 $\pm$ 0.04& (6)\\[2 pt]
        [Ti/H] & 0.03 $\pm$ 0.03& (6)\\[2 pt]
        [$\alpha$/Fe] & $0.05\pm0.05$ & (6) \\[2 pt]
        microturbulence $\xi_t$ [\kms] & 1.07 $\pm$ 0.04& (6)\\[2 pt]
        $v \sin i$ [\kms] & <2 & (7)\\[2 pt]
        $\log g_{\text{spec}}$ & 4.33 $\pm$ 0.11& (5)\\[2 pt]
         $\log g_{\text{iso}}$ & $4.33^{+0.02}_{-0.01} $ & (8)\\[2 pt]
         $M_\ast$ [M$_{\odot}$]&         $0.991^{+0.027}_{-0.020}$   & (8)\\[2 pt]
         $R_\ast$ [R$_{\odot}$] &       $1.1303^{+0.0069}_{-0.0069}$  & (8)\\[2 pt]
         $\rho_\ast$ [$\rho_{\odot}$] &   $0.686^{+0.027}_{-0.022}$  & (8)\\[2 pt]
         Age [Gyr] &                  $7.4^{+0.9}_{-1.1}$  & (8)\\[2 pt]
         \hline
    \end{tabular}

    \begin{tablenotes}[flushleft]
      \small
      \item (1) Calc. from Tycho2 $B_T$ \citep{Hog_2000} in TIC 8.2 \citep{Stassun_2018}.
      \item (2) Calc. from \texttt{Hipparcos} \citep{Perryman_1997} in TIC 8.2.
      \item (3) \citet{Bailer-Jones_2021}.
      \item (4) Calculated based on \textit{Gaia} DR3 -- this work.
      \item (5) ARES+MOOG \& SPC \& CCFPams combined -- this work.
      \item (6) ARES+MOOG -- this work.
      \item (7) SPC -- this work.
      \item (8) \texttt{isochrones} -- this work.
    \end{tablenotes}
  \end{threeparttable}
\label{tab:stellarparameters_1774_other_sources}
\end{table}


\subsection{Stellar activity analysis from spectra}
\label{ss:stellar_activity_from_spectra}

The mean \rhk value for HD~85426 is equal to -4.92, which is comparable to the Sun's mean value and indicates low but not negligible activity. Therefore, we need to thoroughly cross-check our inferences.

RV signals can be produced by planets orbiting the observed star, the star itself modulated by the stellar rotation period and the magnetic cycle \citep[e.g.][]{Lagrange_2010,Meunier_2010b,Cegla_2019,Chaplin_2019,Haywood_2022,Lienhard_2023}, as well as by telluric lines \citep{Cunha_2014,UlmerMoll_2019}, or the instrument itself. The activity indicators are impacted by the same effects, although in slightly differing ways, but not by the planets. Consequently, the comparison of periodic signals in the RVs, expected from planets, and activity indicator time series can help determine whether a signal in the RV time series is due to a planet or one of the other effects.

\begin{figure}
    \centering
    \includegraphics[width=0.95\columnwidth]{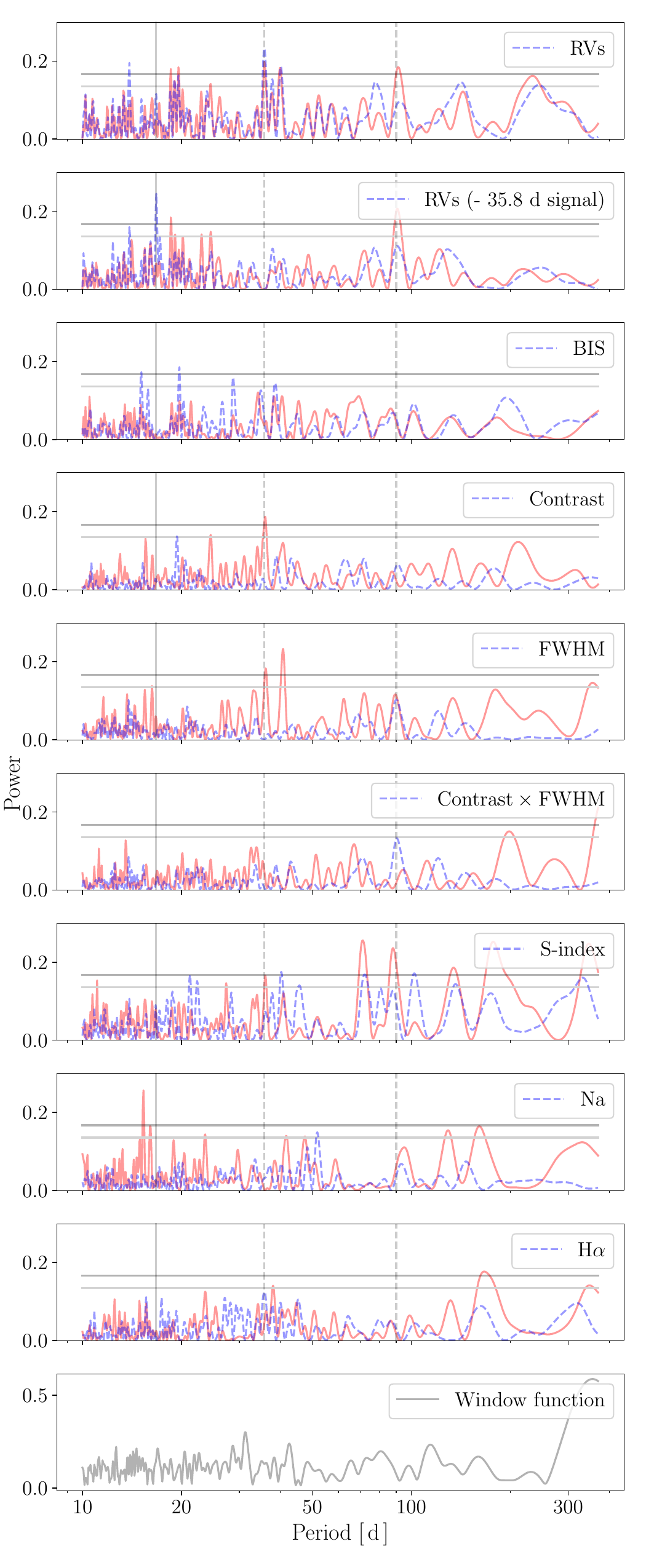}
    \caption{GLS periodograms from DRS (solid red lines) and \texttt{YARARA} (dashed blue lines) for set 1. The solid vertical line indicates the periods of HD~85426~b (16.71 d), and the dashed vertical lines show the periods of the most dominant Keplerian signals at 35.7 and 90 d. The False Alarm Probability of 1 (0.1) per cent is indicated by the light-grey (dark-grey) horizontal line. The top panel shows the periodograms of the two RV sets, whereas the second panel shows the periodograms after removing the most dominant sinusoidal signal at 35.8 d. From the third to the ninth panel from the top, we show the periodograms of the activity indicators, and in the last panel, we show the window function.}
    \label{fig:periodograms1774}
\end{figure}

Periodic signals can be found by analysing the periodograms of time series. The Generalized Lomb-Scargle \citep[GLS;][]{Zechmeister_2009} periodograms of the DRS-derived parameters, as well as those extracted from the \texttt{YARARA}-processed spectra, are shown in Fig. \ref{fig:periodograms1774}. The same data selection (set 1) was applied for both reductions. For the DRS data, we removed the offset between the first and the other seasons by separately subtracting the median from the indicator time series. This procedure was not applied to the RV data because there was no significant offset between the seasons.

\begin{figure}
    \centering
    \includegraphics[width=0.99\linewidth]{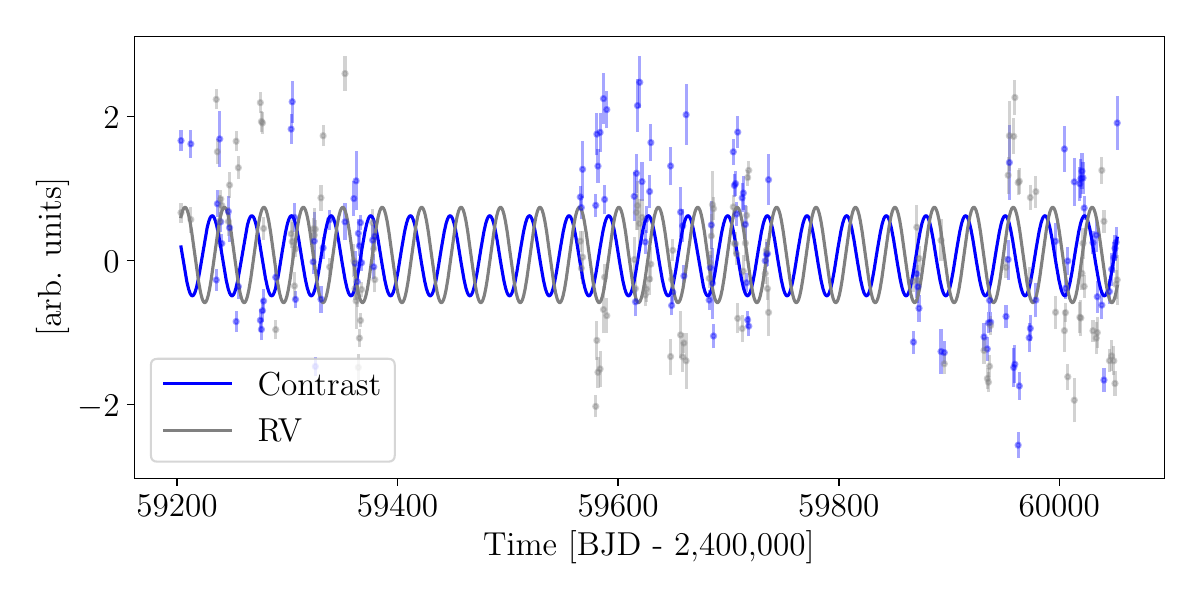}
    \caption{Time series of the DRS RVs and the CCF contrast and best-fitting sinusoidal model. Both time series were normalised independently to match in scatter.}
    \label{fig:contrast_rvs}
\end{figure}

The activity periodograms do not show a dominant periodic signal that is shared among multiple indicators. However, there is a peak at 36.0 d in the periodogram of the DRS CCF contrast that is not present in the YARARA data because it is removed by the stellar activity and PSF correction. The same peak can be seen for the DRS CCF FWHM time series, although it is not the strongest peak in this periodogram. There are a few indications that this variation of CCF contrast and FWHM is not of stellar origin. First, the CCF equivalent width (EW), traced by the product of FWHM and contrast, shows no sign of periodic variation around 36.0 days. Indeed, FWHM and contrast are anticorrelated with a Pearson correlation coefficient of -0.52, after correction for the offset between seasons 1 and 2. A stellar effect resulting from a change in convective flows should directly impact the BIS \citep[e.g.][]{Dravins_1981,Gray_2005} and should also affect the EW by changing the temperature distribution \citep[e.g.][]{Gray_2005}. However, there is no indication of periodic variation in BIS or EW at 36 days.
Secondly, there are studies indicating that there is a lag between stellar activity indicators and the radial velocities. \citet{Collier_Cameron_2019} found that for the Sun, the RVs peak about 1--3 days before the indicators reach their maxima. This corresponds to a lag in phase of about 20 degrees. Similarly, \citet{Burrows_2024} measured a lag of about 40 degrees. For HD~85426, we determined the best fitting sinusoid for both the RVs and the contrast separately and measured the phase lag at the beginning and the end of the time series.
In this way, we find a negative lag between 44 and 90 degrees between the contrast and the RVs, as shown in Fig. \ref{fig:contrast_rvs}, which means that the maxima in RV follow after the maxima in the contrast in time, which is the opposite of the expected behaviour for RV variations linked to stellar activity.
Lastly, \texttt{YARARA} indeed removed the 36.0 d signal from the contrast and FWHM time series, but it did not remove the 35.7 d signal from the RV time series. 

We investigated whether the broadening itself could potentially produce the measured RV variation.
In the DRS pipeline, a CCF is evaluated on a fixed velocity grid with a bin size of about 0.82 \kms \citep{Dumusque_2021} and is subsequently fitted with a Gaussian.
Since CCFs are generally slightly asymmetric \citep[e.g.][]{Gray_2005,Cegla_noise3}, we suspected that the broadening and subsequent binning of the CCF could produce a spurious RV signal. However, we found that the broadening only induces a spurious RV shift with a semi-amplitude of 0.18 \ms, which is too small by a factor of 10 to artificially create the RV signal. For this test, we first created a high-resolution mean CCF. For each of the 127 measurements, we convolved the high-resolution mean-CCF with a Gaussian kernel such that the convolution product, if purely Gaussian, perfectly matched the measured absorption line. We then binned this convolution product to match the velocity grid with a velocity step of 0.82 \kms. We fitted a Gaussian to this binned CCF to extract the RV. This procedure can create a spurious RV shift, but it is too small to produce the measured signal with a semi-amplitude of about 2 \ms.

There are significant peaks in the GLS periodogram of the DRS S-index time series. However, inspection of the residual spectra showed that the Ca II H line was contaminated by ghosts \citep{Cretignier_2021,Dumusque_2021}, which produced this signal. Indeed, the strongest four peaks in the DRS S-index periodogram, in descending order, are at 343 d, 72 d, 177 d, and 89 d, corresponding very closely to the 1-year peak and its harmonics at 73 d, 183 d, and 91 d, respectively. This contamination means that the DRS S-index time series cannot be used to correct for stellar activity. The \texttt{YARARA} S-index time series does not show these clear peaks that we attributed to instrumental contamination. Instead, we see a few peaks that barely surpass the 1 per cent False Alarm Probability around 21 d, 40 d, 72 d, and 102 d. 

We conclude that we cannot deduce the stellar rotation period from the spectra because no periodic signal is sufficiently strong and shared between indicators. 

With older solar-like stars generally exhibiting a magnetic cycle period of the order of 10 years \citep{Olan_2016}, we cannot directly constrain the period of the magnetic cycle with our data. 
However, based on the \texttt{YARARA} S-index data, it appears that we captured the minimum of this cycle, as we see a valley in the S-index time series. Applying Student's t-test, assuming equal variances, to the different observing seasons, we derive a t-statistic of 6.2 (p-value: 0.0002 per cent) for the difference between the S-index values of season 1 and season 2, and a t-statistic of 3.2 (p-value: 0.2 per cent) between seasons 2 and 3. Therefore, the difference in S-index is indeed statistically significant. The p-values do not change significantly if we perform the t-test assuming unequal variances. Note that the extent of the difference in S-index between season 1 and season 2 may be impacted by instrumental changes, even after \texttt{YARARA} correction.

The spectral window function, computed as in \citet{Roberts_1987}, reveals a strong yearly peak due to the seasonality of the data. Another strong peak appears at a period of one day, reflecting that measurements are restricted to nighttime. This peak is not included in the displayed periodogram, as it dominates all other peaks in amplitude. Furthermore, there is a very minor peak at 31.3 days, which may result from observational gaps introduced by the lunar cycle. 
The peak at 31.3 days could indicate that the 35.7 days is produced by aliasing from the true signal of 16.71 d from planet b. However, the 31.3-day period in the window function just appears in season 2, whereas it is absent in the other seasons. The 35.7 d signal in the RVs, on the other hand, persists for all seasons and combinations of seasons, as shown in Section \ref{ss:periodograms}.
This suggests that, aside from the seasonal and nightly sampling, there are no prominent sampling frequencies which could produce artificial peaks in the other periodograms.

\begin{figure}
    \centering
    \includegraphics[width=0.9\columnwidth]{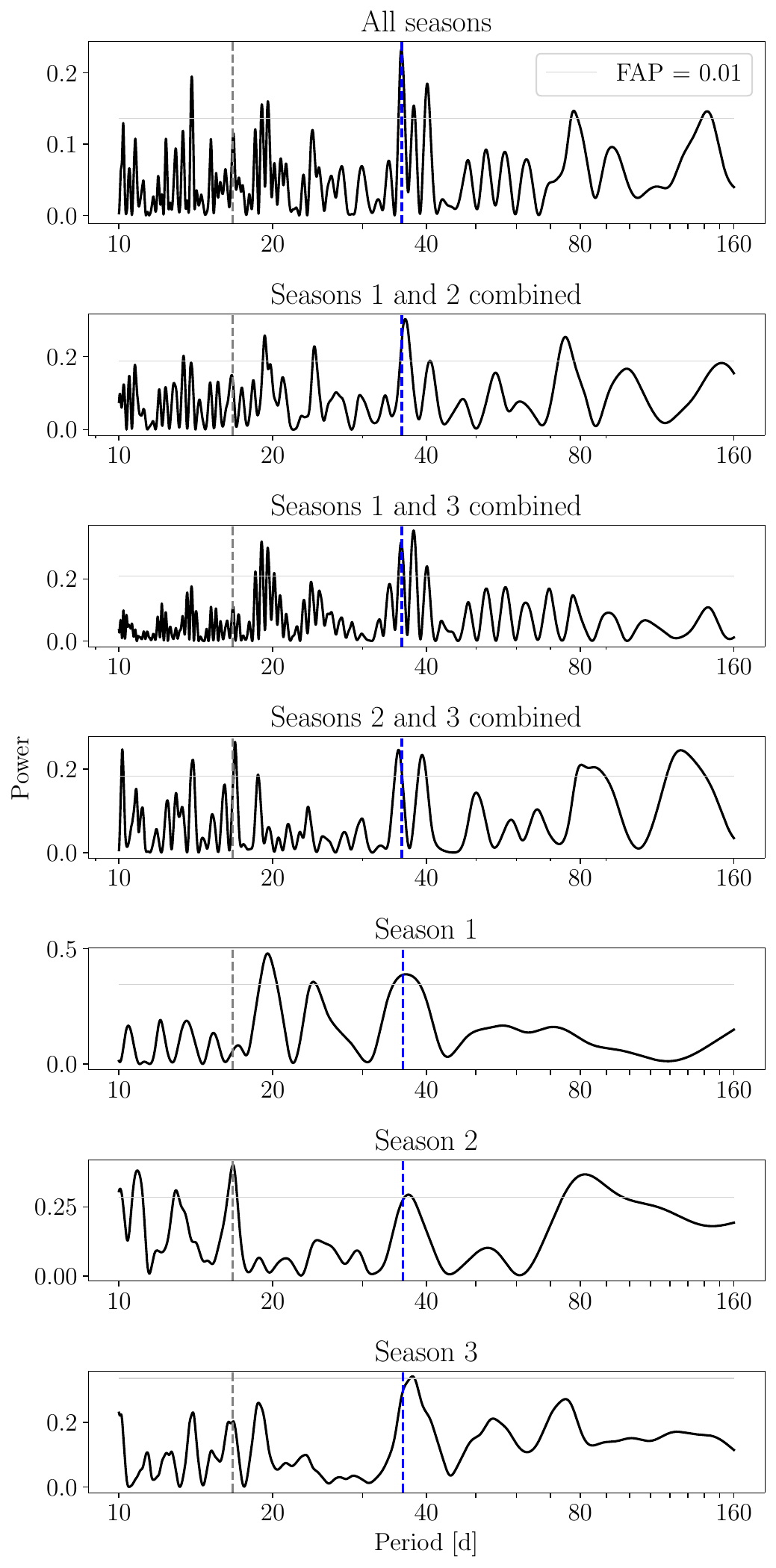}
    \caption{GLS periodograms of the \texttt{YARARA} RVs. The grey vertical line indicates the period of HD~85426~b. The blue vertical line shows the period of planet candidate HD~85426~c. The grey horizontal line indicates the level of 1 per cent False Alarm Probability.}
    \label{fig:periodograms1774_first_all}
\end{figure}

\subsection{Stellar activity analysis from photometry}
The 13.7-day orbit of {\it TESS} induces systematics in the SAP light curves, making it challenging to detect weak stellar signatures with periods longer than the duration of an orbit. HD~85426 is expected to have a rotation period of about twice the duration of a {\it TESS} orbit. Given that there are only two sectors of {\it TESS} data, covering a total of about two rotations of the star, and the star's activity is moderate to low, it is difficult to constrain the rotation period from the {\it TESS} data. 

Indeed, the {\it TESS} PDCSAP light curve shows occasional variations below 500 ppm, but no clear periodic signal beyond 20 days.
In the SAP light curve, which is dominated by instrumental factors, we likewise find no periodogram peak beyond about 20 days that could hint at the rotation period.
The {\it CHEOPS} time series is too short for any meaningful analysis of stellar rotation. We also analysed data from the All Sky Automated Survey (ASAS) cameras \citep{Pojmanski_2002}.
The light curve of the camera $br$ of ASAS, with 443 observations spread over 1225 days, has the least scatter of all ASAS cameras, but still shows a standard deviation of about 18 per cent. The periodogram evaluated up to 100 days showed a forest of peaks without a convincing dominant signal. The WASP archive was investigated but did not contain sufficient high-quality observations for a meaningful result.

\section{Modelling the planetary signals} \label{RVModelling}

The characterisation of the star in the previous Section presents a picture of a star with contaminated activity indicators and RV signatures that are not shared between different reductions, with the additional complications of a series of measurements impacted by an instrumental issue and a change in the instrument between the first and the second observing season. We therefore opted to use various RV cleaning and analysis methods to gain a clearer view of the system. Significant effort was put into avoiding dependence on one single analysis method and testing our conclusions on different subsets of the data to ensure robustness. 
We first refine the planetary parameters using the available photometric information from {\it TESS} and {\it CHEOPS}. These parameters are used as Gaussian priors in the subsequent parts of the analysis when the photometric data are not fitted jointly.

\subsection{Priors from photometry} \label{photometry_priors}

We fitted the six transits with \texttt{Juliet} \citep{Espinoza_2019} with the Nested Sampling package \texttt{dynesty} \citep{Speagle_2020} to derive priors for the independent analysis of the RVs, presented in Table \ref{tab:parameterstransitfit}. The RVs were not included in this fit.  We also did not include the data from sector 48 because the transits of planet b were missed there. We also removed all data that were more than one transit duration away from the observed mid-transit times, thus including windows with a width of about twice the expected transit duration, corresponding to approximately 9.6 hours.

We used a broad uniform prior on the planet's radius, allowing a radius of up to 5 per cent of the star's radius, which is about double the fitted ratio. We centred the reference mid-transit time on the first {\it CHEOPS} transit, which conveniently lies in the centre of the second observing season. The width of this prior was set to the expected transit duration. The width of the uniform prior on the period was also set to the duration of a transit and is consequently based only on the information from the first two {\it TESS} transits. It can be visually checked in the {\it TESS} data that the period prior is sufficiently wide. Similarly, the prior on the reference mid-transit time covers the full {\it CHEOPS} transit.
A $\beta$ prior with the parameters from \citet{Kipping_2013ecc} was set on the orbital eccentricity of the planet, and a uniform prior was set on the argument of periastron. 

We parameterised the quadratic Limb-Darkening (LD) law in $(q_1, q_2)$ \citep{Kipping_2013} with Gaussian priors centred on LD coefficients computed with \textsc{PyLDTk} \citep{Husser2013, Parviainen2015} for {\it CHEOPS} and \textit{TESS}, and set an uncertainty of 0.05 for both coefficients and filters. In the parameterisation used, the mean stellar density is fit. For this prior, we chose a Gaussian distribution centred on the mean stellar density derived from the spectra, listed in Table \ref{tab:stellarparameters_1774_other_sources}, and doubled the uncertainty to avoid depending too strongly on this estimate.

\begin{table}
\centering
\caption{Orbital parameters from photometry using \texttt{Juliet}. The orbital period $P$ and reference mid-transit time $\text{T}_0$ posteriors are used as input for the fits that do not include the photometric data. The impact factor $b$, inclination $i$, orbital eccentricity $e$, and argument of periastron $\omega$ are not used directly.}
\label{tab:parameterstransitfit}
\begin{tabular}{llrr}
Symbol & Value & Fitted/Derived \\ [2 pt]
 \hline \\
P [d]& $16.70988  \pm 0.00003 $ & Fitted\\ [2 pt]
$\text{T}_0$ [BJD - 2,400,000] & $59674.4130  ^{+0.0006}_{-0.0005} $ & Fitted\\ [2 pt]
$b$ & $0.25 ^{+0.23}_{-0.15} $ & Derived \\ [2 pt]
$i$ [deg] & $89.4 ^{+0.3}_{-0.4} $ & Derived\\ [2 pt]
$e$ & $0.08 ^{+0.11}_{-0.06} $ & Fitted\\ [2 pt]
$\omega$ [deg] & $266 ^{+68}_{-64} $ & Fitted\\ [2 pt]
\end{tabular}
\end{table}

\subsection{RV periodogram analysis} \label{ss:periodograms}

The most prominent peak in the GLS periodograms of the DRS and the \texttt{YARARA} RVs (shown in Fig. \ref{fig:periodograms1774}) is located at 35.7 d. Once we have removed the best-fitting sinusoid from the \texttt{YARARA} data, we can see the signal associated with the transiting planet b in the \texttt{YARARA} RVs. This is not the case for the DRS RVs and highlights the importance of proper cleaning and extraction of the RVs. 

To test the coherence of this signal at 35.7 d, we computed the periodograms for all combinations of two seasons and all three seasons individually (cf. Fig. \ref{fig:periodograms1774_first_all}). The peaks are narrower for periodograms that include a longer observing baseline. All periodograms show a peak at 35.7 d; this signal is, therefore, consistent across all seasons and it is the only signal that reaches or surpasses the 1 per cent False Alarm Probability threshold for all seven periodograms.

An additional peak at 38 d is visible in the periodogram of the RVs from the combined seasons 1 and 3. The beat period of 35.7 and 38 d equals the separation in time between these two seasons. This means that they describe a very similar model for seasons 1 and 3, but they are phase-shifted by 180 degrees for season 2. The same periodic signal in the data, therefore, produces this peak. An in-depth analysis, detailed in Appendix \ref{appendix:38or36}, showed that the estimated mass associated with the 38-day signal doubles if we exclude the data of the second season. This is due to the phase being off by 180 degrees in the second season, forcing the fit to converge to a lower amplitude if the data from this season are included. This strongly indicates that 38 d is not the correct period for an outer companion to planet b.

\subsection{General diffusive nested sampling search for planets within the \texttt{YARARA} RVs} \label{s:dns_runs}

To further probe the presence of planetary RV signals, we investigated the preferred RV model in the \texttt{YARARA} data using nested sampling, including the knowledge about the transiting planet via priors. We used \texttt{kima} \citep{Faria_2018}, which utilises the diffusive nested sampling algorithm Dnest4 \citep{brewer2016dnest4}. Dnest4 is expected to be well suited for multi-modal problems, such as the one treated in this study, and computes the model evidence, allowing for model comparison.

For planet b, we used transit-informed Gaussian priors on the mid-transit time ($\mathcal{N}[2,459,674.4130,\ 0.0005^2]$ BJD) and the period ($\mathcal{N}[16.70988,\ 0.00003^2]$ d) based on the parameters estimated in Table \ref{tab:parameterstransitfit}. The priors for the additional unknown planets are shown in Table \ref{tab:priors_kima1} and are similar to those in \citet{John_2023}.
We fitted up to 4 Keplerians to the data. The eccentricity prior was set to the Kumaraswamy distribution \citep{Kumaraswamy_1980} with the listed shape parameters, as in \citet{Standing_2022,John_2023}, and closely resembles the $\beta$ distribution suggested in \citet{Kipping_2013ecc} favouring less eccentric orbits. The Kumaraswamy distribution was implemented in \texttt{kima} for numerical reasons \citep{Faria_2018}. Lastly, we set the number of saves to 100,000 to adequately sample the posterior distributions.

To compare competing models with different numbers of Keplerians, we compute the Bayes factor, i.e. the ratio between the model evidences. These results are shown in Table \ref{tab:evidences_kima}. Following the classification in \citet{Kass_1995}, we find decisive evidence, i.e. $\Delta\ln$Z greater than 4.6, for at least one other planet and strong evidence, i.e. $\Delta\ln$Z greater than 2.3, for a three-Keplerian model for both sets. 
There is strong evidence for including a fourth Keplerian signal for set 0, but insignificant evidence for set 1.

The most likely period for the second Keplerian is 35.8 d, in agreement with the results from the periodogram analysis in Section \ref{ss:periodograms}, and 90 d for the third Keplerian. For the 1-, 2-, and 3-Keplerian fits applied to set 1, we find minimum masses $m\sin i$ for planet b of $6.9^{+1.5}_{-1.5}$, $8.0^{+1.4}_{-1.3}$, and $8.7^{+0.7}_{-1.4}$ $\text{M}_{\oplus}$, respectively. Note that the $\sin i$ term is smaller than 0.01 per cent for transiting planet b and thus the minimum mass values for this planet are equal to the actual masses within the precision quoted in this study.
Applying the probabilistic mass-radius relations described in \citet{Chen_2017}, we expect a mass of $8^{+6}_{-4}\ \text{M}_{\oplus}$ for planet b. Our derived planetary masses are therefore comfortably within the expected range.

We can conclude that there are at least two, and very likely three, detectable planets in the RV time series. The two statistically most favoured models (2 or 3 planets) are investigated in more detail in the following sections.

\begin{table}
\caption{Prior distributions for \texttt{kima} run. $\mathcal{U}$ indicates a uniform distribution, $\mathcal{LU}$ a log-uniform distribution, $\mathcal{MLU}$ a modified log-uniform distribution \citep[e.g.][]{Gregory_2005}, and $\mathcal{K}$ a Kumaraswamy distribution.}
\label{tab:priors_kima1}
\begin{tabular}{llll}

Parameter & Symbol & Unit & Distribution \\ [4pt]
\hline \\
Orbital period & $P$~ & d &  $\mathcal{LU}[1.1,900]$ \\[2 pt]
Orbital Phase & $\phi$~ & deg &  $\mathcal{U}[0,360]$ \\[2 pt]
RV semi-amplitude& $K$ & \ms & $\mathcal{MLU}[0.01,20]$ \\[2 pt]
Eccentricity& $e$& & $\mathcal{K}[0.867, 3.03]$ \\[2 pt]
Argument of periastron& $\omega$ & deg & $\mathcal{U}[0,360]$ \\[2 pt]

\end{tabular}	
\end{table}

\begin{table}
\caption{Evidences (lnZ) and Bayes factors ($\Delta$ lnZ) for models assuming different numbers ($\text{N}_{\text{p}}$) of Keplerians. These models were evaluated using the RV set 0 and set 1.}
\label{tab:evidences_kima}
\centering
\begin{tabular}{cllll}
 & \multicolumn{2}{c}{RV set 0} & \multicolumn{2}{c}{RV set 1}\\
$\text{N}_{\text{p}}$ & lnZ & $\Delta$ lnZ & lnZ & $\Delta$ lnZ \\ [4pt]
\hline \\

1 &	-342.2 &	0.0	&	-327.6 &	0.0 \\[2 pt]
2 &	-335.9 &	6.3	&	-321.0 &	6.6 \\[2 pt]
3 & -333.3 &	2.6	&	-317.1 &	3.9 \\[2 pt]
4 &	-330.1 &    3.2 &	-316.4 &	0.7 \\[2 pt]

\end{tabular}	
\end{table}

\subsection{Multi-Keplerian joint fits including the \texttt{YARARA} RVs} \label{multi_joint_fit_juliet_yarara}

In this Section, we further investigate the case for at least one other planet and extract the respective planetary parameters. The subsequent analyses were computed using \texttt{Juliet} due to its versatility and its ability to jointly model the photometric data described in Sections \ref{photo1} and \ref{photo2}. We used the Nested Sampling package \texttt{dynesty} with 3000 live points to estimate Bayesian posteriors and evidences. To decrease the computation time, we included just the transits of planet b with a margin of one transit duration to either side, as in Section \ref{photometry_priors}. We used uniform priors centred on the best-fit value for the period and the reference time of the inferior conjunction of planet b, i.e. the reference mid-transit time, with a width of 10 $\sigma$ on both sides. The priors on the Limb-Darkening coefficients were set as in Section \ref{photometry_priors}. The dilution factor can be used to account for external sources of contamination and was fixed to 1, which means that we assume that no such source impacts the apparent transit depth. The mflux parameter models the mean out-of-transit flux and was set to a narrow Gaussian prior. Lastly, we chose broad uniform priors for the factors $\text{r}_{1}$ and $\text{r}_{2}$ which parameterise the impact factor and the planet-to-star radius ratio, as in \citep{Espinoza_2019}. A trend was discernible for the first \textit{TESS} transit and was subtracted together with a separate offset from the light curve before phase-folding. The gradient of this trend ($\theta_{\text{TESS}}$) was 0.0004 $\text{d}^{-1}$ and therefore has a very minor impact on the fit and the values of the extracted parameters.
These priors were used in all fits, except when specifically mentioned otherwise.

\subsubsection{Investigating the two most dominant Keplerian signals} \label{inv_2kep}

For the second Keplerian, we initially set a wide log-uniform prior on the orbital period to cover the entire baseline of the data. The posterior again revealed a clear global maximum at 35.8 d, corroborating the results from Sections \ref{ss:periodograms} and \ref{s:dns_runs}, where we found the same dominant signal using two other methods. However, \texttt{Juliet} found other local maxima. The most dominant of these secondary posterior maxima was located at 90 d. 
We thus redefined the prior on the period of the second Keplerian to a uniform prior $\mathcal{U}(34.4,37.4)$. This prior is centred at 35.87 d, as derived from the periodograms, and its width in frequency corresponds to twice the inverse of the baseline of about 850 d. This choice ensures that the periodogram peak and the main peak of the posterior fit comfortably into the prior range.

We then tested the dependence of our results on the eccentricity prior.
For this, we first fitted two models (1 and 2 Keplerians) to the \texttt{YARARA} RVs by setting the eccentricity priors to $\mathcal{U}(0,0.95)$.
For the 1-Keplerian model, we found an orbital eccentricity of $0.2^{+0.08}_{-0.1}$ for planet b. The 2-Keplerian model converged to a lower eccentricity of $0.1^{+0.1}_{-0.07}$, with the eccentricity below 0.3 for all posterior samples. For the second Keplerian, we found a slightly higher value for the eccentricity of $0.21^{+0.17}_{-0.14}$.

In conclusion, the data indicate that the eccentricity of planet b is low and is very likely below about 0.3. This is supported by evidence that the eccentricities of planets in multiple systems tend to be low \citep[e.g.][]{VanEylen_2019}. For subsequent fits, in addition to the results derived using a $\beta$ eccentricity prior, we also derive the main results with a uniform eccentricity prior with an upper limit of 0.3. This serves for comparability with the \texttt{TWEAKS} analysis in Section \ref{s:tweaks}. 

In Table \ref{tab:res_one_two_keplerian}, we show the derived parameters for a 1- and 2-Keplerian model using a $\beta$ prior and a uniform prior on the eccentricity, with the upper limit set to 0.3 as motivated above, while jointly fitting the photometric data. This analysis showed that the mass of planet b depends only insignificantly on whether we model a second Keplerian and on whether we use a uniform or a $\beta$ prior.
Furthermore, we found decisive evidence for including a second Keplerian.
More specifically, the difference between the two models, including one or two Keplerians, in log-evidence is equal to 11.1 for the $\beta$ prior and 10.7 for the uniform prior. This agrees with our conclusions in Section \ref{s:dns_runs}, where we also found that including a second Keplerian was also statistically very strongly preferred.

\begin{figure}
    \centering
    \includegraphics[width=0.99\columnwidth]{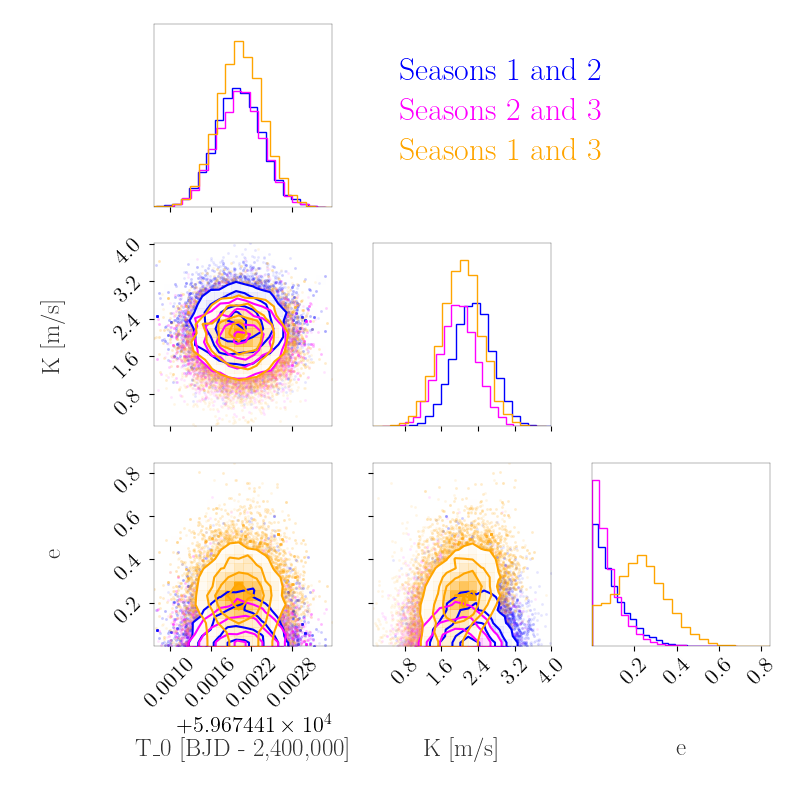}
    \caption{Posterior distributions for the orbital parameters of planet b for all combinations of two seasons.}
    \label{posterior_seasonbyseason2_p1}
\end{figure}

\begin{figure}
    \centering
    \includegraphics[width=0.99\columnwidth]{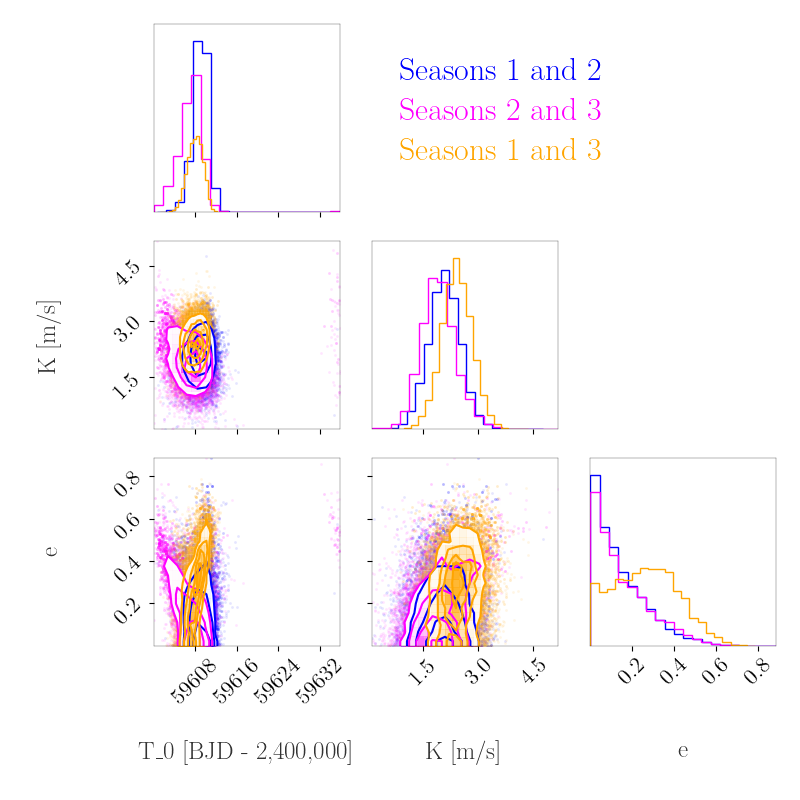}
    \caption{Posterior distributions for the orbital parameters of planet c for all combinations of two seasons.}
    \label{posterior_seasonbyseason2_p2}
\end{figure}

To further probe the periodic signal around 35.7 d, we investigated the signal's stability in time in terms of amplitude and phase.
For this, we again created three separate sets of data, selecting all combinations of two seasons. We then modelled these three sets independently with 2-Keplerian models. We used the transit-informed priors on the conjunction time $\text{T}_\text{0}$ and orbital period of planet b, for computational efficiency, and fixed the period of the second signal to 35.7 d. Fixing this period serves to compare whether the signal shifts in phase when we include different subsets of the data. However, note that the peak in the period posterior to the second Keplerian was always within 35.7 $\pm$ 0.5 d, when not constraining the period, as expected. The posterior distributions of the orbital parameters are shown in Fig. \ref{posterior_seasonbyseason2_p1} and Fig. \ref{posterior_seasonbyseason2_p2}. 
The values for both Keplerians are remarkably consistent, indicating robustness. Specifically, for the second Keplerian signal, the inferior conjunction times $T_0$ are very consistent (2,459,498.3 $\pm$ 1.1, 2,459,499.6 $\pm$ 1.2, 2,459,498.6 $\pm$ 1.3) given the long period of 35.7 d. Furthermore, while slightly more variable, the semi-amplitudes are consistent within the error bars. The extracted minimum masses associated with the second Keplerian correspond to values between 9 and 12 $\text{M}_{\oplus}$, in very good agreement with the value derived from the entire RV time series.

With the remarkable coherence of the second Keplerian in the RV time series and given the decisive statistical preference in favour of including a second Keplerian, we consider it warranted to deduce that the Keplerian signal is caused by a previously unknown planet, HD~85426~c. Even in the case that this signal turned out to be spurious, it is warranted to include it in the fit because of its stability over time. We show the priors and posteriors for the 2-Keplerian model with the $\beta$ prior set on the orbital eccentricity in Table \ref{tab:allresults_2pmodel}.

\subsubsection{A third planetary signal} \label{thirdsignal}

We have shown that including a second Keplerian is very strongly preferred over a 1-planet solution and that the extracted parameters are coherent in time. In this Section, we test whether there may be an additional planet and how the modelling of this planet would change our other inferences. This is due to evidence found in Section \ref{s:dns_runs} that a 3-Keplerian model is a very good fit to the RV data.

\begin{figure}
    \centering
    \includegraphics[width=0.99\linewidth]{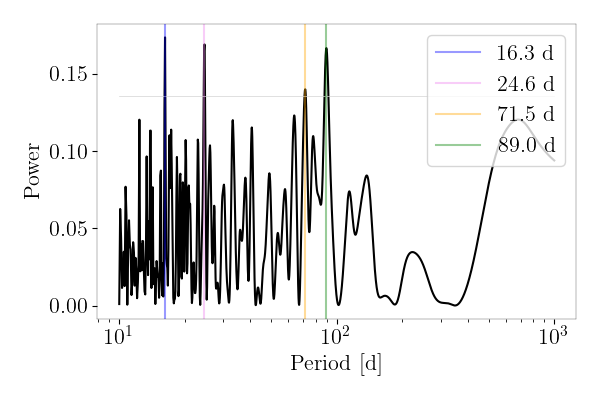}
    \caption{Periodogram of RVs with the contribution of planets b and c removed. The horizontal grey line indicates the 1 per cent FAP value.}
    \label{fig:resid_periodogram}
\end{figure}

Since multiple approaches with broad log-uniform priors recover the 35.7 d signal, and have shown its strong coherence in time, we consider it warranted to posit that this signal is real and not caused by the interplay of other Keplerians. 
If we subtract the RV signatures of planets b and c, according to the 2-Keplerian fit, from the RVs and recompute the GLS periodogram we find four periodic signals with False Alarm Probability below 1 per cent: 16.3 d, 24.6 d, 71.5 d, and 89 d, cf. Fig. \ref{fig:resid_periodogram}.

The reliability of an RV periodogram analysis after subtracting a 2-Keplerian fit is limited due to uneven data sampling and the potential partial absorption of a third Keplerian signal into the existing fit. However, the periodogram can suggest candidate periods that can be compared statistically in a subsequent step.

The 16.3 d period cannot be attributed to a planet because it would strongly interact with planet b with its orbital period of 16.71 days. 
Two of the remaining periods (71.5 d and 89 d) are yearly aliases. A periodogram peak analysis following \citet{Dawson_2010} did not clearly favour one of the two periods.
To compare the three candidate periods, we modelled the RV data using the same priors for the first two Keplerians, while applying three different priors on the third Keplerian in three separate model runs. More specifically, we set the prior on the period of the third Keplerian to  $\mathcal{U}(23.8,25.2)$~d,  $\mathcal{U}(66.0,78.1)$~d, and $\mathcal{U}(80.6,99.4)$~d, respectively. 
The range to either side of the central value is again set to the inverse of the baseline in frequency space. The width of the prior on the conjunction time was set to the maximum period for each case. The other priors were set identically to the previous 2-Keplerian model in Section \ref{multi_joint_fit_juliet_yarara} with the $\beta$ prior on the eccentricity.
The period of 89 d was decisively favoured over the 24.5 d period with a $\Delta\ln$Z of 4.8 and strongly favoured over the 71.5 d period with $\Delta\ln$Z equal to 3.3. 

In fact, we can also find the 89 d signal by setting the log-uniform prior $\mathcal{LU}(1.1,850)$ d on this Keplerian's period. This run clearly also favoured the period of 89 d for the third Keplerian. However, the posterior distribution of the reference conjunction time of the third Keplerian showed multiple modes, separated by multiples of the favoured period of about 89 d. We therefore needed to rerun the model restricting that prior distribution to $\mathcal{U}$[59500.0, 59590.0] d to avoid artificially inflating the error bars of the reference conjunction time.

We conclude that the 89 d period is clearly preferred, in agreement with the findings from Section \ref{s:dns_runs}.
The relevant results of the model with the broad log-uniform prior $\mathcal{LU}(1.1,850)$ d applied to the period of the third Keplerian are shown in Table \ref{tab:res_one_two_keplerian} and the full priors and posteriors are shown in Table \ref{tab:allresults_3pmodel}. There is strong statistical evidence for this model as compared to the 2-Keplerian solution, with $\Delta\ln$Z being 3.3 in favour of the 3-Keplerian model when applying the $\beta$ prior on the orbital eccentricities. The True Inclusion Probability \citep{Hara_2022} for the 89 d signal is 75 per cent, adding weight to the hypothesis that there is a detectable planet with a period of about 89 d. We find $\Delta\ln$Z to be about 3.9 in favour of the 3-Keplerian model if we use the uniform priors on the eccentricity, suggesting strong evidence in favour of this model.

The mass estimates for the modelled signals depend marginally, but are within the 1-sigma uncertainties, on how many other Keplerians we model and whether we choose a restricted uniform prior or a $\beta$ prior (cf. Table \ref{tab:res_one_two_keplerian}). The eccentricities for all three Keplerians are low and the arguments of periastron, although hard to constrain given the low eccentricities, are consistent across the different models. This shows that we have extracted robust orbital parameters.

\begin{figure}
    \centering
    \includegraphics[width=0.99\columnwidth]{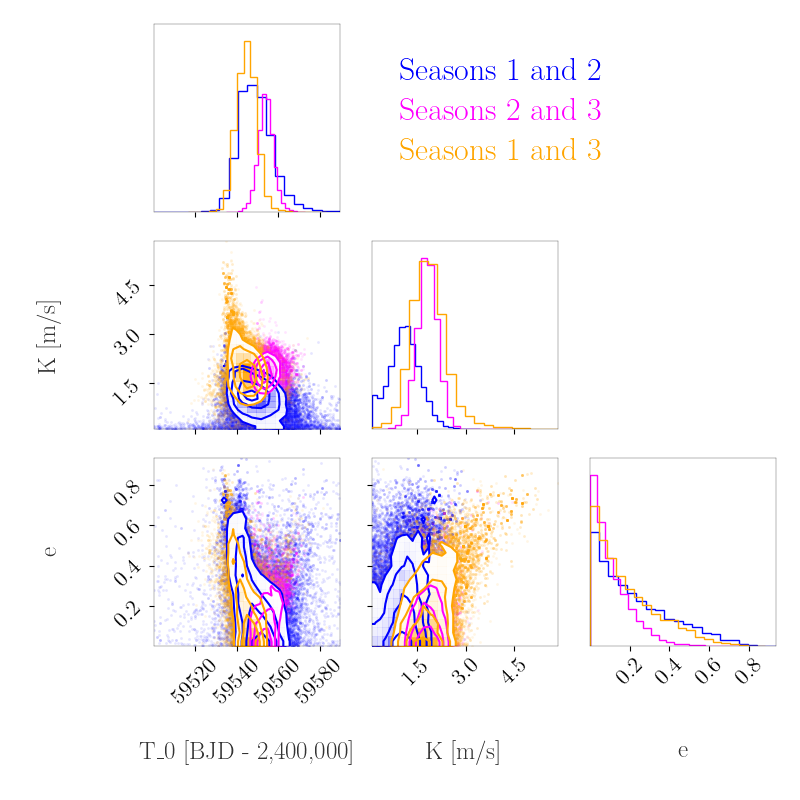}
    \caption{Posterior distributions for the orbital parameters of planet candidate d for all combinations of two seasons.}
    \label{posterior_seasonbyseason2_p3_90}
\end{figure}

Analogously to the approach in Section \ref{inv_2kep}, we investigated the stability of this third Keplerian in terms of semi-amplitude and phase. This is again achieved by fitting a 3-Keplerian model to the data in three different runs, excluding one season at a time. We set normal priors on the period and time of conjunction for planets b and c using the results from the 2-Keplerian fit and $\beta$ priors on the eccentricities. We fixed the period of the third Keplerian to 89 d and set a uniform prior with a width of 90 days on the time of conjunction. The posteriors for the third Keplerian at 89 d are shown in Fig. \ref{posterior_seasonbyseason2_p3_90}. The times of conjunction for all three runs align well given the long orbital period (2,459,548.4 $\pm$ 7.7, 2,459,544.6 $\pm$ 4.8, 2,459,553.9 $\pm$ 3.6). The semi-amplitudes align well too; however, the signal amplitude is less constrained if the first season of data is included.
The amplitudes and times of conjunction of this third Keplerian align slightly less well compared to the values we extracted for planet c, shown in Fig. \ref{posterior_seasonbyseason2_p2}. This is, however, expected because the orbital period is 2.5 times longer, the semi-amplitude is lower, and there are more parameters to fit. Given these considerations, the signal coherence is still a good indicator for the stability of the signal at 89 d. Nevertheless, long-term observations are necessary to further solidify this detection.

Lastly, since there is limited evidence for non-circular orbits, we also ran our model constraining the orbits to be circular. We chose the same priors as in Table \ref{tab:allresults_3pmodel}, apart from the eccentricity and the argument of periastron, which we set to zero. This produced posterior distributions that were very similar to those obtained with the $\beta$ or the uniform eccentricity prior. 
For example, the minimum masses of planets b, c, and d converged to  $8.5^{+1.3}_{-1.2}$, $10.7^{+1.5}_{-1.4}$, and $10.3^{+2.3}_{-2.4}$ $\text{M}_{\oplus}$, respectively. The differences between these masses and those produced using the other two eccentricity priors are negligible.

\subsubsection{Analysis of the favoured model} \label{sss:favoured_model}
In this Section, we examine the properties of the 3-Keplerian model derived using the $\beta$ eccentricity prior. This model was selected because it is statistically preferred over the 2-Keplerian models. The selection of the eccentricity prior has a very minor effect on the derived parameters. However, since the $\beta$ prior is more commonly used in RV analyses, we have chosen this model for further analysis.

\begin{figure}
    \centering
    \includegraphics[width=0.95\columnwidth]{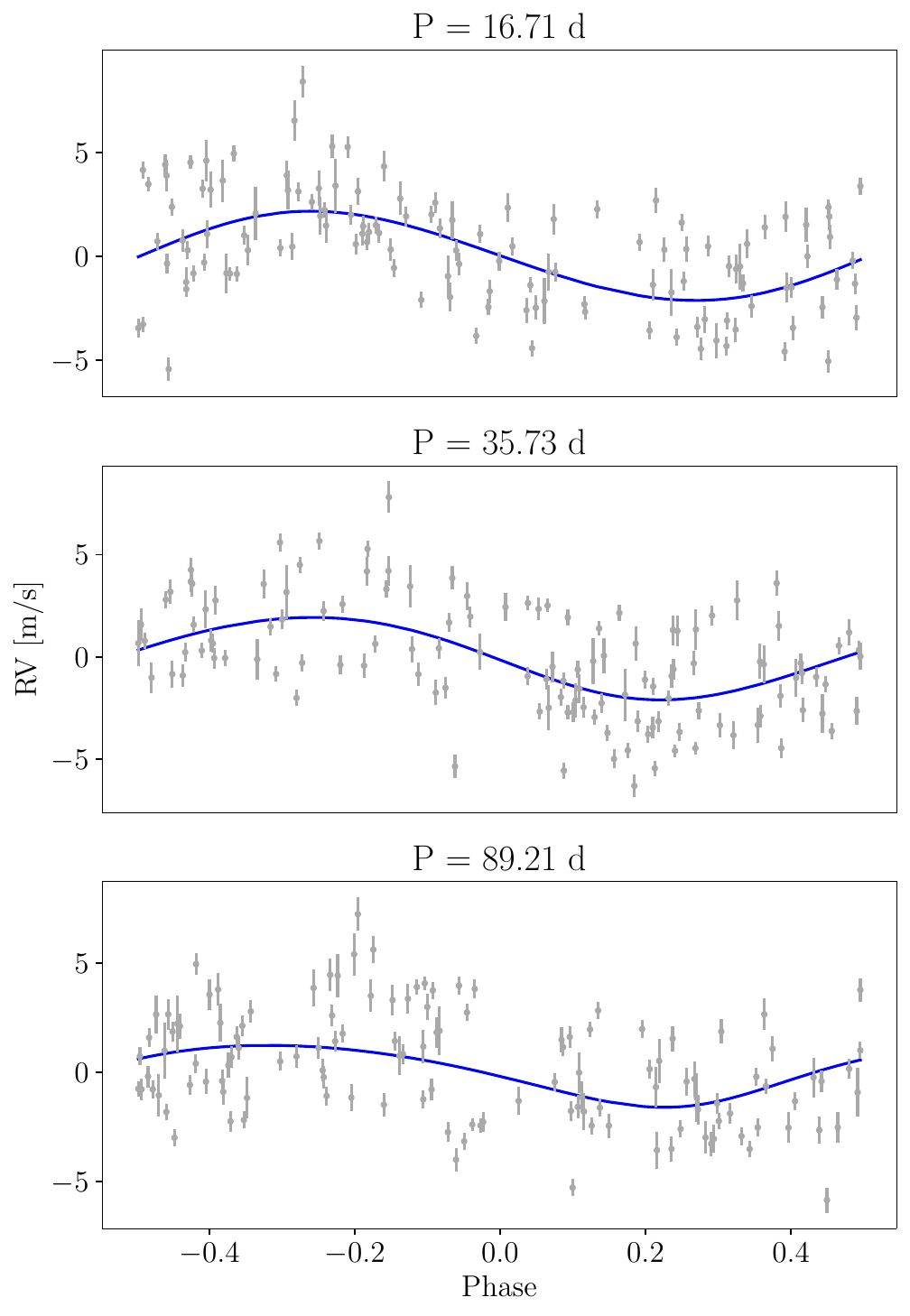}
    \caption{Phase-folded RV curves for all three signals present in the HD~85426 data. The best-fitting model is shown by the blue solid line. The RV measurements, with the contribution of the two other signals subtracted, are shown in grey. The orbital periods are displayed above each panel.}
    \label{fig:phase_fold}
\end{figure}

The phase-folded RV time series are shown in Fig. \ref{fig:phase_fold}. As apparent in this Figure, the phases of all three signals have been sampled appropriately, and the RVs agree well with the model. 

\begin{figure}
    \centering
    \includegraphics[width=0.95\columnwidth]{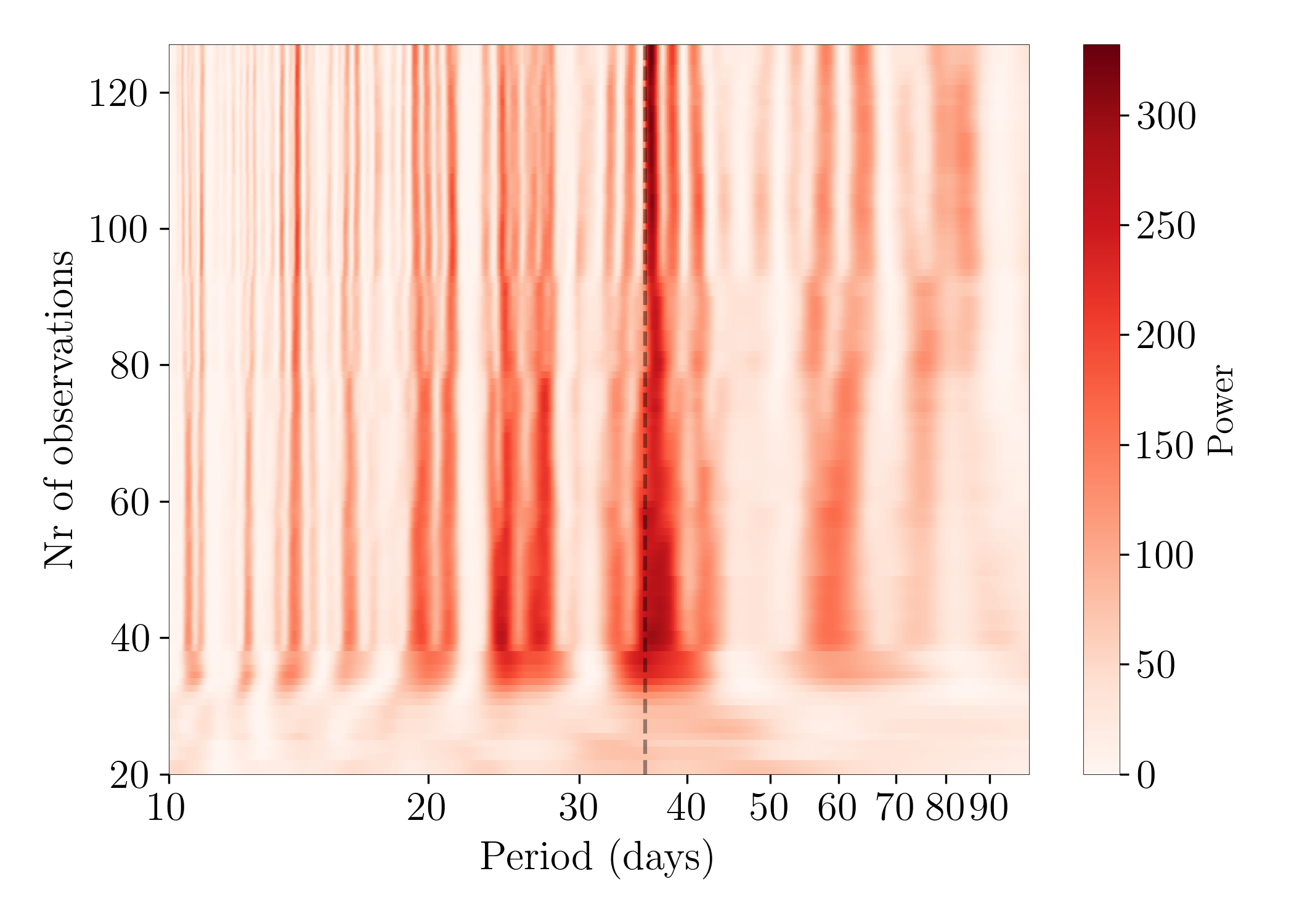}
    \includegraphics[width=0.95\columnwidth]{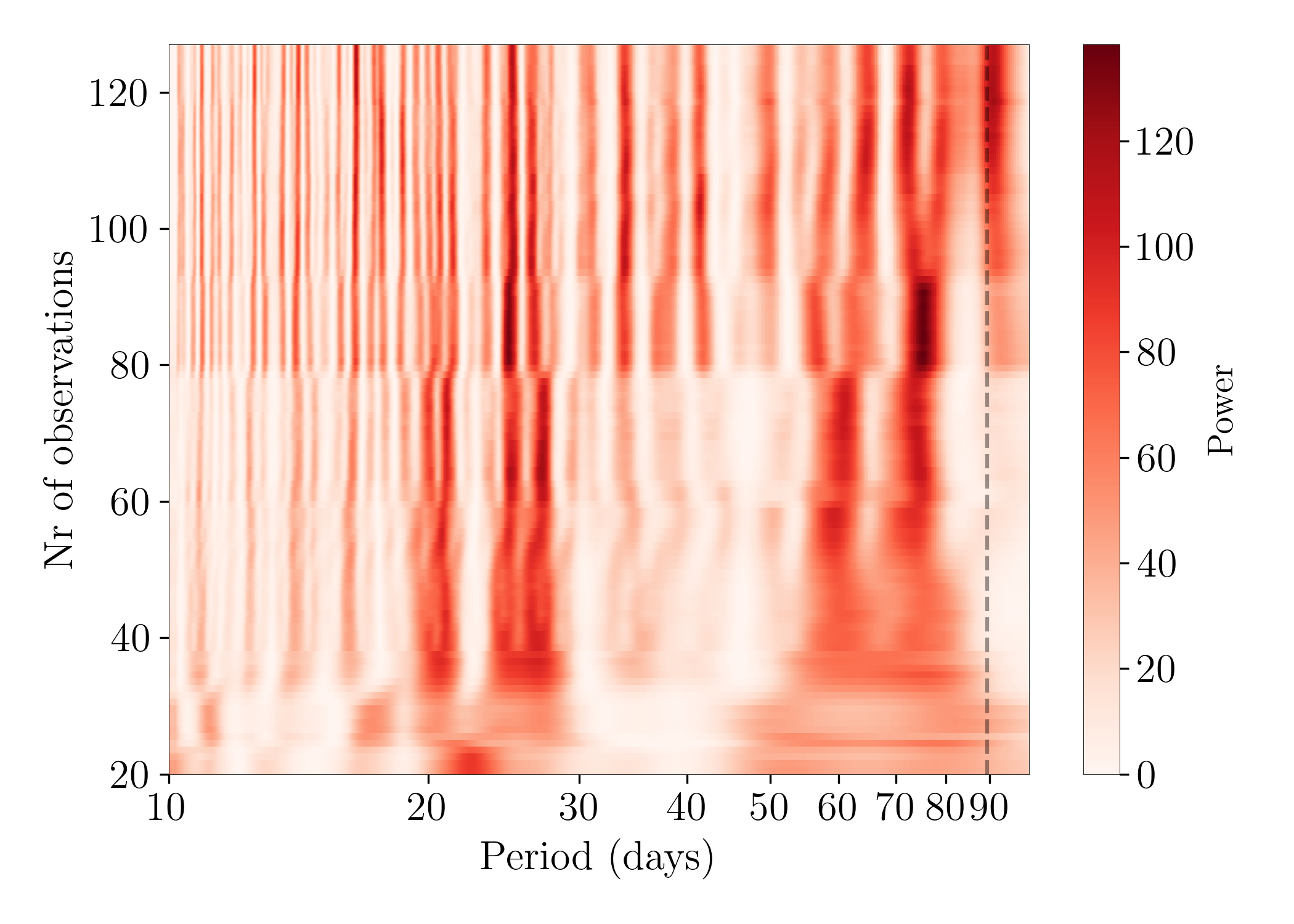}
    \caption{In the top (bottom) panel, we show the stacked BGLS periodogram for the \texttt{YARARA} RVs with planets b and d (b and c) removed. The orbital period of planet c (d) is indicated by the dashed vertical line.}
    \label{fig:bgls}
\end{figure}

We computed the stacked Bayesian GLS (BGLS) periodograms \citep{Mortier_2017} for two cases displayed in Fig. \ref{fig:bgls}: (1) the RVs with the signals from planet b and candidate d removed, and (2) RVs with the signals from planets b and c removed, such that the signal of only one planet is expected to remain in the RVs. The stacked BGLS periodogram is generated by computing the BGLS periodogram \citep{Mortier_2015} for the first $i$ observations and stacking these periodograms.
This serves to investigate whether the power of the signal in the periodogram increases as we include more data, as expected for a real signal, or whether the signal is generated by a strong artefact in the time series and its power subsides with the inclusion of more data points. For the RV signal of planet c, we find very good agreement with this premise. The case is less clear for planet candidate d, as discussed in Section \ref{thirdsignal} and shown in Fig. \ref{fig:resid_periodogram} and \ref{posterior_seasonbyseason2_p3_90}. However, it is still in line with a real signal, given the complications of having seasonal data and a comparably long period of about 89 days.

\begin{figure}
    \centering
    \includegraphics[width=0.99\columnwidth]{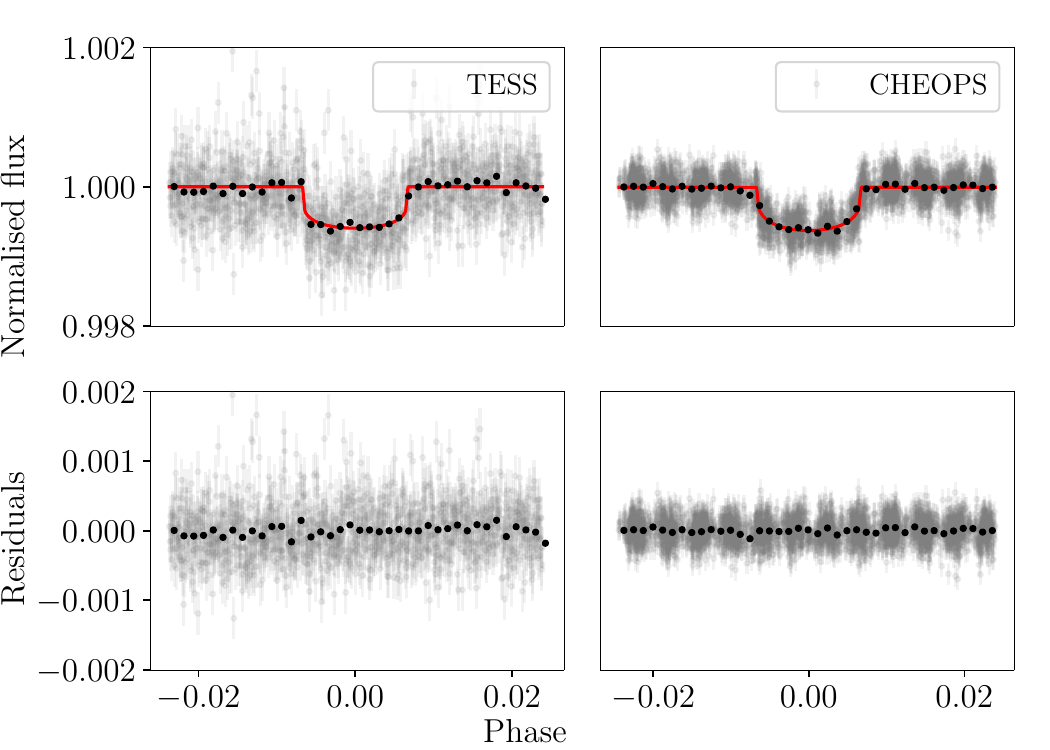}
    \caption{{\it TESS} (left) and {\it CHEOPS} (right) light curves phase-folded with the period of planet b with best-fit transit model from joint fit with \texttt{Juliet} (red). The flux values are plotted in grey. The cadence for {\it TESS} was 2 minutes and 1 minute for {\it CHEOPS}. The phase-folded light curves binned in 30-minute bins are overplotted in black.}
    \label{phase_folded_transits}
\end{figure}

We show the phase-folded transits in Fig. \ref{phase_folded_transits}. The radius posterior of planet b shows a slight correlation with the eccentricity, with the radius estimate increasing with eccentricity.
This produces a slightly asymmetric uncertainty estimate. However, the derived radius is perfectly consistent with the independently extracted value assuming circular orbits in Section \ref{TTVs}.

\begin{figure}
    \centering
    \includegraphics[width=0.95\linewidth]{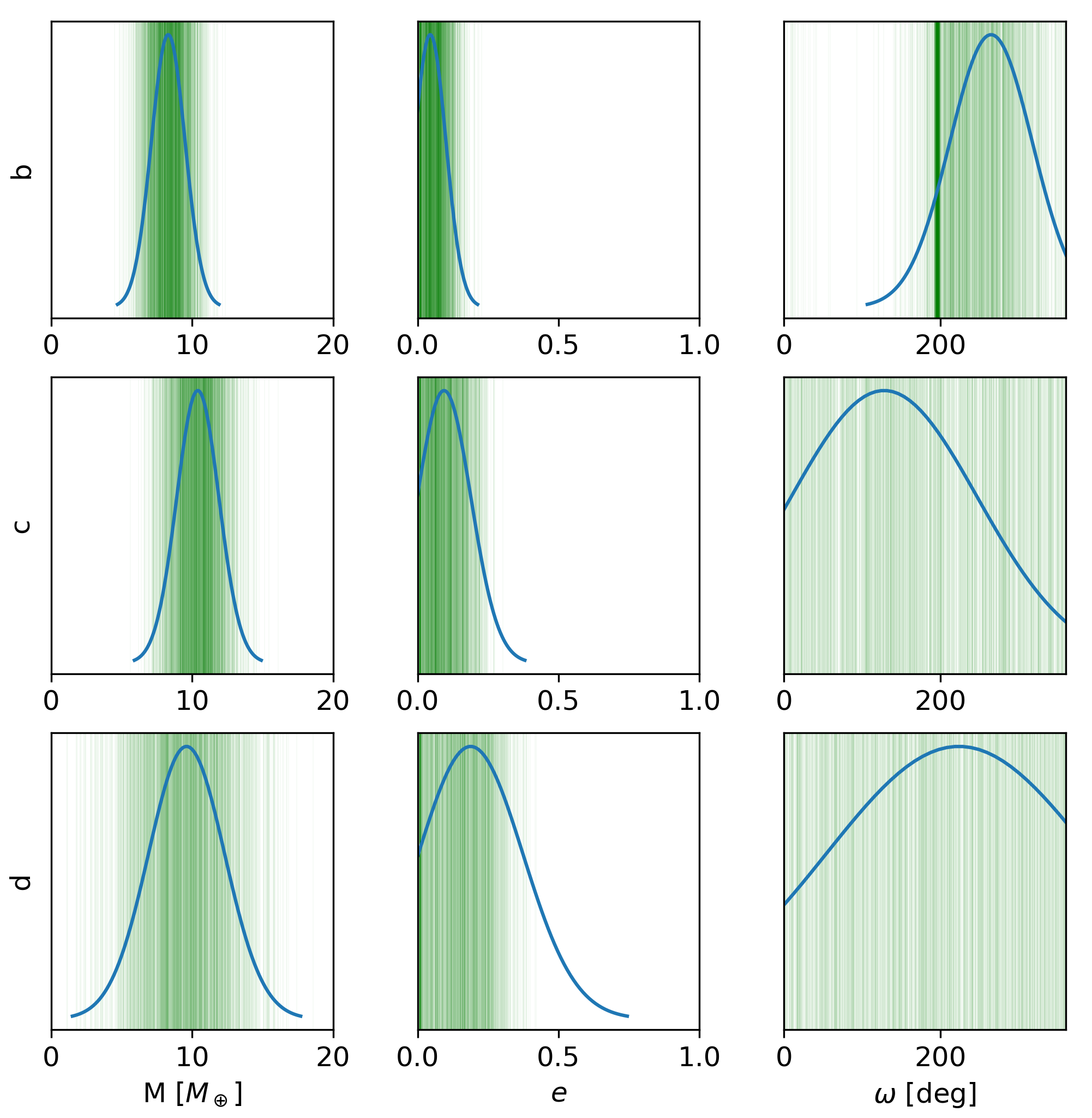}
    \caption{Posteriors of the orbital elements corresponding to the 3-Keplerian solution with $\beta$ eccentricity prior. The initial values of the stable orbital solutions are indicated with green vertical lines.}
    \label{fig:genga_sim_173691}
\end{figure}

We tested the stability of the 3-planet configuration derived with the $\beta$ eccentricity prior using a CPU version of the hybrid symplectic integrator \texttt{GENGA} \citep{Genga_2014,Genga_2022}. Of the 2300 simulated systems, 1471 systems survived without collisions for at least 10 million years, corresponding to the maximal duration of the simulation.
The initial values of the stable solutions are all consistent with the derived posteriors, with a slight preference for lower eccentricities for planet candidate d than expected from the posterior, see Fig. \ref{fig:genga_sim_173691}. The argument of periastron is set to 200 degrees if the eccentricity is equal to zero, which explains the accumulation of these values, specifically for planet b with its tight eccentricity posterior close to zero. We conclude that the true orbital parameters of the system, which we expect to correspond to one of the stable solutions, are fully consistent with the derived posteriors.

By propagating the impact factor of planet b of $0.24^{+0.17}_{-0.13}$ within 1 $\sigma$, we infer that all objects in a coplanar orbit with planet b and periods below 64 d are expected to transit. Therefore, we would expect planet c to transit if it were in a perfectly coplanar orbit with planet b. Based on the RV data, if planet c were in a transiting orbit, it would have passed between Earth and its host star during the second half of \textit{TESS} sector 21,  cf. Fig \ref{fig:tesslc}. However, no transits are discernible in the \textit{TESS} light curve at the expected time of transit or generally in the light curve, despite the expected high signal-to-noise ratio of HD~85426 c's transits. Therefore, planet c's inclination must be less than 88.6$\degr$, as compared to the derived value of 89.4$\degr$ for planet b.

To further explore the stability of the system, we also performed dynamical stability analyses following \citet{Correia_2005,Couetdic_2010}. These analyses also found that the system is indeed stable and indicated that the non-transiting planets may be in a stable 5:2 resonance. 
By varying the inclinations of planets c and d with the longitude of the ascending node set to 0$\degr$, we find a broad region of stability ranging from 20$\degr$ to 160$\degr$ for both non-transiting planets. This implies that the true masses of planets c and d are likely less than approximately 30 $\text{M}_{\oplus}$.

\begin{table*}

\caption{Joint photometry and RV (\texttt{YARARA}, set 1) fits using one or two Keplerians using \texttt{Juliet}.}
\label{tab:res_one_two_keplerian}
\begin{tabular}{lllrrrrrr}
   & & & \multicolumn{3}{c}{$\beta$ eccentricity prior} & \multicolumn{3}{c}{$\mathcal{U}$[0,0.3] eccentricity prior} \\[4 pt]
Parameter & Symbol & Unit \:\:& b & c & d \:\: & b  & c & d\\ [4pt]
\hline \\
\multicolumn{9}{c}{\textbf{One Keplerian}}\\[3pt]

Orbital period        & $P$        & d         & $16.70988^{+0.00003}_{-0.00002}$   &    &   & $16.70989^{+0.00003}_{-0.00003}$ & & \\[2 pt]
RV semi-amplitude     & $K$        & \ms          & $1.9^{+0.3}_{-0.4}$                &    &   & $1.9^{+0.4}_{-0.4}$              & & \\[2 pt]
Eccentricity          & $e$        &              & $0.12^{+0.08}_{-0.08}$              &    &   & $0.17^{+0.09}_{-0.09}$           & & \\[2 pt]
Argument of periastron& $\omega$   & deg          & $307^{+27}_{-50}$            &    &   & $308^{+25}_{-25}$          & &\\[2 pt]

Minimum Mass          & $m\sin(i)$ & M$_{\oplus}$ & $7.3^{+1.3}_{-1.4}$                &    &   & $7.5^{+1.4}_{-1.6}$              & &  \\[2 pt]

log-evidence (lnZ)    & & & \multicolumn{3}{c}{30244.5 $\pm$ 0.7} & \multicolumn{3}{c}{30244.8 $\pm$ 0.7} \\[4 pt]

\multicolumn{9}{c}{\textbf{Two Keplerians}}\\[3pt]

Orbital period        & $P$        & d         & $16.70988^{+0.00003}_{-0.00002}$   &  $35.7^{+0.1}_{-0.1}$    &    & $16.70988^{+0.00002}_{-0.00002}$ & $35.7^{+0.1}_{-0.1}$ &        \\[2 pt]
RV semi-amplitude     & $K$        & \ms          & $2.0^{+0.3}_{-0.4}$                &  $2.1^{+0.4}_{-0.4}$     &    & $2.0^{+0.4}_{-0.3}$              & $2.1^{+0.4}_{-0.3}$    &        \\[2 pt]
Eccentricity          & $e$        &              & $0.05^{+0.07}_{-0.04}$             &  $0.13^{+0.1}_{-0.08}$   &    & $0.14^{+0.06}_{-0.07}$           &  $0.14^{+0.1}_{-0.09}$ &        \\[2 pt]
Argument of periastron& $\omega$   & deg          & $273^{+42}_{-94}$            &  $112^{+204}_{-76}$ &   & $290.0^{+30.0}_{-30.0}$          & $90.0^{+200.0}_{-60.0}$&         \\[2 pt]

Minimum Mass          & $m\sin(i)$ & M$_{\oplus}$ & $8.1^{+1.4}_{-1.4}$                &$10.4^{+1.8}_{-1.8}$        &  & $8.0^{+1.4}_{-1.3}$              & $10.7^{+1.8}_{-1.8}$ &         \\[2 pt]

log-evidence (lnZ)    & & & \multicolumn{3}{c}{30255.6 $\pm$ 0.7} & \multicolumn{3}{c}{30255.5 $\pm$ 0.7} \\[4 pt]

\multicolumn{9}{c}{\textbf{Three Keplerians}}\\[3pt]

Orbital period        & $P$        & d         & $16.70988^{+0.00002}_{-0.00002}$   &  $35.73^{+0.09}_{-0.1}$    &   $89.2^{+1.8}_{-2.5}$     & $16.70988^{+0.00002}_{-0.00002}$ & $35.73^{+0.11}_{-0.09}$ & $89.5^{+1.3}_{-1.2}$        \\[2 pt]
RV semi-amplitude     & $K$        & \ms          & $2.2^{+0.3}_{-0.4}$                &  $2.0^{+0.3}_{-0.3}$     &      $1.4^{+0.4}_{-0.4}$      & $2.0^{+0.3}_{-0.3}$              & $2.2^{+0.3}_{-0.3}$    &  $1.4^{+0.3}_{-0.4}$       \\[2 pt]
Eccentricity          & $e$        &              & $0.05^{+0.07}_{-0.03}$             &  $0.06^{+0.07}_{-0.04}$   &     $0.15^{+0.23}_{-0.11}$       & $0.05^{+0.05}_{-0.03}$          &  $0.12^{+0.1}_{-0.08}$ &  $0.15^{+0.1}_{-0.1}$      \\[2 pt]
Argument of periastron& $\omega$   & deg          & $282^{+36}_{-46}$             &  $143^{+148}_{-95}$ &    $208^{+98}_{-150}$          & $261^{+42}_{-61}$         & $76^{+81}_{-50}$ & $80^{+129}_{-51}$        \\[2 pt]

Minimum Mass          & $m\sin(i)$ & M$_{\oplus}$ & $8.5^{+1.3}_{-1.4}$                &$10.3^{+1.6}_{-1.5}$        & $9.5^{+2.2}_{-2.4}$     & $8.1^{+1.1}_{-1.1}$              & $10.9^{+1.6}_{-1.7}$ & $9.8^{+2.3}_{-2.9}$       \\[2 pt]

log-evidence (lnZ)    & & & \multicolumn{3}{c}{30258.9 $\pm$ 0.8} & \multicolumn{3}{c}{30259.4 $\pm$ 0.9} \\[4 pt]

\end{tabular}
\end{table*}

\begin{table*} 
\caption{Prior and posterior distributions for the 3-Keplerian joint run. $\mathcal{U}$ indicates a uniform distribution, $\mathcal{LU}$ a log-uniform distribution, and $\mathcal{\beta}$ a beta distribution.}
\label{tab:allresults_3pmodel}
\begin{tabular}{llllc}

Parameter & Symbol & Unit & Prior distribution & Posterior \\ [4pt]
\hline 
\multicolumn{5}{c}{\emph{Fitted parameters}}\\[2 pt]
\multicolumn{5}{l}{\textbf{Planet b}}\\[2 pt]
Orbital period                 & $\text{P}_{\text{b}}$ & d               & $\mathcal{U}$[16.70959, 16.71019] & $16.70988^{+0.00002}_{-0.00002}$ \\[2 pt]
Reference conjunction time         & ${\text{T}_{\text{0}}}_{\text{b}}$ & d               & $\mathcal{U}$[59674.408, 59674.418] &$59674.4131^{+0.0005}_{-0.0005}$ \\[2 pt]
RV semi-amplitude              & $\text{K}_{\text{b}}$ & \ms             & $\mathcal{LU}$[0.1, 10.0] & $2.2^{+0.3}_{-0.4}$ \\[2 pt]
Eccentricity                   & $\text{ecc}_{\text{b}}$ &                 & $\mathcal{\beta}$[0.867, 3.03] & $0.05^{+0.07}_{-0.03}$ \\[2 pt]
Argument of periastron         & $\omega_{\text{b}}$  & deg             & $\mathcal{U}$[0.0, 360.0] & $282^{+36}_{-46}$ \\[2 pt]
$\text{r}_{1_\text{b}}$                           & $\text{r}_{1_\text{b}}$                      &                 & $\mathcal{U}$[0.0, 1.0] & $0.5^{+0.1}_{-0.1}$ \\[2 pt]
$\text{r}_{2_\text{b}}$                           & $\text{r}_{2_\text{b}}$&                 & $\mathcal{U}$[0.0, 1.0] & $0.0225^{+0.0003}_{-0.0002}$\\[2 pt]
\multicolumn{5}{l}{\textbf{Planet c}}\\[2 pt]
Orbital period                 & $\text{P}_{\text{c}}$ & d               & $\mathcal{U}$[34.4, 37.4] & $35.73^{+0.09}_{-0.1}$ \\[2 pt]
Reference conjunction time         & ${\text{T}_{\text{0}}}_{\text{c}}$ & d               & $\mathcal{U}$[59600.0, 59637.4] & $59608^{+1}_{-1}$\\[2 pt]
RV semi-amplitude              & $\text{K}_{\text{c}}$ & \ms             & $\mathcal{LU}$[0.1, 10.0] & $2.0^{+0.3}_{-0.3}$\\[2 pt]
Eccentricity                   & $\text{ecc}_{\text{c}}$ &                 & $\mathcal{\beta}$[0.867, 3.03] & $0.06^{+0.07}_{-0.04}$ \\[2 pt]
Argument of periastron         & $\omega_{\text{c}}$  & deg             & $\mathcal{U}$[0.0, 360.0] & $143^{+148}_{-95}$ \\[2 pt]
\multicolumn{5}{l}{\textbf{Planet candidate d}}\\[2 pt]
Orbital period                 & $\text{P}_{\text{d}}$ & d               & $\mathcal{LU}$[1.1, 850.0]& $89^{+1}_{-1}$  \\[2 pt]
Reference conjunction time         & ${\text{T}_{\text{0}}}_{\text{d}}$ & d               & $\mathcal{U}$[59500.0, 59590.0] & $59548^{+7}_{-11}$ \\[2 pt]
RV semi-amplitude              & $\text{K}_{\text{d}}$ & \ms             & $\mathcal{LU}$[0.1, 10.0] & $1.4^{+0.4}_{-0.4}$ \\[2 pt]
Eccentricity                   & $\text{ecc}_{\text{d}}$ &                 & $\mathcal{\beta}$[0.867, 3.03] & $0.15^{+0.23}_{-0.11}$ \\[2 pt]
Argument of periastron         & $\omega_{\text{d}}$  & deg             & $\mathcal{U}$[0.0, 360.0] &$208^{+98}_{-150}$ \\[2 pt]
\multicolumn{5}{l}{\textbf{Stellar and instrumental}}\\[2 pt]
Mean RV HARPS-N                & $\mu_{\text{HARPS-N}}$ & \ms             & $\mathcal{U}$[-5.0, 5.0] & $0.3^{+0.2}_{-0.2}$ \\[2 pt]
Quadratic ld coefficient       & $\text{q1}_{\text{TESS}}$ &                 & $\mathcal{N}$[0.33, $0.05^2$] & $0.31^{+0.04}_{-0.04}$ \\[2 pt]
Quadratic ld coefficient       & $\text{q2}_{\text{TESS}}$ &                 & $\mathcal{N}$[0.36, $0.05^2$] & $0.36^{+0.04}_{-0.04}$ \\[2 pt]
Quadratic ld coefficient       & $\text{q1}_{\text{CHEOPS}}$ &                 & $\mathcal{N}$[0.45, $0.05^2$] & $0.46^{+0.04}_{-0.04}$ \\[2 pt]
Quadratic ld coefficient       & $\text{q2}_{\text{CHEOPS}}$ &                 & $\mathcal{N}$[0.41, $0.05^2$] & $0.39^{+0.03}_{-0.04}$ \\[2 pt]
Stellar density                & $\rho$               & kg/$\text{m}^3$ & $\mathcal{N}$[966.0, $70.0^2$] & $973^{+52}_{-43}$ \\[2 pt]
dilution-\textit{TESS}                 &                      &                 & Fixed 1.0 &  \\[2 pt]
mflux-\textit{TESS}                     &                      &                 & $\mathcal{N}$[0.0, $0.01^2$] & $5^{+2}_{-2} 10^{-5}$ \\[2 pt]
Offset-\textit{TESS}                    & $\phi_{\text{TESS}}$ &                 & $\mathcal{U}$[-0.001, 0.001] & $-0.00005^{+0.00003}_{-0.00003}$ \\[2 pt]
Gradient-\textit{TESS}                    & $\theta_{\text{TESS}}$ & $\text{d}^{-1}$                & $\mathcal{U}$[-0.001, 0.001] & $0.00038^{+0.00007}_{-0.00006}$ \\[2 pt]
Scatter \textit{TESS}                   & $\sigma_{\text{TESS}}$ & ppm                 & $\mathcal{LU}$[1.0, 500.0] & $272^{+11}_{-11}$ \\[2 pt]
dilution-{\it CHEOPS}               &                      &                 & Fixed 1.0 &  \\[2 pt]
mflux-{\it CHEOPS}                   &                      &                 & $\mathcal{N}$[0.0, $0.01^2$] & $7^{+3}_{-3} 10^{-6}$ \\[2 pt]
Scatter {\it CHEOPS}                 & $\sigma_{\text{CHEOPS}}$ & ppm                 & $\mathcal{LU}$[1.0, 500.0] & $85^{+4}_{-4}$ \\[2 pt]
Scatter HARPS-N                & $\sigma_{\text{HARPS-N}}$ & \ms                & $\mathcal{LU}$[0.1, 10.0] & $2.3^{+0.2}_{-0.2}$ \\[4 pt]
\multicolumn{5}{c}{\emph{Derived parameters}}\\[2 pt]
\multicolumn{5}{l}{\textbf{Planet b}}\\[2 pt]
Radius &  $R_{\text{b}}$ & $\text{R}_{\oplus}$  && $2.78^{+0.05}_{-0.04}$\\[2 pt]
Impact parameter & b &  && $0.24^{+0.17}_{-0.13}$\\[2 pt]
Scaled semi-major axis & a/$R_{\ast}$ &  && $24.3^{+0.4}_{-0.4}$\\[2 pt]
Inclination &  $i_{\text{b}}$ & \text{deg}  && $89.4^{+0.3}_{-0.4}$\\[2 pt]
Transit duration &  $\text{T}_{\text{14}}$ & h  & & $5.41^{+0.03}_{-0.03}$  \\[2 pt]
Minimum mass & $m_{\text{b}}\sin(i_{\text{b}})$ & M$_{\oplus}$ &    &$8.5^{+1.3}_{-1.4} $  \\[2 pt]
Equilibrium temperature (black body) & $T_{\text{eq}}$ & K  && $824^{+11}_{-11}$\\[2 pt]
\multicolumn{5}{l}{\textbf{Planet c}}\\[2 pt]
Minimum mass planet c& $m_{\text{c}}\sin(i_{\text{c}})$ & M$_{\oplus}$    &  & $10.3^{+1.6}_{-1.5} $   \\[2 pt]
Scaled semi-major axis & $\text{a}/\text{R}_\ast$ &&& $40.3^{+0.7}_{-0.6}$ \\ [0.2pt]
Equilibrium temperature (black body) &$\text{T}_{\text{eq}}$& K && $640^{+9}_{-9}$\\ [0.2pt]
\multicolumn{5}{l}{\textbf{Planet candidate d}}\\[2 pt]
Minimum mass planet candidate d& $m_{\text{c}}\sin(i_{\text{c}})$ & M$_{\oplus}$    &  & $9.5^{+2.2}_{-2.4} $   \\[2 pt]
Scaled semi-major axis & $\text{a}/\text{R}_\ast$ &&& $74^{+2}_{-3}$ \\ [0.2pt]
Equilibrium temperature (black body) &$\text{T}_{\text{eq}}$& K && $472^{+12}_{-10}$\\ [0.2pt]
\end{tabular}	
\end{table*}

\subsection{Gaussian Process regression applied on DRS RVs}

In this section, we briefly describe attempts to extract the planetary signals directly from the DRS data (set 1). The fits presented so far in this study were based on the assumption that there is a negligible amount of residual stellar activity and instrumental noise in the RVs after postprocessing with \texttt{YARARA}. 
To test our inferences, we attempted to extract the planetary signals directly from the DRS RVs. The latter contain the unfiltered stellar signal manifesting as correlated noise, which we attempted to account for with Gaussian Process \citep[GP;][]{gp_2006} regression, as is commonly done in RV analyses \citep[e.g.][]{Haywood_2014,Rajpaul_2015,Espinoza_2020,Dalal_2024}. 
We did not use the multidimensional GP approach \citep[e.g.][]{Rajpaul_2020,Barragan_2022} due to the contamination of the DRS activity indices.
Instead, we employed a one-dimensional GP model, as in e.g. \citet{Espinoza_2020} or \citet{Dalal_2024}, fitting the GP simultaneously with the Keplerians to the RV data. For this, we tested the exp-sine-squared kernel described in \citet{Haywood_2014} and the exponential kernel defined in \texttt{celerite} \citep{celerite}. 
While we obtained planetary mass estimates for planet b consistent with the masses derived in the previous sections, we found that the GP kernel’s decay timescales converged to 2 to 3 days. These short timescales are atypical for stellar activity. Therefore, other noise components are interfering with the modelling, and the stellar contribution to the RVs has not been properly accounted for. Given the limited number of RVs with uneven sampling, we conclude that the results from the GP runs cannot contribute to a meaningful analysis of this RV dataset, and we explored other methods to test our results.

\subsection{TWEAKS analysis} \label{s:tweaks}

Our results from the previous sections rely on \texttt{YARARA} as an RV post-processing pipeline. However, cleaning stellar spectra and properly extracting RVs is a very challenging task. No current method is expected to perfectly disentangle RV contributions from the planets, the star, the instrument, or variations due to the atmosphere of the Earth \citep{Zhao_2022}. 

To test our previously obtained results, we ran \texttt{TWEAKS} on the DRS 3.0.1 data. The inner workings of this pipeline are explained further in Section \ref{sss:tweaks_descr}.
We used set 1 to include the same data as in the \texttt{YARARA}-based analysis above. Since there is ample evidence for at least one other signal, we ran \texttt{TWEAKS} modelling two, three, and four Keplerians in three separate runs, with the parameters of planet b being informed by the transits, i.e. we used Gaussian priors on the conjunction time and period, as in Section \ref{s:dns_runs}.

\subsubsection{A different solution from \texttt{TWEAKS}} \label{sss:different_solution}
In the first run, we set the same uninformative priors as in Section \ref{s:dns_runs} on the orbital parameters of the non-transiting planets. For a 2-Keplerian fit, \texttt{TWEAKS} favoured a companion to planet b with a period of about 70 d in an eccentric orbit. This signal persisted if we allowed more Keplerians, with these runs favouring an additional Keplerian with a period of about 38 d.
The mass associated with planet b converged to a value between 4 and 5 $\text{M}_{\oplus}$ and is thus significantly below our previous estimates, which led us to cross-check this solution in the \texttt{YARARA} RVs.

We could recreate the architecture favored by \texttt{TWEAKS} with the \texttt{YARARA} data by setting a uniform prior $\mathcal{U}[36.6,38.8]$~d on the period of the second Keplerian, excluding the period of 35.7~d that was otherwise preferred, but allowing the solution found by \texttt{TWEAKS}.
In this way, we found a very similar solution involving three Keplerians with periods 16.7 d, 38 d, and 71 d. 
As in the solution suggested by \texttt{TWEAKS}, we found a large eccentricity for the third Keplerian of $0.53^{+0.11}_{-0.09}$.
The mass estimate for planet b converged to $7.5^{+1.7}_{-1.2}$ $\text{M}_{\oplus}$, which is consistent with our other estimates but shows an intrinsic difference between the masses inferred by \texttt{TWEAKS} and the masses derived from the \texttt{YARARA} RVs. 
As shown in Appendix \ref{appendix:38or36}, the 38 d solution is not stable in time for the \texttt{YARARA} data, implying a phase offset in season 2, and raising first doubts about the validity of this solution. However, one could imagine a real effect, such as periastron precession, or an inadequate model fooled by the superposition of Keplerians to produce the discrepancy in mass estimates. We, therefore, performed N-body simulations with \texttt{GENGA} drawing from the posteriors of the 3-Keplerian fit to test the stability of this orbital solution. We expect to observe a stable planetary system, as it is very unlikely to observe an unstable system given the age of the star.

\begin{figure}
    \centering
    \includegraphics[width=0.85\linewidth]{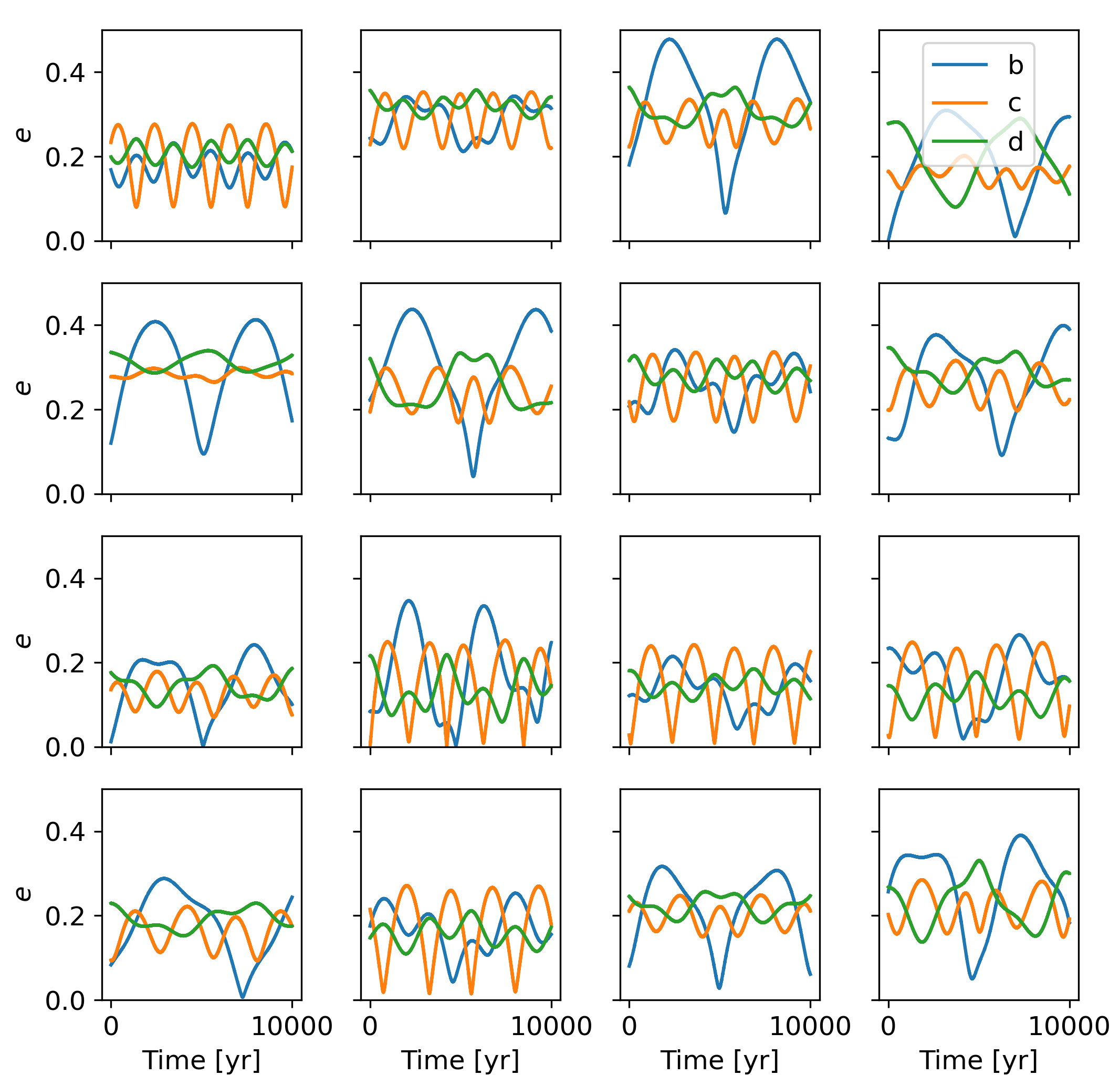}
    \includegraphics[width=0.85\linewidth]{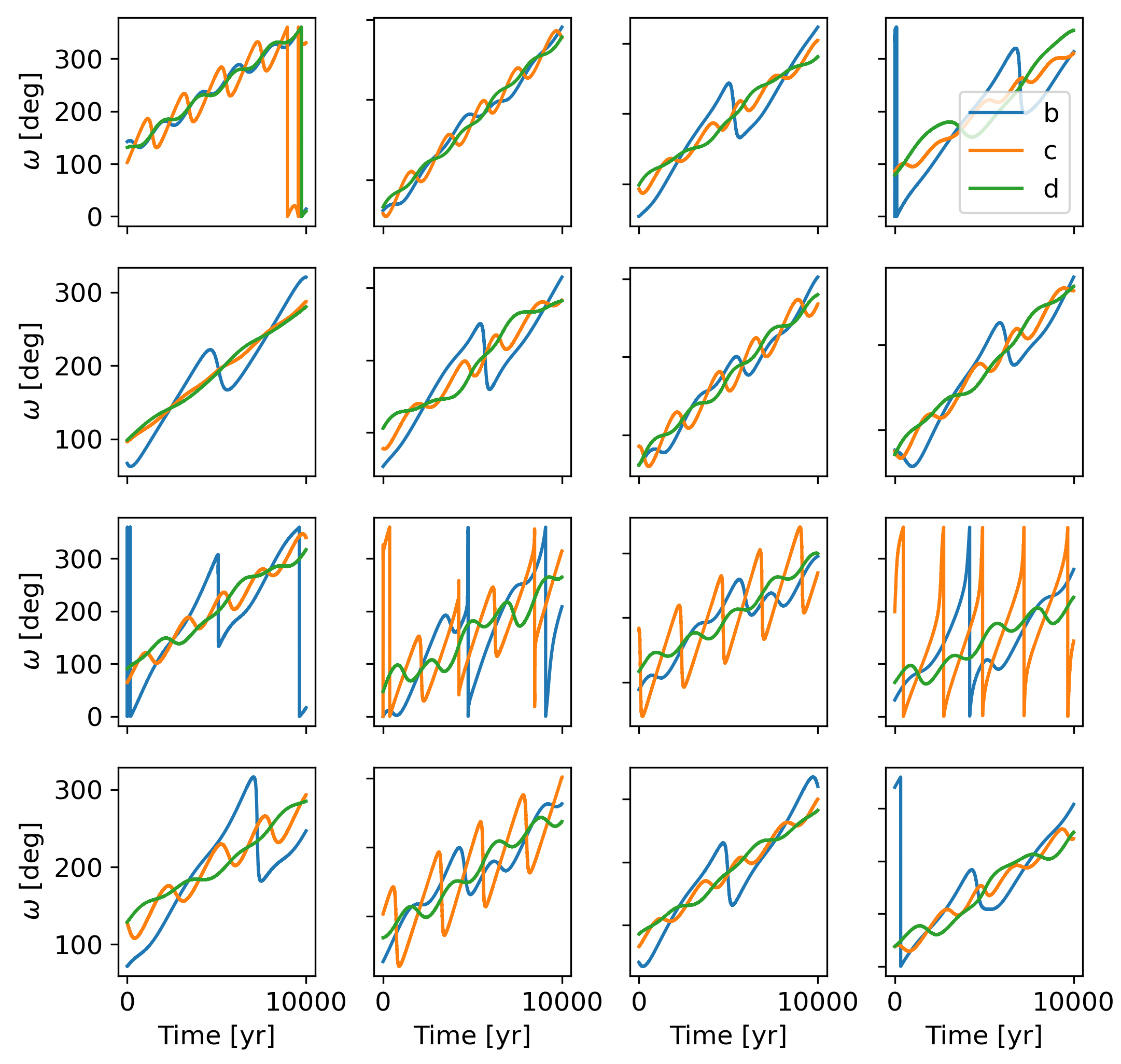}
    \includegraphics[width=0.85\linewidth]{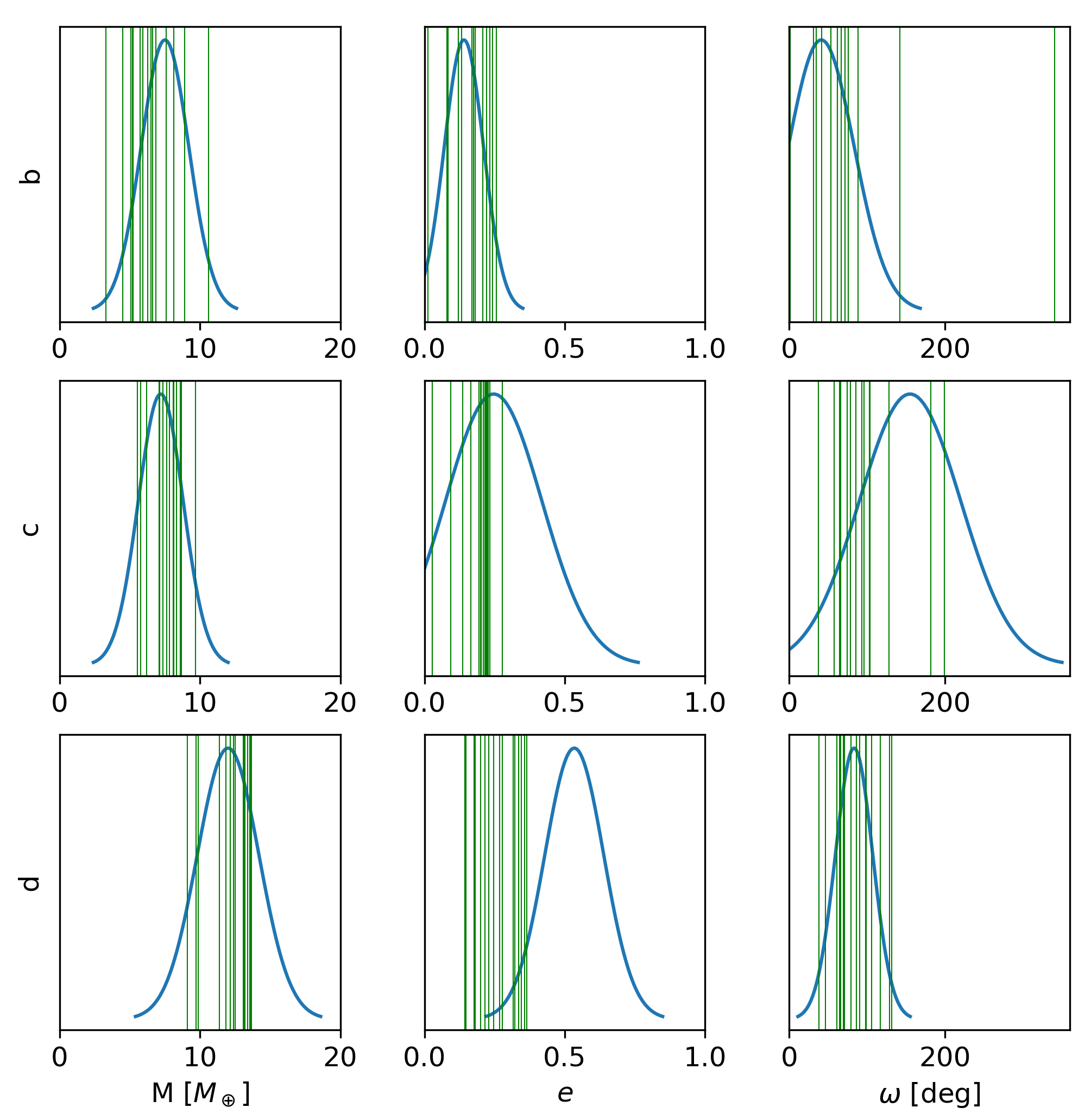}
    \caption{Eccentricity (top panel) and argument of periastron (middle panel) of the 16 stable solutions associated with initial periods of 16.71 (b), 38 (c), and 71 d (d) for the first 10,000 years of the simulation. The posteriors with initial values of the stable solutions indicated by green vertical lines are shown in the bottom panel.}
    \label{fig:genga_sol}
\end{figure}

We found that 16 out of 13,000 simulated systems did not lead to collisions and were stable throughout the duration of the simulation, which was again set to 10 million years. Note that more simulations were performed in this run compared to Section \ref{multi_joint_fit_juliet_yarara}, where we simulated a total of 1651 systems, because most of the simulated systems in this configuration led to collisions very early on, reducing the computational time required per simulated system.
We show the eccentricities and arguments of periastron of the 16 stable solutions for the first 10,000 years, after which they continue in a similar manner, in Fig. \ref{fig:genga_sol}. All stable solutions show strong variations in eccentricity and argument of periastron, with some solutions showing distinctively similar features, indicating that there are some islands of stability in the parameter space. As a first conclusion, we find that the orbital elements evolve with periods of the order of 1000 years and therefore we can assume the eccentricity and argument of periastron to be constant over the duration of our observations of about 3 years. 
It is particularly striking that the eccentricity of the third Keplerian of $0.53^{+0.11}_{-0.09}$ coming from the best fit to the RVs is not compatible within about two $\sigma$ with any of the stable solutions.

We conclude that the orbital configuration associated with the best-fit parameters is not stable because none of the simulated systems with values near the peak of the posterior survived in the simulation. 
Therefore, if we are indeed dealing with a system with three planets with periods of 16.7, 38, and 71 d, we expect to observe it during a time of its existence when the eccentricities are near or below 0.3.
Therefore, we reran \texttt{TWEAKS} excluding the solutions with high eccentricities, which we have shown to be unstable, but still including the stable solutions.

\subsubsection{Second \texttt{TWEAKS} run}

In this run, we restricted the eccentricity to a uniform prior $\mathcal{U}[0,0.3]$ as motivated in the previous Section \ref{sss:different_solution} and in Section \ref{multi_joint_fit_juliet_yarara} where we found that the eccentricity of all three modelled best-fit Keplerians is expected to be below 0.3.

Running \texttt{TWEAKS} with this new prior excluding the dynamically unstable solutions, \texttt{TWEAKS} favoured the same solution as the analysis on the \texttt{YARARA} RVs. More specifically, we found that the median period of the second Keplerian in the 2-Keplerian model was equal to $35.8^{+0.1}_{0.1}$ d. For a 3-Keplerian model, we found a period of $35.8^{+0.1}_{-0.1}$ d for the first outer companion and $90.0^{+0.9}_{-0.9}$ d for the second companion of planet b. These results again corroborate our findings in the previous Section regarding the non-transiting planet c and the likely existence of planet candidate d.

The log-evidence for \texttt{TWEAKS} increases from -339.0 (two Keplerians) to -334.2 (three Keplerians) and therefore clearly favoured the 3-Keplerian model. Following \citet{Kass_1995}, a difference in log-evidence greater than 4.6 can be interpreted as decisive evidence. For \texttt{TWEAKS}, we also tested modelling four Keplerians. This increased the log-evidence to -328.8, which is again decisively favoured over the 3-Keplerian model. The 4-Keplerian model also favoured including a $35.8^{+0.1}_{-0.1}$ and a $89.9^{+0.9}_{-0.8}$ d Keplerian, as found in the previous run and the previous sections, but did not converge to a unique, well-constrained solution for this fourth Keplerian. Therefore, we consider the 3-Keplerian model to be the best-suited. 

The False Inclusion Probability \citep[FIP;][]{Hara_2022} value for planet c with a period of $35.8^{+0.1}_{-0.1}$ d is 30 per cent in the 3-Keplerian model, and we find the FIP of the signal at $90.0^{+0.9}_{-0.9}$ to be 36 per cent. These values are comparable to what we found in the analysis using \texttt{Juliet} applied to the \texttt{YARARA} RVs including the photometric data in Section \ref{multi_joint_fit_juliet_yarara}.

\begin{figure}
    \centering
    \includegraphics[width=0.99\columnwidth]{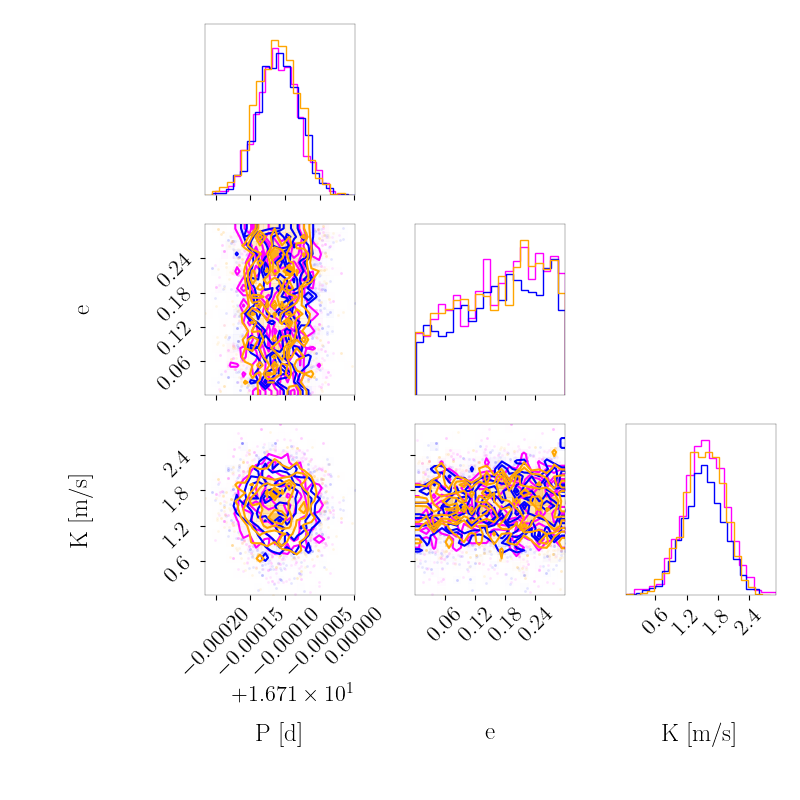}
    \caption{Posteriors of planet b's semi-amplitude $K$, eccentricity $e$, and period $P$ for a 1- (magenta), 2- (cyan), and 3-Keplerian model (blue).}
    \label{fig:tweaks_mass posterior}
\end{figure}

The orbital parameter estimates for planet b are practically independent of whether we model up to three other Keplerians. We derive minimum masses of $6.0^{+1.6}_{-1.6}$, $6.0^{+1.5}_{-1.6}$, $6.1^{+1.5}_{-1.5}$ $\text{M}_{\oplus}$ for the three cases, respectively. Note that this model cannot constrain the eccentricity of the planet's orbit, as visible in Fig. \ref{fig:tweaks_mass posterior}.
For planet c, we find a minimum mass of $11.5^{+1.9}_{-2.0}$ $\text{M}_{\oplus}$, and the minimum mass associated with the 90-d signal is $13.3^{+3.0}_{-2.5}$ $\text{M}_{\oplus}$. These values are marginally higher than what we found in Section \ref{multi_joint_fit_juliet_yarara}, where we found a minimum mass for planet c of $10.3^{+1.6}_{-1.5}$ ($10.9^{+1.6}_{-1.7}$) $\text{M}_{\oplus}$ and $9.5^{+2.2}_{-2.4}$ ($9.8^{+2.3}_{-2.9}$) $\text{M}_{\oplus}$ for the 90-d signal. 

\section{Long period massive planets} \label{lpps}

We can use the three years of RV data to draw conclusions about the existence of outer planets. For instance, a Jupiter-mass planet with an orbital period of a few years in a transiting orbit would produce a noticeable trend in the RVs.
After subtracting the RV signal of the three planets from the \texttt{YARARA} RVs, we find a maximum offset between any two seasons of 1.44 \ms and no large gradients within the seasons. Therefore, we can exclude the existence of such a planet. 

\begin{figure}
    \centering
    \includegraphics[width=0.99\columnwidth]{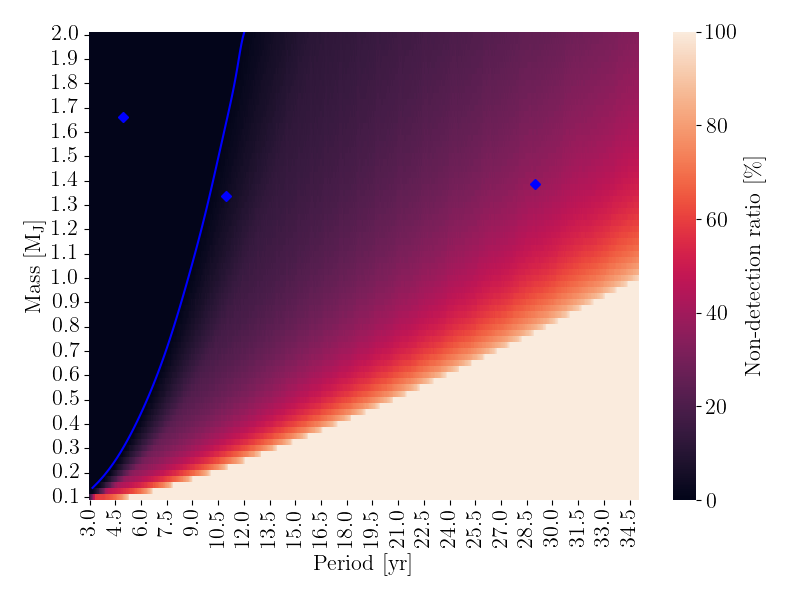}
    \caption{Non-detection ratio of the RV signature of massive outer planets on edge-on circular orbit. Planets with parameters to the left of the blue line have a non-detection ratio of zero. The blue diamonds indicate the upper mass limits for companions at different distances to the star from \citet{Kervella_2022}. The quoted mass values correspond to $m\sin(i)$ for the RV-based analysis.}
    \label{fig:massive_outer_planets}
\end{figure}

We applied a very simple criterion to explore the sensitivity of our analysis to outer massive planets. We assumed circular edge-on orbits and simulated the RV signature of planets with orbital periods between 3 and 35 years and minimum masses between 0.1 and 2 $\text{M}_{\text{J}}$ on a grid of phases from 0 to 360 degrees. If the RV signal of such a planet produced an offset between any two seasons greater than twice the largest offset we measure, we deduced that it cannot exist. We show the fraction of non-detections, i.e. planets that would produce a signature smaller than the threshold of 2.88 \ms, for all tested parameters in Fig. \ref{fig:massive_outer_planets}. We also show the limits to the mass of outer planets derived from astrometry from \texttt{Hipparcos} and {\it Gaia} in \citet{Kervella_2022} who do not find a significant indication of any outer planets. This means that we can rule out the existence of, for example, a Jupiter-mass planet with an orbital period smaller than about 9 years, assuming circular edge-on orbits. Taking into account the $\sin(i)$ factor, we can rule out planets with masses greater than 1 $M_\text{J}$ with orbital inclinations between 30 and 150 degrees with orbital periods smaller than about 6 years. From astrometry, we can also rule out planets with masses greater than the values from \citet{Kervella_2022} indicated in Fig. \ref{fig:massive_outer_planets}.

\section{Transit timing variations} \label{TTVs}

We analysed the two {\it TESS} transits and the four {\it CHEOPS} transits in the search for Transit Timing Variations (TTVs) that could hint at the existence of planets other than HD~85426~b orbiting this star. First, we removed outliers from the {\it CHEOPS} light curves \citep[default aperture size of 25~px, DRP v14][]{Hoyer2020A&A...635A..24H} by removing any flux point deviating by more than 5$\sigma$ from the biweight curve (window-length equal to a {\it CHEOPS} orbit of 98.77 minutes) using the \texttt{wotan} package \citep{Hippke_2019}.

Next, we detrended the {\it CHEOPS} light curves by fitting, using PyDE \citep{Parviainen_2016}, four parameters to each visit (constant flux, linear and quadratic term in time, background), and 13 parameters common to all visits (contamination, smearing, $\mathrm{d}x$, $\mathrm{d}x^2$, $\mathrm{d}y$, $\mathrm{d}y^2$, $\mathrm{d}x\mathrm{d}y$, and 3 harmonics of the sine and cosine of the roll angle) to the out-of-transit data, extrapolating to the in-transit data.  We removed outliers from PDCSAP {\it TESS} photometry with a procedure similar to {\it CHEOPS} light curves,  but applying \textsc{wotan}-biweight curve with a window-length of 1.3 d and an asymmetric clipping ($5\sigma$ below, $3\sigma$ above). It is important to note that the weight-flattened light curves were used solely to remove outliers, and not for computing the {\it CHEOPS} detrending and subsequent analysis.

We then fit a 2-planet model, i.e. the conservative solution, to the \texttt{YARARA} RVs (set 1) and the photometric data (\textit{TESS} data portioned around each transit time, spanning around three transit durations) with PyORBIT \citep{Malavolta_2016, Malavolta2018}. We assumed circular orbits in this analysis, which is well justified by our earlier findings. We fitted for the stellar density ($\rho_\star \sim \mathcal{N}$[0.69, $0.05^2$]), period ($P_\mathrm{b} \sim \mathcal{U}$[14.0, 18.5]~d and $P_\mathrm{c} \sim \mathcal{U}$[30, 40]~d) and RV semi-amplitude ($K$) for both planets, planet-to-star radius ($R_\mathrm{b}/R_\star$), impact parameter ($b$), and reference mid-transit time ($T_{0,\mathrm{ref, b}}$) only for b. The limb darkening coefficients were chosen the same as in Section \ref{multi_joint_fit_juliet_yarara}. For the {\it TESS} portion, we added a linear trend to take out-of-transit slopes into account. An RV offset and jitter (in base-2 log-scale) have been included in the analysis.

\begin{figure}
    \centering
    \includegraphics[width=0.9\columnwidth]{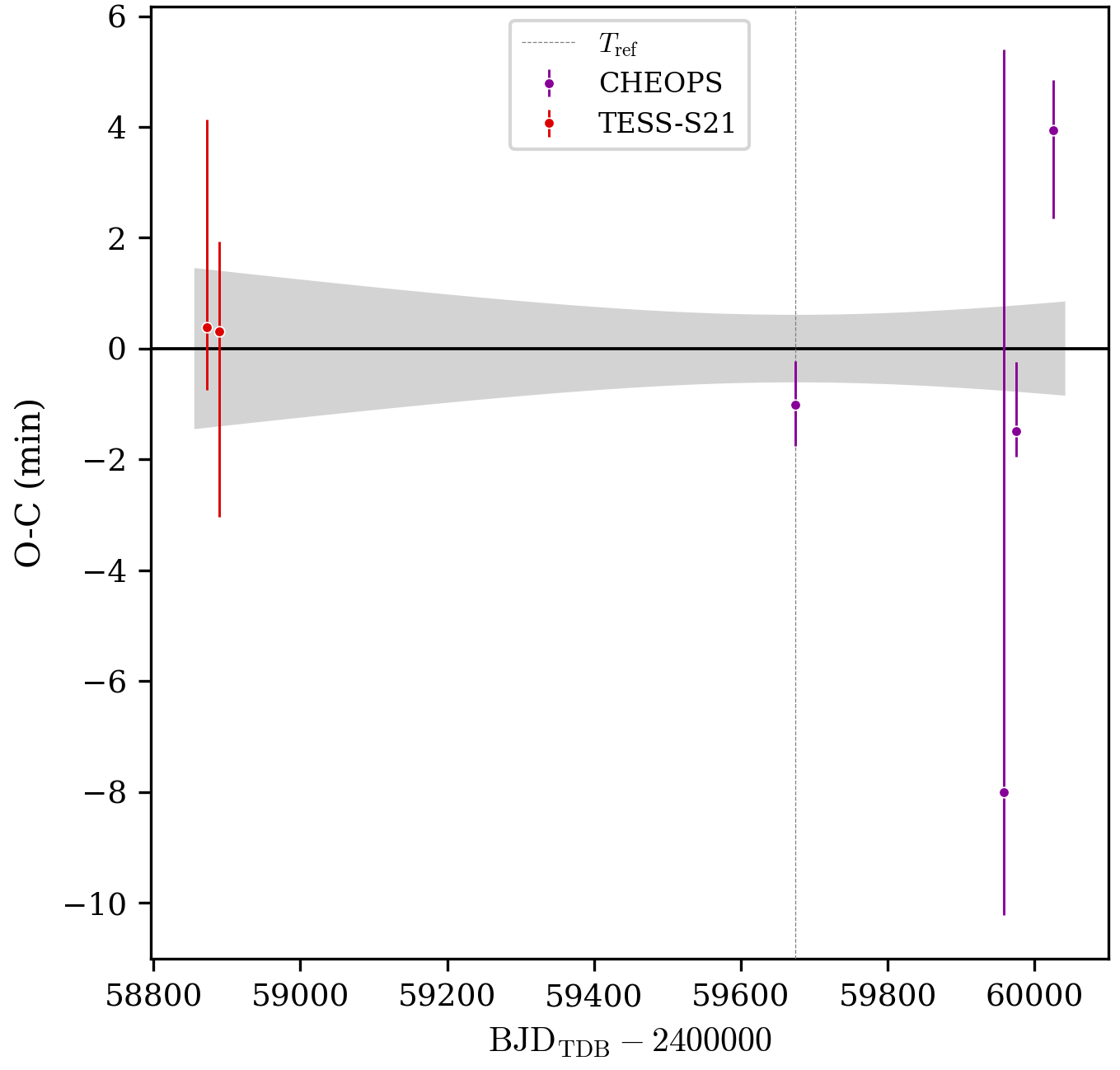}
    \caption{Transit timing variations (observed minus computed mid-transit time, O-C) of HD~85426~b, as observed in the {\it TESS} (red) and {\it CHEOPS} (purple) transits. The shaded region indicates the uncertainty of the computed mid-transit time.}
    \label{fig:ttvs}
\end{figure}

We combined \textsc{PyDE} and \textsc{emcee} for 100,000 generations and 500,000 steps, respectively. We applied a conservative thinning factor of 100 and discarded the first 200,000 steps as burn-in  (after checking convergence through the autocorrelation function, Gelman-Rubin statistics, and visual inspections of the chains). Using the Maximum-a-Posteriori (MAP) value of the period and of the transit reference time posteriors, we could then compute the expected mid-transit times of the individual transits (C). The observed (O) mid-transit times were computed by fitting the transits individually, fixing the period of planet b to the one resulting from the previous analysis. We show O-C for all transits in Fig. \ref{fig:ttvs}. 

The ingress and egress were missed for the penultimate observed transit, resulting in comparably large error bars. A faint hint of TTV is observable, but more observations are necessary to confirm this and deduce the parameters of a potential perturber.

Based on the best-fit solution from Table \ref{tab:allresults_3pmodel}, the expected TTV semi-amplitude integrated over 5 years with \texttt{TRADES} \citep{Borsato_2014} is 2.97 min. These TTVs are small and comparable in magnitude to the uncertainties on the transit times. Sampling randomly 100 times from the posterior Keplerian solutions yields a semi-amplitude of $3.27^{+7.74}_{-2.00}$ min, suggesting that there may be measurable TTVs depending on the system's true configuration. This implies that additional transit observations may, but are not guaranteed to, provide further insight into the system's configuration. Note that these estimates depend on, for example, the unknown inclination of planet c and planet candidate d. For this analysis, we randomly sampled these values from those derived for planet b.

The PyORBIT fit also represents an independent analysis of the data, with the {\it CHEOPS} data being detrended differently to the \texttt{pycheops}-generated light curves that we used in the previous parts of this study. The planetary radius converged to $2.77^{+0.03}_{-0.03}$ $\text{R}_{\oplus}$. The radius of $2.78^{+0.05}_{-0.04}$ $\text{R}_{\oplus}$ that we derived in Section \ref{multi_joint_fit_juliet_yarara}, as given in Table \ref{tab:allresults_3pmodel}, agrees very well with this result. 

\section{Internal structure modelling} \label{InteriorComposition}
\begin{figure}
    \centering
    \includegraphics[width=0.99\columnwidth]{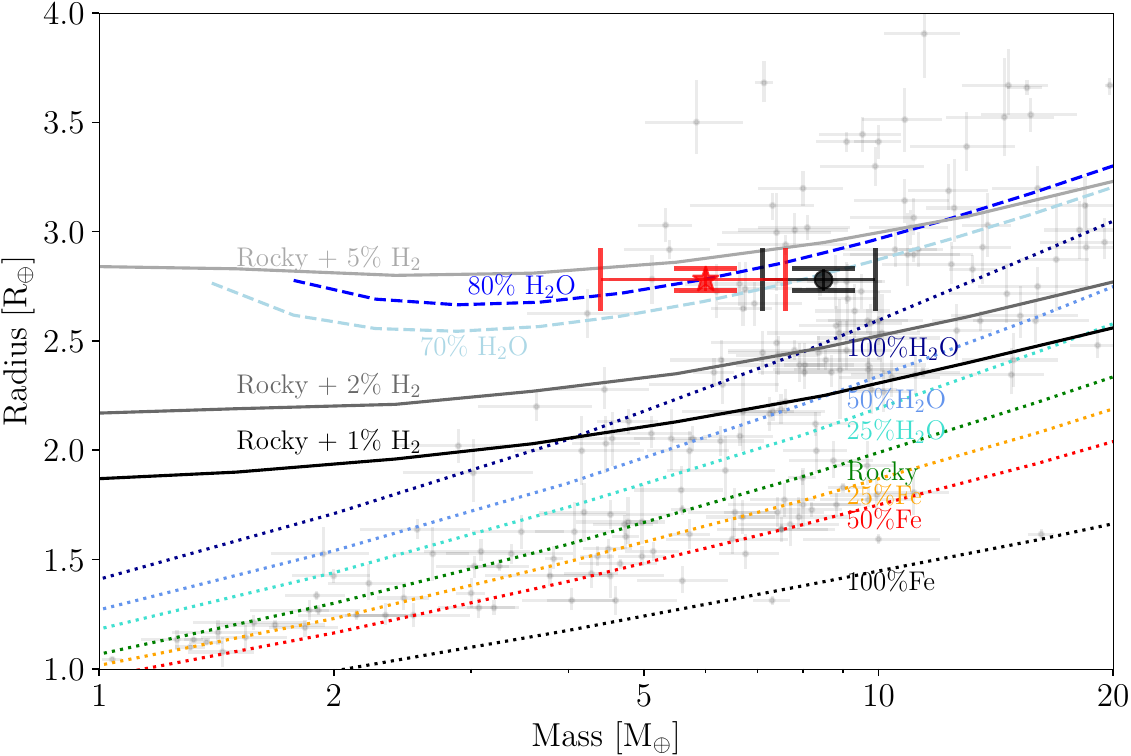}
    \caption{Mass-Radius diagram for HD~85426~b. The \texttt{TWEAKS} mass is highlighted by the red star, and the mass extracted from the \texttt{YARARA} RVs with a 3-Keplerian model is marked by the black marker. Other confirmed planets with mass uncertainties below 20 per cent and radius uncertainties below 10 per cent are indicated by light grey markers. The dotted lines show the planetary composition results from \citet{Zeng_2019}. The mass-radius estimates from \citet{Lopez_Fortney_2014} (10 Gyr, solar metallicity, 10 F$_{\oplus}$) are shown with grey solid lines. The dashed lines show the compositional tracks from \citet{Aguichine_2021} at an irradiation temperature of 800 K for a core to core+mantle mass fraction of 20 per cent.}
    \label{fig:mr_plot}
\end{figure}

We put HD~85426~b in context with the known exoplanets with precisely measured masses and compositional models in Fig. \ref{fig:mr_plot}. The shown compositional curves from \citet{Lopez_Fortney_2014} are based on a model including a rocky core and a layer of H–He. The models from \citet{Aguichine_2021} assume an irradiated ocean world with varying water mass fractions. We also show the compositional results from \citet{Zeng_2019} assuming no volatiles.

This analysis suggests that HD~85426~b is consistent with a water world structure, as well as a rocky core with a 2 per cent H/He atmosphere. 
However, the existence of sub-Neptunes with water mass fractions as high as 70 to 80 per cent, as needed to explain the measured bulk density of HD~85426~b, is being contested by ab-initio simulations \citep{Luo_2024} and global equilibrium chemistry models \citep{Werlen_2025} due to magma ocean-atmosphere interaction.
The conclusions based on the bulk density hold for both the \texttt{TWEAKS} and \texttt{YARARA}-derived results. The compositional degeneracy in this parameter space is also noted and further discussed in e.g. \citet{Palethorpe_2024}.

To further investigate the internal structure of the transiting sub-Neptune HD~85426~b, we used the publicly available internal structure modelling framework \texttt{plaNETic}\footnote{\url{https://github.com/joannegger/plaNETic}} \citep{Egger+2024}. This framework is based on the planetary structure model BICEPS \citep{Haldemann+2024} but uses a neural network as a fast surrogate model in a full grid accept-reject sampling scheme instead of classical Bayesian inference. This allows for a fast yet robust characterisation of the planet's interior. The planet is self-consistently modelled as three layers: (i) an inner core of iron and sulphur, (ii) a mantle of oxidised silicon, magnesium, and iron, and (iii) a volatile envelope composed of uniformly mixed water and H/He.

To account for the intrinsic degeneracy of the problem, we ran six models with varying priors. These priors influence the results to some extent, reflecting the sensitivity of interior structure modelling to initial assumptions.
All priors were motivated by current planet formation theory, with two different priors for the water content in the volatile layer (compatible with a formation scenario outside or inside the iceline, respectively) and three for the composition of the core and mantle. More specifically, we first assumed the planetary Si/Mg/Fe ratios to match those of the host star exactly \citep[e.g.][]{Thiabaud+2015}, secondly that the planet is iron-enriched compared to the host star \citep{Adibekyan+2021}, and lastly we modelled the planet independently of the stellar Si/Mg/Fe ratios by sampling the molar fractions of Si, Mg and Fe uniformly from the simplex on which they add up to unity (with an upper limit of 0.75 for Fe). These priors, along with the model itself, are described in more detail in \cite{Egger+2024}.

\begin{figure*}
    \centering
    \includegraphics[width=\textwidth]{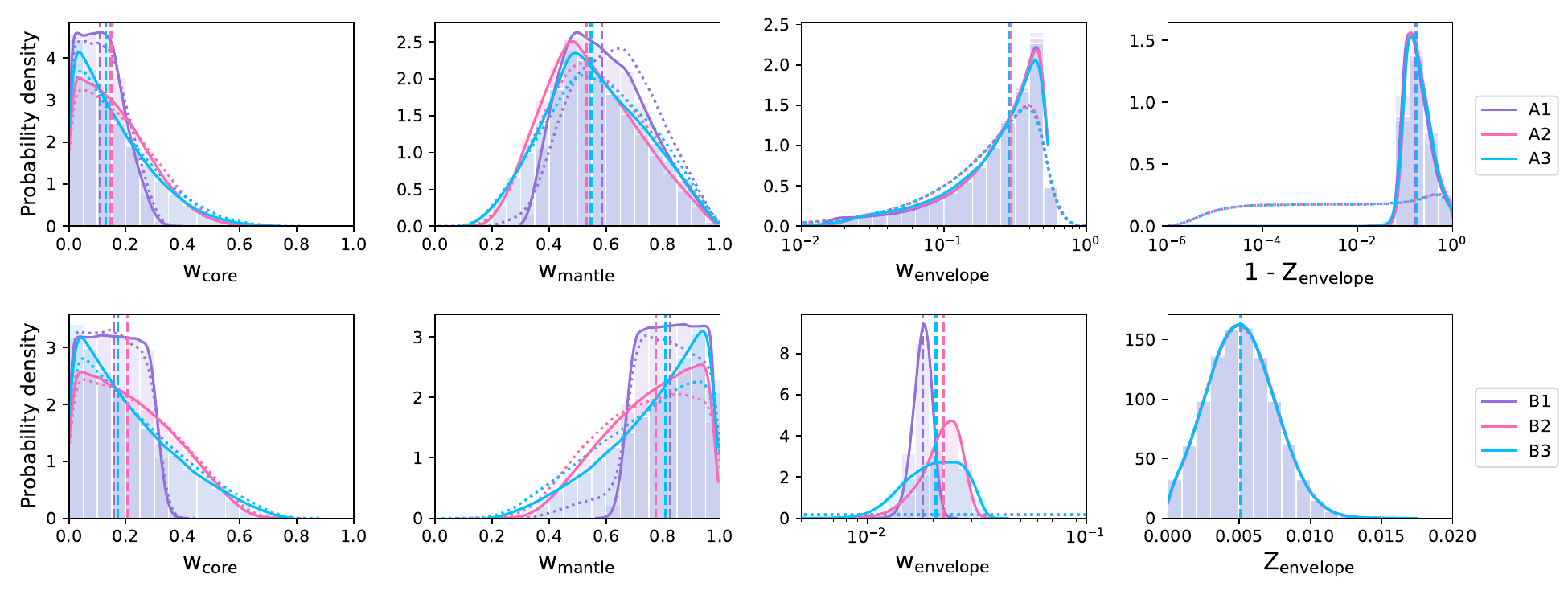}
    \caption{Posterior distributions of the mass fraction of the inner core ($\text{w}_\text{core}$), the mantle ($\text{w}_\text{mantle}$), and the envelope layer ($\text{w}_\text{envelope}$), as well as the mass fraction of water in the envelope ($\text{Z}_\text{envelope}$). For the top (bottom) panels, formation beyond (inside) the iceline was assumed. For the distributions labelled as A1 (B1), we assumed that the planetary Si/Mg/Fe ratios matched the stellar ones. The planet was assumed to be iron-enriched for A2 (B2). The A3 (B3) distributions show the results when the elemental abundances of Si, Mg and Fe are sampled uniformly from a simplex, disregarding the stellar abundances. The planetary mass and radius from the 3-Keplerian fit from \texttt{YARARA} with the $\beta$ eccentricity prior was used as an input. The posteriors are visually almost indistinguishable from the results when using the masses extracted using \texttt{TWEAKS}.}
    \label{fig:intstruct}
\end{figure*}

The resulting posterior distributions for the mass fractions of the inner core, mantle, and volatile layers, as well as the water mass fraction in the volatile layer, are visualised in Figure \ref{fig:intstruct}. The layer mass fractions of the inner core and mantle are close to identical with the priors in all six cases and are therefore not constrained by the data. Similarly, for the mass fraction of the volatile layer in the case of the water-rich prior, a large number of combinations of envelope mass fractions and metallicities are compatible with the data. The median value for the water mass fraction in the envelope, in this scenario, is around 80 per cent for all models. For the water-poor prior corresponding to a formation scenario inside the iceline, we find that the data constrain the envelope mass fraction very well, with median values of around 2 per cent.

To compute the results displayed in Fig. \ref{fig:intstruct} we used the mass ($8.5^{+1.3}_{-1.4}$ $\text{M}_{\oplus}$), radius ($2.78^{+0.05}_{-0.04}$ $\text{R}_{\oplus}$), and semi-amplitude ($2.2^{+0.3}_{-0.4}$ \ms) of HD~85426~b extracted using the 3-Keplerian fit, with priors as shown in Table \ref{tab:allresults_3pmodel}.
Furthermore, we used the stellar parameters, that is, age, mass, radius, effective temperature, and abundances, shown in Table \ref{tab:stellarparameters_1774_other_sources}. The results are visually hardly distinguishable from those extracted using the mass from the \texttt{TWEAKS} run. The detailed results for both masses are also shown in Table \ref{tab:intstruct_YARARA} for \texttt{YARARA} and in Table \ref{tab:intstruct_TWEAKS} for \texttt{TWEAKS}.

Effects related to geophysical evolution have not been included in the internal structure modelling so far, cf. discussion in \citet{Haldemann+2024,Egger+2024}. For example, sub-Neptunes may have magma oceans that can interact with the atmosphere through dissolution and outgassing and may store large fractions of water \citep{Kite_2020, Dorn_2021}. These effects will become relevant once more is known about the planet and the degeneracy of the problem can be lifted.

\section{Suitability for atmospheric follow-up} \label{sec:atmo_follow_up}

Using the stellar radius, apparent J magnitude, and planetary parameters (radius, mass, and equilibrium temperature), we estimate a Transmission Spectroscopy Metric \citep[TSM;][]{Kempton_2018} of 115 for the lower mass estimate of 6 $\text{M}_{\oplus}$ coming from \texttt{TWEAKS}. The TSM decreases to 82 if we use the mass estimate of 8.5 $\text{M}_{\oplus}$ based on the \texttt{YARARA} RVs.
Therefore, both TSM values are near or above the threshold of 90 chosen in \citep{Kempton_2018} for the selection of high-quality atmospheric characterisation targets.
For the emission spectroscopy metric \citep[ESM;][]{Kempton_2018}, we calculated a value of 5.5, which is below the threshold of 7.5 for ESM suggested in \citep{Kempton_2018}.

Transmission spectra of sub-Neptune atmospheres often show muted or absent spectral features \citep[e.g.][]{Kreidberg_2014,Guo_2020,Gao_2023,Wallack_2024}. 
This is hypothesised to be due to high-altitude aerosols and a high mean molecular weight atmosphere \citep{Kreidberg_2014,Gao_2023}. \citet{Brande_2024} suggested that the attenuation of atmospheric features in sub-Neptune spectra is strongest between $\sim$500 and $\sim$700 K due to efficient aerosol production, as found in \citet{Morley_2015,Gao_2020}. Planetary atmospheres would thus be expected to be clearer at cooler temperatures below 500 K and at hotter temperatures above 700 K. 
For example, \citet{Wallack_2024} measured a featureless spectrum for TOI-836~c at a zero albedo equilibrium temperature of 665 K. 
\citet{Davenport_2025} analysed JWST transmission spectra of TOI-421~b, a planet with an equilibrium temperature of about 920 K orbiting a Sun-like star, and detected a low mean molecular weight and no significant aerosol coverage. This supports the hypothesis that the atmospheres of at least some hot sub-Neptunes may not be dominated by hydrocarbon hazes or clouds. 
Setting the bond albedo $A_{\text{B}}$ to 0.3, as in \citet{Brande_2024}, we calculate an equilibrium temperature of $754^{+10}_{-10}$ K for HD~85426~b, rather than the $824^{+11}_{-11}$ K that we computed assuming zero albedo.
HD~85426~b is therefore in the parameter space region where we could expect the transition to haze-free, low mean molecular weight atmospheres. This suggests that HD~85426~b is an interesting candidate for transmission spectroscopy. 
With a K magnitude of 6.7 and a J magnitude of 7.1 \citep{Cutri_2003}, HD~85426 is very bright but still a suitable target for transmission spectroscopy with JWST.
Since the bulk density could not constrain the internal composition, transmission spectroscopy may enable us to conclude whether HD~85426~b formed beyond or within the iceline, depending on the mean molecular weight of the atmosphere found.

Of the planets known to date (9 September 2025), there are 114 planets with a mass between 5 and 10 $\text{M}_{\oplus}$ and a radius and mass measured to a precision better than 25 per cent\footnote{NASA Exoplanet Archive, \url{https://exoplanetarchive.ipac.caltech.edu}.}. 17 of these planets have orbital periods greater than 16.7 d. With HD~85426~b, we add another point to this sparsely populated parameter space.

Among the first set of 114 planets in the parameter space defined above, the TSM of just 14 planets exceeds our most conservative estimate of 82. Therefore, HD~85426~b is an interesting target for further study, specifically because it may lie at the boundary where sub-Neptune transmission spectra display measurable features due to a lower prevalence of high-altitude aerosols.

\section{Discussion and summary} \label{Discussion}

Two independent stellar activity mitigation techniques, combined with various modelling approaches and data selections, yielded mass estimates for the transiting planet b ranging from 6 to 9 $\text{M}_{\oplus}$. We found evidence for two planetary companions to planet b and tested the dependence of planet b's derived parameters on the inclusion of these in the modelling. 

\subsection{Non-transiting planets}
Both main methods independently showed decisive evidence for another planet, planet c, with an orbital period of 35.7 d and a minimum mass of about 10 $\text{M}_{\oplus}$.
A prominent peak near 36.0 days in the DRS contrast and FWHM periodograms initially cast doubt on the planetary nature of planet c.
However, the RV signal with period 35.7 d survived two independent stellar activity mitigation techniques and shows remarkable stability in time for phase and amplitude (cf. Fig. \ref{posterior_seasonbyseason2_p2}), whereas the variation of contrast and FWHM disappear after post-processing with \texttt{YARARA} and is not consistent with the variations in the other activity indicators or the expected phase lag between the RVs and activity indicators. We thus regard this similarity as coincidental. This target, being bright and solar-like, consequently also represents a challenging test bed for activity mitigation techniques.

There is also strong evidence for the existence of a planet with an orbital period near 89 d for both stellar activity mitigation techniques, planet candidate d. More observations are needed to ultimately confirm the stability of this signal.

If planet c were transiting, a transit would have been expected during the second half of \texttt{TESS} sector 21. However, no transits apart from those from HD~85426~b are discernible in the \texttt{TESS} light curves. HD~85426~c is therefore expected to have an inclination smaller than 88.6$\degr$. For planet candidate d, we cannot definitively rule out a transiting orbit due to the uncertainty of the time of conjunction, resulting in a probability slightly below 50 per cent for a transit in \texttt{TESS} sector 48.

The minimum masses associated with planet c and planet candidate d are about 10 $\text{M}_{\oplus}$ based on the \texttt{YARARA} RVs and slightly larger with \texttt{TWEAKS} ($11.5^{+1.9}_{-2.0}$ $\text{M}_{\oplus}$ for planet c and $13.3^{+3.0}_{-2.5}$ $\text{M}_{\oplus}$ for planet candidate d). This means that with \texttt{TWEAKS} we get a lower mass for planet b but slightly higher masses for planets c and d.

\subsection{Planet b}

For the analysis based on the \texttt{YARARA} data, we consider the results with a 3-Keplerian model applied as our main result because these are statistically favoured. The derived results do not depend significantly on the number of modelled Keplerians, though.

For the \texttt{TWEAKS} analysis, we find the same planetary companions as in the \texttt{YARARA} analysis if we presuppose the planetary system to be stable. The assumption of stability is warranted given the age of the system. The 3-Keplerian model is also decisively favoured over models that include fewer Keplerians for \texttt{TWEAKS}. Including four Keplerians, although statistically favoured, did not produce well-constrained parameters for the fourth Keplerian. 
Also for \texttt{TWEAKS}, we find that the derived masses for planet b were practically independent of the number of included Keplerians in our tests.

For our main results, we therefore derive masses for planet b of $8.5^{+1.3}_{-1.4}$ $\text{M}_{\oplus}$ for \texttt{YARARA} and $6.0^{+1.5}_{-1.6}$ $\text{M}_{\oplus}$ with \texttt{TWEAKS}.
The mass estimate for planet b remains in the cited mass bracket of 6 and 9 $\text{M}_{\oplus}$ for all analyses and depends more on the stellar activity mitigation technique than on the number of modelled Keplerians or the choice of the eccentricity prior.

By jointly fitting the \texttt{YARARA} RVs and the photometric data from {\it TESS} and {\it CHEOPS}, we derived a radius of $2.78^{+0.05}_{-0.04}$ $\text{R}_{\oplus}$ for planet b. This result is in agreement with an independent second analysis that was based on the conservative approach of including just two Keplerians and setting their orbits to circular, converging to a radius estimate of $2.77^{+0.03}_{-0.03}$ $\text{R}_{\oplus}$. This second analysis also searched for transit timing variations, but just found a faint hint of the latter.

We determined the internal structure of planet b. These results are largely dependent on the chosen priors and do not depend significantly on whether we adopt the mass extracted from the \texttt{YARARA} data or the mass from the \texttt{TWEAKS} analysis.

Finally, we found that HD~85426~b is of high value for atmospheric follow-up observations, with a TSM between 82 and 115, and may help to shed light on the potential transition between hazy atmospheres producing featureless atmospheres and clear atmospheres with low mean molecular weight.

\newpage
\section*{Acknowledgements}
{\it CHEOPS} is an ESA mission in partnership with Switzerland with important contributions to the payload and the ground segment from Austria, Belgium, France, Germany, Hungary, Italy, Portugal, Spain, Sweden, and the United Kingdom. The {\it CHEOPS} Consortium would like to gratefully acknowledge the support received by all the agencies, offices, universities, and industries involved. Their flexibility and willingness to explore new approaches were essential to the success of this mission. {\it CHEOPS} data analysed in this article will be made available in the {\it CHEOPS} mission archive (\url{https://cheops.unige.ch/archive_browser/}).
This paper made use of data collected by the TESS mission and are publicly available from the Mikulski Archive for Space Telescopes (MAST) operated by the Space Telescope Science Institute (STScI). Funding for the TESS mission is provided by NASA’s Science Mission Directorate. We acknowledge the use of public TESS data from pipelines at the TESS Science Office and at the TESS Science Processing Operations Center. Resources supporting this work were provided by the NASA High-End Computing (HEC) Program through the NASA Advanced Supercomputing (NAS) Division at Ames Research Center for the production of the SPOC data products.
This work has made use of data from the European Space Agency (ESA) mission
{\it Gaia} (\url{https://www.cosmos.esa.int/gaia}), processed by the {\it Gaia}
Data Processing and Analysis Consortium (DPAC,
\url{https://www.cosmos.esa.int/web/gaia/dpac/consortium}). Funding for the DPAC
has been provided by national institutions, in particular the institutions
participating in the {\it Gaia} Multilateral Agreement.
This research has made use of data and/or services provided by the International Astronomical Union's Minor Planet Center.
This publication makes use of The Data \& Analysis Center for Exoplanets (DACE), which is a facility based at the University of Geneva (CH) dedicated to extrasolar planets data visualisation, exchange and analysis. DACE is a platform of the Swiss National Centre of Competence in Research (NCCR) PlanetS, federating the Swiss expertise in Exoplanet research. The DACE platform is available at https://dace.unige.ch.
This research has made use of the NASA Exoplanet Archive, which is operated by the California Institute of Technology, under contract with the National Aeronautics and Space Administration under the Exoplanet Exploration Program.
A.M. acknowledges funding from a UKRI Future Leader Fellowship, grant number MR/X033244/1 and a UK Science and Technology Facilities Council (STFC) small grant ST/Y002334/1. 
ACC acknowledges support from STFC consolidated grant number ST/V000861/1 and UKRI/ERC Synergy Grant EP/Z000181/1 (REVEAL).
B.S.L acknowledges funding from a UKRI Future Leader Fellowship (grant number MR/X033244/1).
R.D.H. is funded by the UK Science and Technology Facilities Council (STFC)'s Ernest Rutherford Fellowship (grant number ST/V004735/1).
LBo, GBr, VNa, IPa, GPi, RRa, GSc, VSi, and TZi acknowledge support from CHEOPS ASI-INAF agreement n. 2019-29-HH.0. 
LBo acknowledges financial support from the Bando Ricerca Fondamentale INAF 2023, Mini-Grant:
``Decoding the dynamical properties of planetary systems observed by TESS and CHEOPS through TTV analysis with parallel computing''.
TZi acknowledges support from NVIDIA Academic Hardware Grant Program for the use of the Titan V GPU card and the Italian MUR Departments of Excellence grant 2023-2027 ``Quantum Frontiers''.
This work has been carried out within the framework of the NCCR PlanetS supported by the Swiss National Science Foundation under grants 51NF40\_182901 and 51NF40\_205606. 
TWi acknowledges support from the UKSA and the University of Warwick. 
A.De. 
PM acknowledges support from STFC research grant number ST/R000638/1. 
S.G.S. acknowledge support from FCT through FCT contract nr. CEECIND/00826/2018 and POPH/FSE (EC). 
The Portuguese team thanks the Portuguese Space Agency for the provision of financial support in the framework of the PRODEX Programme of the European Space Agency (ESA) under contract number 4000142255. 
YAl acknowledges support from the Swiss National Science Foundation (SNSF) under grant 200020\_192038. 
ACMC acknowledges support from the FCT, Portugal, through the CFisUC projects UIDB/04564/2020 and UIDP/04564/2020, with DOI identifiers 10.54499/UIDB/04564/2020 and 10.54499/UIDP/04564/2020, respectively. 
A.C., A.D., B.E., K.G., and J.K. acknowledge their role as ESA-appointed CHEOPS Science Team Members. 
DB, EP, EV, IR and RA acknowledge financial support from the Agencia Estatal de Investigación of the Ministerio de Ciencia e Innovación MCIN/AEI/10.13039/501100011033 and the ERDF “A way of making Europe” through projects PID2021-125627OB-C31, PID2021-125627OB-C32, PID2021-127289NB-I00, PID2023-150468NB-I00 and PID2023-149439NB-C41. 
from the Centre of Excellence “Severo Ochoa'' award to the Instituto de Astrofísica de Canarias (CEX2019-000920-S), the Centre of Excellence “María de Maeztu” award to the Institut de Ciències de l’Espai (CEX2020-001058-M), and from the Generalitat de Catalunya/CERCA programme. 
SCCB acknowledges the support from Fundação para a Ciência e Tecnologia (FCT) in the form of work contract through the Scientific Employment Incentive program with reference 2023.06687.CEECIND and DOI 10.54499/2023.06687.CEECIND/CP2839/CT0002. 
ABr was supported by the SNSA. 
C.B. and A.S. acknowledge support from the Swiss Space Office through the ESA PRODEX program. 
P.E.C. is funded by the Austrian Science Fund (FWF) Erwin Schroedinger Fellowship, program J4595-N. 
This project was supported by the CNES. 
This work was supported by FCT - Funda\c{c}\~{a}o para a Ci\^{e}ncia e a Tecnologia through national funds and by FEDER through COMPETE2020 through the research grants UIDB/04434/2020, UIDP/04434/2020, 2022.06962.PTDC. 
O.D.S.D. is supported in the form of work contract (DL 57/2016/CP1364/CT0004) funded by national funds through FCT. 
B.-O. D. acknowledges support from the Swiss State Secretariat for Education, Research and Innovation (SERI) under contract number MB22.00046. 
This project has received funding from the Swiss National Science Foundation for project 200021\_200726. It has also been carried out within the framework of the National Centre of Competence in Research PlanetS supported by the Swiss National Science Foundation under grant 51NF40\_205606. The authors acknowledge the financial support of the SNSF. 
MF and CMP gratefully acknowledge the support of the Swedish National Space Agency (DNR 65/19, 174/18). 
DG gratefully acknowledges financial support from the CRT foundation under Grant No. 2018.2323 “Gaseous or rocky? Unveiling the nature of small worlds”. 
M.G. is an F.R.S.-FNRS Senior Research Associate. 
MNG is the ESA CHEOPS Project Scientist and Mission Representative. BMM is the ESA CHEOPS Project Scientist. KGI was the ESA CHEOPS Project Scientist until the end of December 2022 and Mission Representative until the end of January 2023. All of them are/were responsible for the Guest Observers (GO) Programme. None of them relay/relayed proprietary information between the GO and Guaranteed Time Observation (GTO) Programmes, nor do/did they decide on the definition and target selection of the GTO Programme. 
CHe acknowledges financial support from the Österreichische Akademie 1158 der Wissenschaften and from the European Union H2020-MSCA-ITN-2019 1159 under Grant Agreement no. 860470 (CHAMELEON). Calculations were performed using supercomputer resources provided by the Vienna Scientific Cluster (VSC). 
K.W.F.L. was supported by Deutsche Forschungsgemeinschaft grants RA714/14-1 within the DFG Schwerpunkt SPP 1992, Exploring the Diversity of Extrasolar Planets. 
This work was granted access to the HPC resources of MesoPSL financed by the Region Ile de France and the project Equip@Meso (reference ANR-10-EQPX-29-01) of the programme Investissements d'Avenir supervised by the Agence Nationale pour la Recherche. 
A.L. and J.K. acknowledge support of the Swiss National Science Foundation under grant number  TMSGI2\_211697. 
M.L. acknowledges support of the Swiss National Science Foundation under grant number PCEFP2\_194576. 
This work was also partially supported by a grant from the Simons Foundation (PI Queloz, grant number 327127). 
NCSa acknowledges funding by the European Union (ERC, FIERCE, 101052347). Views and opinions expressed are however those of the author(s) only and do not necessarily reflect those of the European Union or the European Research Council. Neither the European Union nor the granting authority can be held responsible for them. 
GyMSz acknowledges the support of the Hungarian National Research, Development and Innovation Office (NKFIH) grant K-125015, a PRODEX Experiment Agreement No. 4000137122, the Lend\"ulet LP2018-7/2021 grant of the Hungarian Academy of Science and the support of the city of Szombathely. 
V.V.G. is an F.R.S-FNRS Research Associate. 
JV acknowledges support from the Swiss National Science Foundation (SNSF) under grant PZ00P2\_208945. 
NAW acknowledges UKSA grant ST/R004838/1.

\section*{Data Availability}
The raw and detrended photometric \texttt{CHEOPS} time series data, as well as the radial velocity measurements, will be made available in electronic form at CDS upon publication.



\bibliographystyle{mnras}
\bibliography{example} 



\vspace{1cm}
\noindent $^{1}$ ETH Zurich, Department of Physics, Wolfgang-Pauli-Strasse 2, CH-8093 Zurich, Switzerland \\
$^{2}$ Cavendish Laboratory, JJ Thomson Avenue, Cambridge CB3 0HE, UK \\
$^{3}$ School of Physics \& Astronomy, University of Birmingham, Edgbaston, Birmingham B15 2TT, UK \\
$^{4}$ Centre for Exoplanet Science, SUPA School of Physics and Astronomy, University of St Andrews, North Haugh, St Andrews KY16 9SS, UK \\
$^{5}$ Sub-department of Astrophysics, University of Oxford, Keble Rd, Oxford OX13RH, UK \\
$^{6}$ INAF, Osservatorio Astronomico di Padova, Vicolo dell'Osservatorio 5, 35122 Padova, Italy \\
$^{7}$ Space Research and Planetary Sciences, Physics Institute, University of Bern, Gesellschaftsstrasse 6, 3012 Bern, Switzerland \\
$^{8}$ Space sciences, Technologies and Astrophysics Research (STAR) Institute, Université de Liège, Allée du 6 Août 19C, 4000 Liège, Belgium \\
$^{9}$ Astrobiology Research Unit, Université de Liège, Allée du 6 Août 19C, B-4000 Liège, Belgium \\
$^{10}$ Department of Physics, University of Warwick, Gibbet Hill Road, Coventry CV4 7AL, United Kingdom \\
$^{11}$ Observatoire astronomique de l'Université de Genève, Chemin Pegasi 51, 1290 Versoix, Switzerland \\
$^{12}$ Center for Space and Habitability, University of Bern, Gesellschaftsstrasse 6, 3012 Bern, Switzerland \\
$^{13}$ Center for Astrophysics | Harvard \& Smithsonian, 60 Garden Street, Cambridge, MA 02138, USA \\
$^{14}$ Dipartimento di Fisica e Astronomia "Galileo Galilei", Università degli Studi di Padova, Vicolo dell'Osservatorio 3, 35122 Padova, Italy \\
$^{15}$ Astrophysics Group, Lennard Jones Building, Keele University, Staffordshire, ST5 5BG, United Kingdom \\
$^{16}$ Instituto de Astrofisica e Ciencias do Espaco, Universidade do Porto, CAUP, Rua das Estrelas, 4150-762 Porto, Portugal \\
$^{17}$ ETH Zurich, Institute for Particle Physics and Astrophysics, Wolfgang-Pauli-Str. 27, 8093 Zurich \\
$^{18}$ University of Zurich, Department of Astrophysics, Winterthurerstrasse 190, CH-8057, Zurich, Switzerland \\
$^{19}$ DTU Space, Technical University of Denmark, Elektrovej 328, 2800 Kgs. Lyngby, Denmark \\
$^{20}$ Astrophysics Group, University of Exeter, Exeter EX4 2QL, UK \\
$^{21}$ Institute of Space Research, German Aerospace Center (DLR), Rutherfordstraße 2, 12489 Berlin, Germany \\
$^{22}$ INAF – Osservatorio Astrofisico di Torino, Via Osservatorio 20, 10025 Pino Torinese, Italy \\
$^{23}$ CFisUC, Departamento de Física, Universidade de Coimbra, 3004-516 Coimbra, Portugal \\
$^{24}$ Space Research Institute, Austrian Academy of Sciences, Schmiedlstrasse 6, A-8042 Graz, Austria \\
$^{25}$ Instituto de Astrofísica de Canarias, Vía Láctea s/n, 38200 La Laguna, Tenerife, Spain \\
$^{26}$ Departamento de Astrofísica, Universidad de La Laguna, Astrofísico Francisco Sanchez s/n, 38206 La Laguna, Tenerife, Spain \\
$^{27}$ Admatis, 5. Kandó Kálmán Street, 3534 Miskolc, Hungary \\
$^{28}$ Depto. de Astrofísica, Centro de Astrobiología (CSIC-INTA), ESAC campus, 28692 Villanueva de la Cañada (Madrid), Spain \\
$^{29}$ Departamento de Fisica e Astronomia, Faculdade de Ciencias, Universidade do Porto, Rua do Campo Alegre, 4169-007 Porto, Portugal \\
$^{30}$ Department of Astronomy, Stockholm University, AlbaNova University Center, 10691 Stockholm, Sweden \\
$^{31}$ INAF, Osservatorio Astrofisico di Torino, Via Osservatorio, 20, I-10025 Pino Torinese To, Italy \\
$^{32}$ Centre for Mathematical Sciences, Lund University, Box 118, 221 00 Lund, Sweden \\
$^{33}$ Aix Marseille Univ, CNRS, CNES, LAM, 38 rue Frédéric Joliot-Curie, 13388 Marseille, France \\
$^{34}$ ARTORG Center for Biomedical Engineering Research, University of Bern, Bern, Switzerland \\
$^{35}$ ELTE Gothard Astrophysical Observatory, 9700 Szombathely, Szent Imre h. u. 112, Hungary \\
$^{36}$ Observatoire astronomique de l'Universit\'e de Gen\`eve, Chemin Pegasi 51, 1290, Versoix, Switzerland \\
$^{37}$ SRON Netherlands Institute for Space Research, Niels Bohrweg 4, 2333 CA Leiden, Netherlands \\
$^{38}$ Centre Vie dans l’Univers, Faculté des sciences, Université de Genève, Quai Ernest-Ansermet 30, 1211 Genève 4, Switzerland \\
$^{39}$ Leiden Observatory, University of Leiden, PO Box 9513, 2300 RA Leiden, The Netherlands \\
$^{40}$ Department of Space, Earth and Environment, Chalmers University of Technology, Onsala Space Observatory, 439 92 Onsala, Sweden \\
$^{41}$ Dipartimento di Fisica, Università degli Studi di Torino, via Pietro Giuria 1, I-10125, Torino, Italy \\
$^{42}$ National and Kapodistrian University of Athens, Department of Physics, University Campus, Zografos GR-157 84, Athens, Greece \\
$^{43}$ Department of Astrophysics, University of Vienna, Türkenschanzstrasse 17, 1180 Vienna, Austria \\
$^{44}$ European Space Agency (ESA), European Space Research and Technology Centre (ESTEC), Keplerlaan 1, 2201 AZ Noordwijk, The Netherlands \\
$^{45}$ Institute for Theoretical Physics and Computational Physics, Graz University of Technology, Petersgasse 16, 8010 Graz, Austria \\
$^{46}$ NASA Ames Research Center, Moffett Field, CA 94035, USA \\
$^{47}$ Konkoly Observatory, Research Centre for Astronomy and Earth Sciences, 1121 Budapest, Konkoly Thege Miklós út 15-17, Hungary \\
$^{48}$ ELTE E\"otv\"os Lor\'and University, Institute of Physics, P\'azm\'any P\'eter s\'et\'any 1/A, 1117 Budapest, Hungary \\
$^{49}$ IMCCE, UMR8028 CNRS, Observatoire de Paris, PSL Univ., Sorbonne Univ., 77 av. Denfert-Rochereau, 75014 Paris, France \\
$^{50}$ Institut d'astrophysique de Paris, UMR7095 CNRS, Université Pierre \& Marie Curie, 98bis blvd. Arago, 75014 Paris, France \\
$^{51}$ Fundación Galileo Galilei – INAF (Telescopio Nazionale Galileo), Rambla Jos\'e Ana Fern\`andez P\'erez 7, 38712 Breña Baja (La Palma), Canary Islands, Spain \\
$^{52}$ European Space Agency, ESA - European Space Astronomy Centre, Camino Bajo del Castillo s/n, 28692 Villanueva de la Cañada, Madrid, Spain \\
$^{53}$ INAF, Osservatorio Astrofisico di Catania, Via S. Sofia 78, 95123 Catania, Italy \\
$^{54}$ SUPA, Institute for Astronomy, University of Edinburgh, Blackford Hill, Edinburgh EH9 3HJ, UK \\
$^{55}$ Centre for Exoplanet Science, University of Edinburgh, Edinburgh EH9 3FD, UK \\
$^{56}$ Weltraumforschung und Planetologie, Physikalisches Institut, University of Bern, Gesellschaftsstrasse 6, 3012 Bern, Switzerland \\
$^{57}$ Institut fuer Geologische Wissenschaften, Freie Universitaet Berlin, Malteserstrasse 74-100,12249 Berlin, Germany \\
$^{58}$ Institut de Ciencies de l'Espai (ICE, CSIC), Campus UAB, Can Magrans s/n, 08193 Bellaterra, Spain \\
$^{59}$ Institut d'Estudis Espacials de Catalunya (IEEC), 08860 Castelldefels (Barcelona), Spain \\
$^{60}$ HUN-REN-ELTE Exoplanet Research Group, Szent Imre h. u. 112., Szombathely, H-9700, Hungary \\
$^{61}$ Leiden Observatory, University of Leiden, Einsteinweg 55, 2333 CA Leiden, The Netherlands \\
$^{62}$ Institute of Astronomy, University of Cambridge, Madingley Road, Cambridge, CB3 0HA, United Kingdom


\appendix

\section{Crossing object} \label{appendix:otherobj}

\begin{figure}
    \centering
    \includegraphics[width=0.98\columnwidth]{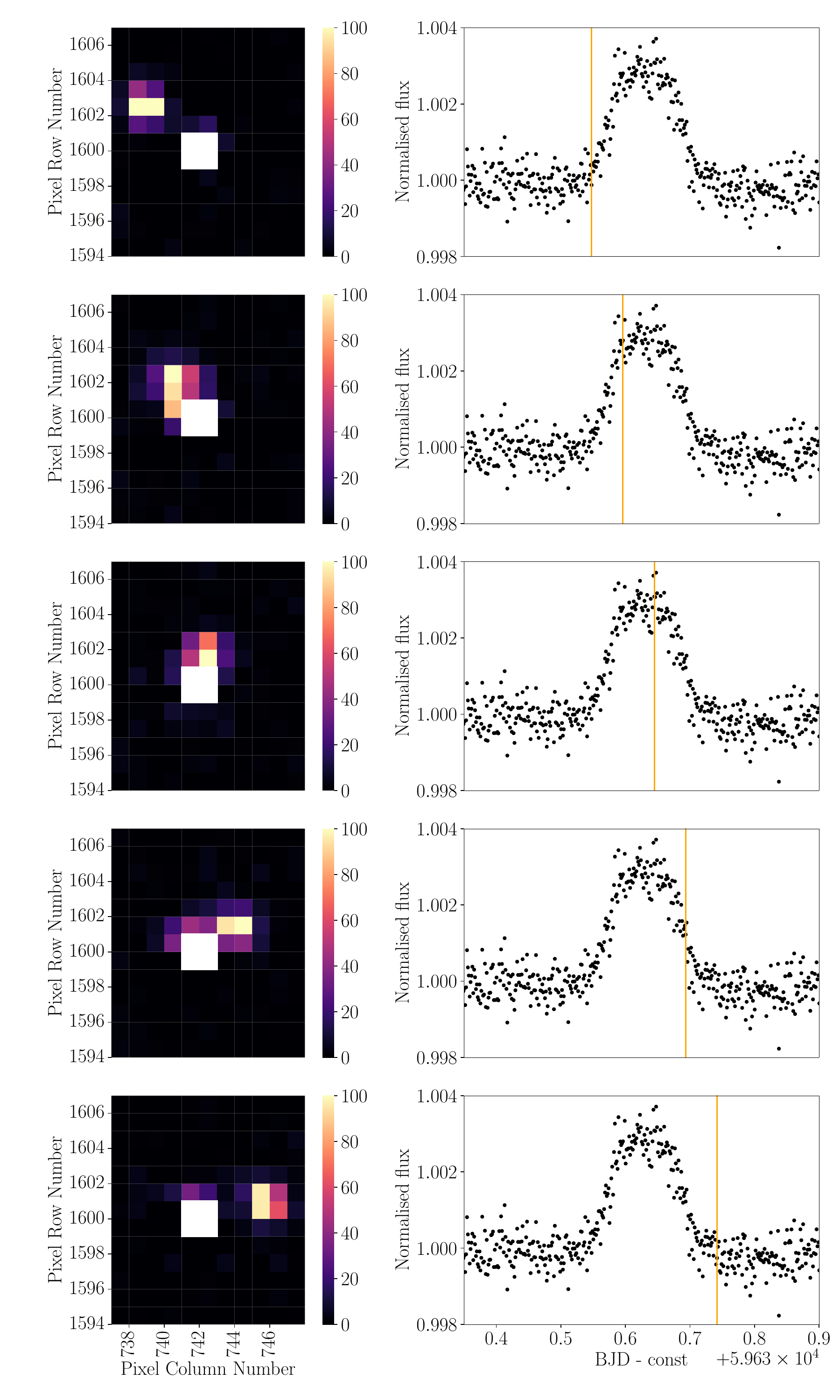}
    \caption{\textit{Left panels}: Difference between {\it TESS} target pixel file flux of HD~85426 at time $t$ and a randomly chosen target pixel file that showed no flux anomaly. The four pixels with the highest contribution from the target star are masked with the white square. \textit{Right panels}: Flux time series of HD~85426. The vertical orange line indicates the timestamp $t$ of the observation.} 
    \label{fig:otherobj}
\end{figure}

In Fig. \ref{fig:otherobj}, we show the origin of the increase in flux in the {\it TESS} light curve of HD~85426 observed in sector 48 around BJD 2,459,630. The five panels show observations taken 70 minutes apart. The time of the observation of each panel is indicated in the right column by a vertical orange line. On the left, we show the flux difference between the target pixel file at the indicated time and a target pixel file that shows no secondary object. The target pixel files are dominated by the flux of the star itself. Thus, we show the differential target pixel files. Clearly, there is a secondary object that crosses very close to the centroid of HD~85426 and increases the photon count while it is near the aperture associated with the target. The constraint that the partial collection of photons in the respective aperture leads to an apparent flux increase of about 0.3 per cent, together with the time and coordinates of the crossing event, enabled us to conclude that the crossing object is the asteroid 581 Tauntonia.

\section{38 or 35.7 d for planet c?} \label{appendix:38or36}
We tested which of these two periods is the true period by fitting a two-planet model to all data (case a) and seasons 1 and 3 combined (case b) setting a narrow prior centred at 35.7 d and, in a separate run, a narrow period prior centred at 38 d. In both cases, the prior for the period of planet b is informed by the {\it TESS} and {\it CHEOPS} transit fit. 
The true period of planet c should produce consistent results for both cases. For the wrong period, however, the estimated semi-amplitude associated with the signal is expected to depend significantly on whether season 2 is included, because the model will need to accommodate a phase offset of 180 degrees for season 2. We found that a narrow prior centred at 38 d produces vastly different results for cases a and b. The associated semi-amplitude dropped from 3 \ms to 1.5 \ms after including season 2. The narrow prior centred at 35.7, on the other hand, produced consistent results with associated semi-amplitudes around 2.8 \ms for both cases.

\section{Posteriors for 2-Keplerian model} \label{A:2keplerian_posterior}

\begin{table*} 
\caption{Prior and posterior distributions for the 2-Keplerian joint run. $\mathcal{U}$ indicates a uniform distribution, $\mathcal{LU}$ a log-uniform distribution, and $\mathcal{\beta}$ a beta distribution.}
\label{tab:allresults_2pmodel}
\begin{tabular}{llllc}

Parameter & Symbol & Unit & Prior distribution & Posterior \\ [4pt]
\hline 
\multicolumn{5}{c}{\emph{Fitted parameters}}\\[2 pt]
\multicolumn{5}{l}{\textbf{Planet b}}\\[2 pt]
Orbital period                 & $\text{P}_{\text{b}}$ & d               & $\mathcal{U}$[16.70959, 16.71019] & $16.70988^{+0.00003}_{-0.00002}$ \\[2 pt]
Reference conjunction time         & ${\text{T}_{\text{0}}}_{\text{b}}$ & d               & $\mathcal{U}$[59674.408, 59674.418] & $59674.4132^{+0.0005}_{-0.0005}$ \\[2 pt]
RV semi-amplitude              & $\text{K}_{\text{b}}$ & \ms             & $\mathcal{LU}$[0.1, 10.0] & $2.0^{+0.3}_{-0.4}$ \\[2 pt]
Eccentricity                   & $\text{ecc}_{\text{b}}$ &                 & $\mathcal{\beta}$[0.867, 3.03] & $0.05^{+0.07}_{-0.04}$ \\[2 pt]
Argument of periastron         & $\omega_{\text{b}}$  & deg             & $\mathcal{U}$[0.0, 360.0] & $273^{+42}_{-94}$ \\[2 pt]
$\text{r}_{1_\text{b}}$                           &     $\text{r}_{1_\text{b}}$                 &                 & $\mathcal{U}$[0.0, 1.0] & $0.5^{+0.1}_{-0.1}$ \\[2 pt]
$\text{r}_{2_\text{b}}$                           & $\text{r}_{2_\text{b}}$ &                 & $\mathcal{U}$[0.0, 1.0] & $0.0225^{+0.0004}_{-0.0002}$ \\[2 pt]
\multicolumn{5}{l}{\textbf{Planet c}}\\[2 pt]
Orbital period                 & $\text{P}_{\text{c}}$ & d               & $\mathcal{U}$[34.4, 37.4] & $35.7^{+0.1}_{-0.1}$ \\[2 pt]
Reference conjunction time         & ${\text{T}_{\text{0}}}_{\text{c}}$ & d               & $\mathcal{U}$[59600.0, 59637.4] & $59608^{+1}_{-1}$ \\[2 pt]
RV semi-amplitude              & $\text{K}_{\text{c}}$ & \ms             & $\mathcal{LU}$[0.1, 10.0] & $2.1^{+0.4}_{-0.4}$ \\[2 pt]
Eccentricity                   & $\text{ecc}_{\text{c}}$ &                 & $\mathcal{\beta}$[0.867, 3.03] & $0.11^{+0.13}_{-0.08}$ \\[2 pt]
Argument of periastron         & $\omega_{\text{c}}$  & deg             & $\mathcal{U}$[0.0, 360.0] & $112^{+204}_{-76}$ \\[2 pt]
\multicolumn{5}{l}{\textbf{Stellar and instrumental}}\\[2 pt]
Mean RV HARPS-N                & $\mu_{\text{HARPS-N}}$ & \ms             & $\mathcal{U}$[-5.0, 5.0] & $0.2^{+0.2}_{-0.2}$ \\[2 pt]
Quadratic ld coefficient       & $\text{q1}_{\text{TESS}}$ &                 & $\mathcal{N}$[0.33, $0.05^2$] & $0.32^{+0.05}_{-0.05}$ \\[2 pt]
Quadratic ld coefficient       & $\text{q2}_{\text{TESS}}$ &                 & $\mathcal{N}$[0.36, $0.05^2$] & $0.36^{+0.05}_{-0.05}$ \\[2 pt]
Quadratic ld coefficient       & $\text{q1}_{\text{CHEOPS}}$ &                 & $\mathcal{N}$[0.45, $0.05^2$] & $0.47^{+0.04}_{-0.04}$ \\[2 pt]
Quadratic ld coefficient       & $\text{q2}_{\text{CHEOPS}}$ &                 & $\mathcal{N}$[0.41, $0.05^2$] & $0.4^{+0.04}_{-0.05}$ \\[2 pt]
Stellar density                & $\rho$               & kg/$\text{m}^3$ & $\mathcal{N}$[966.0, $70.0^2$] & $959^{+68}_{-61}$ \\[2 pt]
dilution-\textit{TESS}                 &                      &                 & Fixed 1.0 &  \\[2 pt]
mflux-\textit{TESS}                     &                      &                 & $\mathcal{N}$[0.0, $0.01^2$] & $6^{+2}_{-2} 10^{-5}$\\[2 pt]
Offset-\textit{TESS}                    & $\phi_{\text{TESS}}$ &                 & $\mathcal{U}$[-0.001, 0.001] & $5^{+4}_{-4} 10^{-5}$\\[2 pt]
Gradient-\textit{TESS}                    & $\theta_{\text{TESS}}$ & $\text{d}^{-1}$                & $\mathcal{U}$[-0.001, 0.001] & $37^{+8}_{-8} 10^{-5}$\\[2 pt]
Scatter \textit{TESS}                   & $\sigma_{\text{TESS}}$ & ppm                & $\mathcal{LU}$[1.0, 500.0] & $273^{+12}_{-13}$ \\[2 pt]
dilution-{\it CHEOPS}               &                      &                 & Fixed 1.0 &  \\[2 pt]
mflux-{\it CHEOPS}                   &                      &                 & $\mathcal{N}$[0.0, $0.01^2$] & 
$7^{+3}_{-4} 10^{-6}$\\[2 pt]
Scatter {\it CHEOPS}                 & $\sigma_{\text{CHEOPS}}$ & ppm                 & $\mathcal{LU}$[1.0, 500.0] & $84^{+5}_{-5}$ \\[2 pt]
Scatter HARPS-N                & $\sigma_{\text{HARPS-N}}$ & \ms                & $\mathcal{LU}$[0.1, 10.0] & $2.5^{+0.2}_{-0.2}$ \\[4 pt]
\multicolumn{5}{c}{\emph{Derived parameters}}\\[2 pt]
\multicolumn{5}{l}{\textbf{Planet b}}\\[2 pt]
Radius &  $R_{\text{b}}$ & $\text{R}_{\oplus}$  && $2.78^{+0.05}_{-0.04}$\\[2 pt]
Impact parameter & b &  && $0.22^{+0.20}_{-0.15}$\\[2 pt]
Scaled semi-major axis & a/$R_{\ast}$ &  && $24.2^{+0.6}_{-0.5}$\\[2 pt]
Inclination &  $i_{\text{b}}$ & \text{deg}  && $89.5^{+0.4}_{-0.5}$\\[2 pt]
Transit duration &  $\text{T}_{\text{14}}$ & h  & & $5.42^{+0.03}_{-0.03}$  \\[2 pt]
Minimum mass & $m_{\text{b}}\sin(i_{\text{b}})$ & M$_{\oplus}$ &    &$8.1^{+1.4}_{-1.4} $  \\[2 pt]
Equilibrium temperature (black body) & $T_{\text{eq}}$ &  && $826^{+13}_{-13}$\\[2 pt]
\multicolumn{5}{l}{\textbf{Planet c}}\\[2 pt]
Minimum mass planet c& $m_{\text{c}}\sin(i_{\text{c}})$ & M$_{\oplus}$    &  & $10.4^{+1.9}_{-1.8} $   \\[2 pt]
Scaled semi-major axis & $\text{a}/\text{R}_\ast$ &&& $40.1^{+0.9}_{-0.9}$ \\ [0.2pt]
Equilibrium temperature (black body) &$\text{T}_{\text{eq}}$& K && $641^{+10}_{-10}$\\ [0.2pt]
\end{tabular}	
\end{table*}

\section{Full posteriors for internal structure modelling}

\begin{table*}
\renewcommand{\arraystretch}{1.5}
\caption{Results of the internal structure modelling for HD~85426~b (YARARA)}
\centering
\begin{tabular}{r|ccc|ccc}
\hline \hline
Water prior &              \multicolumn{3}{c|}{Formation outside iceline (water-rich)} & \multicolumn{3}{c}{Formation inside iceline (water-poor)} \\
Si/Mg/Fe prior &           Stellar (A1) &       Iron-enriched (A2) &      Free (A3) &
                           Stellar (B1) &       Iron-enriched (B2) &      Free (B3) \\
\hline
w$_\textrm{core}$ [\%] &        $11_{-7}^{+8}$ &    $15_{-10}^{+14}$ &    $13_{-9}^{+16}$ &
                           $16_{-11}^{+11}$ &    $20_{-14}^{+19}$ &    $17_{-13}^{+21}$ \\
w$_\textrm{mantle}$ [\%] &      $58_{-13}^{+17}$ &    $53_{-15}^{+19}$ &    $55_{-16}^{+20}$ &
                           $83_{-11}^{+11}$ &    $77_{-19}^{+14}$ &    $81_{-22}^{+13}$ \\
w$_\textrm{envelope}$ [\%] &    $29.7_{-18.4}^{+15.8}$ &    $29.5_{-18.4}^{+16.0}$ &    $28.7_{-18.2}^{+16.5}$ &
                           $1.8_{-0.2}^{+0.2}$ &    $2.2_{-0.5}^{+0.4}$ &    $2.1_{-0.6}^{+0.7}$ \\
\hline
Z$_\textrm{envelope}$ [\%] &        $83.6_{-17.5}^{+6.6}$ &    $82.1_{-19.1}^{+7.1}$ &    $82.5_{-18.6}^{+7.1}$ &
                           $0.5_{-0.2}^{+0.2}$ &    $0.5_{-0.2}^{+0.2}$ &    $0.5_{-0.2}^{+0.2}$ \\
\hline
x$_\textrm{Fe,core}$ [\%] &     $90.3_{-6.3}^{+6.6}$ &    $90.4_{-6.4}^{+6.5}$ &    $90.3_{-6.4}^{+6.5}$ &
                           $90.3_{-6.4}^{+6.5}$ &    $90.4_{-6.4}^{+6.5}$ &    $90.3_{-6.4}^{+6.5}$ \\
x$_\textrm{S,core}$ [\%] &      $9.7_{-6.6}^{+6.3}$ &    $9.6_{-6.5}^{+6.4}$ &    $9.7_{-6.5}^{+6.4}$ &
                           $9.7_{-6.5}^{+6.4}$ &    $9.6_{-6.5}^{+6.4}$ &    $9.7_{-6.5}^{+6.4}$ \\
\hline
x$_\textrm{Si,mantle}$ [\%] &   $39_{-5}^{+6}$ &    $35_{-9}^{+9}$ &    $33_{-23}^{+30}$ &
                           $39_{-5}^{+6}$ &    $35_{-9}^{+8}$ &    $35_{-24}^{+30}$ \\
x$_\textrm{Mg,mantle}$ [\%] &   $44_{-6}^{+7}$ &    $40_{-10}^{+9}$ &    $37_{-25}^{+31}$ &
                           $44_{-6}^{+7}$ &    $40_{-10}^{+9}$ &    $36_{-25}^{+30}$ \\
x$_\textrm{Fe,mantle}$ [\%] &   $16_{-10}^{+9}$ &    $25_{-16}^{+18}$ &    $21_{-15}^{+25}$ &
                           $16_{-10}^{+9}$ &    $24_{-16}^{+18}$ &    $20_{-15}^{+24}$ \\
\hline
\end{tabular}
\label{tab:intstruct_YARARA}
\end{table*}
\renewcommand{\arraystretch}{1.0}

\begin{table*}
\renewcommand{\arraystretch}{1.5}
\caption{Results of the internal structure modelling for HD~85426~b (TWEAKS)}
\centering
\begin{tabular}{r|ccc|ccc}
\hline \hline
Water prior &              \multicolumn{3}{c|}{Formation outside iceline (water-rich)} & \multicolumn{3}{c}{Formation inside iceline (water-poor)} \\
Si/Mg/Fe prior &           Stellar (A1) &       Iron-enriched (A2) &      Free (A3) &
                           Stellar (B1) &       Iron-enriched (B2) &      Free (B3) \\
\hline
w$_\textrm{core}$ [\%] &        $11_{-7}^{+8}$ &    $15_{-11}^{+15}$ &    $13_{-10}^{+16}$ &
                           $16_{-11}^{+11}$ &    $21_{-14}^{+19}$ &    $18_{-13}^{+22}$ \\
w$_\textrm{mantle}$ [\%] &      $59_{-14}^{+16}$ &    $53_{-15}^{+19}$ &    $55_{-17}^{+20}$ &
                           $83_{-11}^{+11}$ &    $77_{-19}^{+15}$ &    $80_{-22}^{+13}$ \\
w$_\textrm{envelope}$ [\%] &    $28.9_{-18.0}^{+16.0}$ &    $28.7_{-17.9}^{+16.3}$ &    $28.1_{-17.8}^{+16.6}$ &
                           $1.8_{-0.2}^{+0.2}$ &    $2.1_{-0.5}^{+0.4}$ &    $2.0_{-0.5}^{+0.6}$ \\
\hline
Z$_\textrm{envelope}$ [\%] &        $80.7_{-17.6}^{+6.7}$ &    $79.3_{-18.9}^{+7.2}$ &    $79.6_{-18.6}^{+7.2}$ &
                           $0.5_{-0.2}^{+0.2}$ &    $0.5_{-0.2}^{+0.2}$ &    $0.5_{-0.2}^{+0.2}$ \\
\hline
x$_\textrm{Fe,core}$ [\%] &     $90.3_{-6.4}^{+6.5}$ &    $90.3_{-6.4}^{+6.5}$ &    $90.3_{-6.4}^{+6.5}$ &
                           $90.3_{-6.4}^{+6.6}$ &    $90.3_{-6.4}^{+6.5}$ &    $90.3_{-6.4}^{+6.6}$ \\
x$_\textrm{S,core}$ [\%] &      $9.7_{-6.5}^{+6.4}$ &    $9.7_{-6.5}^{+6.4}$ &    $9.7_{-6.5}^{+6.4}$ &
                           $9.7_{-6.6}^{+6.4}$ &    $9.7_{-6.5}^{+6.4}$ &    $9.7_{-6.6}^{+6.4}$ \\
\hline
x$_\textrm{Si,mantle}$ [\%] &   $39_{-5}^{+6}$ &    $35_{-9}^{+9}$ &    $33_{-23}^{+30}$ &
                           $39_{-5}^{+6}$ &    $35_{-9}^{+9}$ &    $34_{-24}^{+29}$ \\
x$_\textrm{Mg,mantle}$ [\%] &   $44_{-6}^{+7}$ &    $40_{-10}^{+9}$ &    $37_{-25}^{+31}$ &
                           $44_{-6}^{+7}$ &    $40_{-10}^{+9}$ &    $36_{-25}^{+30}$ \\
x$_\textrm{Fe,mantle}$ [\%] &   $16_{-10}^{+9}$ &    $25_{-17}^{+18}$ &    $22_{-16}^{+24}$ &
                           $16_{-10}^{+9}$ &    $24_{-16}^{+18}$ &    $21_{-15}^{+24}$ \\
\hline
\end{tabular}
\label{tab:intstruct_TWEAKS}
\end{table*}
\renewcommand{\arraystretch}{1.0}


\bsp	
\label{lastpage}
\end{document}